\documentclass[a4paper,11pt]{article}
\pdfoutput=1 % if your are submitting a pdflatex (i.e. if you have
\usepackage{jheparxiv} % for details on the use of the package, please
\usepackage[T1]{fontenc} % if needed

\usepackage{Settings}
\usepackage{amssymb}
\usepackage{comment}
\usepackage{blkarray}% http://ctan.org/pkg/blkarray
\usepackage{todonotes}
\usepackage[colorlinks = true,
            linkcolor = blue,
            urlcolor  = blue,
            citecolor = blue,
            anchorcolor = blue]{hyperref}

\usepackage{blindtext}%
\usepackage{etoolbox}%
\let\oldbibliography\bibliography% Store \bibliography in \oldbibliography
\renewcommand{\bibliography}[1]{{%
  \let\chapter\section% Copy \section over \chapter
  \oldbibliography{#1}}}
\title{Bethe-Salpeter equation for classical gravitational bound states}

\author[1]{Tim Adamo} 
\author[2]{\& Riccardo Gonzo}

\affiliation[1]{School of Mathematics and Maxwell Institute for Mathematical Sciences \\ University of Edinburgh, EH9 3FD, UK}
\affiliation[2]{Higgs Centre for Theoretical Physics, School of Physics and Astronomy \\ University of Edinburgh, EH9 3FD, UK}

\emailAdd{t.adamo@ed.ac.uk}
\emailAdd{rgonzo@ed.ac.uk}

\abstract{The Bethe-Salpeter equation is a non-perturbative, relativistic and covariant description of two-body bound states. We derive the classical Bethe-Salpeter equation for two massive point particles (with or without spin) in a bound gravitational system. This is a recursion relation which involves two-massive-particle-irreducible diagrams in the space of classical amplitudes, defined by quotienting out by symmetrization over internal graviton exchanges. In this context, we observe that the leading eikonal approximation to two-body scattering arises directly from unitarity techniques with a coherent state of virtual gravitons. More generally, we solve the classical Bethe-Salpeter equation analytically at all orders by exponentiating the classical kernel in impact parameter space. We clarify the connection between this classical kernel and the Hamilton-Jacobi action, making manifest the analytic continuation between classical bound and scattering observables. Using explicit analytic resummations of classical (spinless and spinning) amplitudes in momentum space, we further explore the relation between poles with bound state energies and residues with bound state wavefunctions. Finally, we discuss a relativistic analogue of Sommerfeld enhancement which occurs for bound state cross sections.}

\begin{document}

\maketitle
\flushbottom

\section{Introduction}

An exciting new era of gravitational wave physics has just begun, opening a unique window on the binary dynamics of extremely massive compact objects like black holes and neutron stars. The need for new, precision-frontier calculations has in turn spurred new theoretical developments on the two-body problem in general relativity -- see~\cite{Bjerrum-Bohr:2022blt,Buonanno:2022pgc,Goldberger:2022ebt,Goldberger:2022rqf} for recent overviews. Scattering amplitudes-inspired techniques, rooted in the post-Minkowskian (PM) expansion, have proven to be a powerful tool for the study of the long-distance inspiral phase of gravitational two-body dynamics, pushing the state-of-the-art to 4PM order in the spinless case~\cite{Neill:2013wsa,Cheung:2018wkq,Bern:2019crd,Bern:2019nnu,Kalin:2020fhe,Jakobsen:2021smu,DiVecchia:2021bdo,Bjerrum-Bohr:2021din,Bjerrum-Bohr:2021wwt,Brandhuber:2021eyq,Bern:2021dqo,Bern:2021yeh,DiVecchia:2022piu,Dlapa:2022lmu} and to 2PM (and partially 3PM) order in the spinning case~\cite{Guevara:2018wpp,Guevara:2019fsj,Bern:2020buy,Chung:2020rrz,Liu:2021zxr,Jakobsen:2021lvp,Chen:2021kxt,Jakobsen:2022fcj,Menezes:2022tcs,Bern:2022kto,Aoude:2022trd,Aoude:2022thd,Jakobsen:2022zsx}.

While scattering amplitudes in quantum field theory (QFT) are naturally connected to the classical scattering problem for hyperbolic orbits, novel methods have been developed to provide a dictionary between `boundary' (i.e., scattering) and bound dynamics~\cite{Kalin:2019rwq,Kalin:2019inp,Cho:2021arx,Saketh:2021sri}. This is possible because the full classical dynamics is encoded in a set of differential equations, and the type of motion (scattering vs. bound) is entirely captured by the initial conditions used to solve these equations. Hence, in the absence of radiative modes the classical potential extracted from scattering amplitudes can be used to describe classical bound states through the Hamiltonian~\cite{Neill:2013wsa,Cheung:2018wkq,Cristofoli:2019neg}, or alternatively scattering observables like the scattering angle can be analytically continued to bound observables like periastron advance.  Such analytic continuation is a major recent achievement in the description of classical bound states, at least for eccentric orbits, and very much relies on the analytic properties of the functions appearing in the Hamilton-Jacobi action for the binary problem. 

\medskip

\begin{figure}[h]
\centering
\includegraphics[scale=0.8]{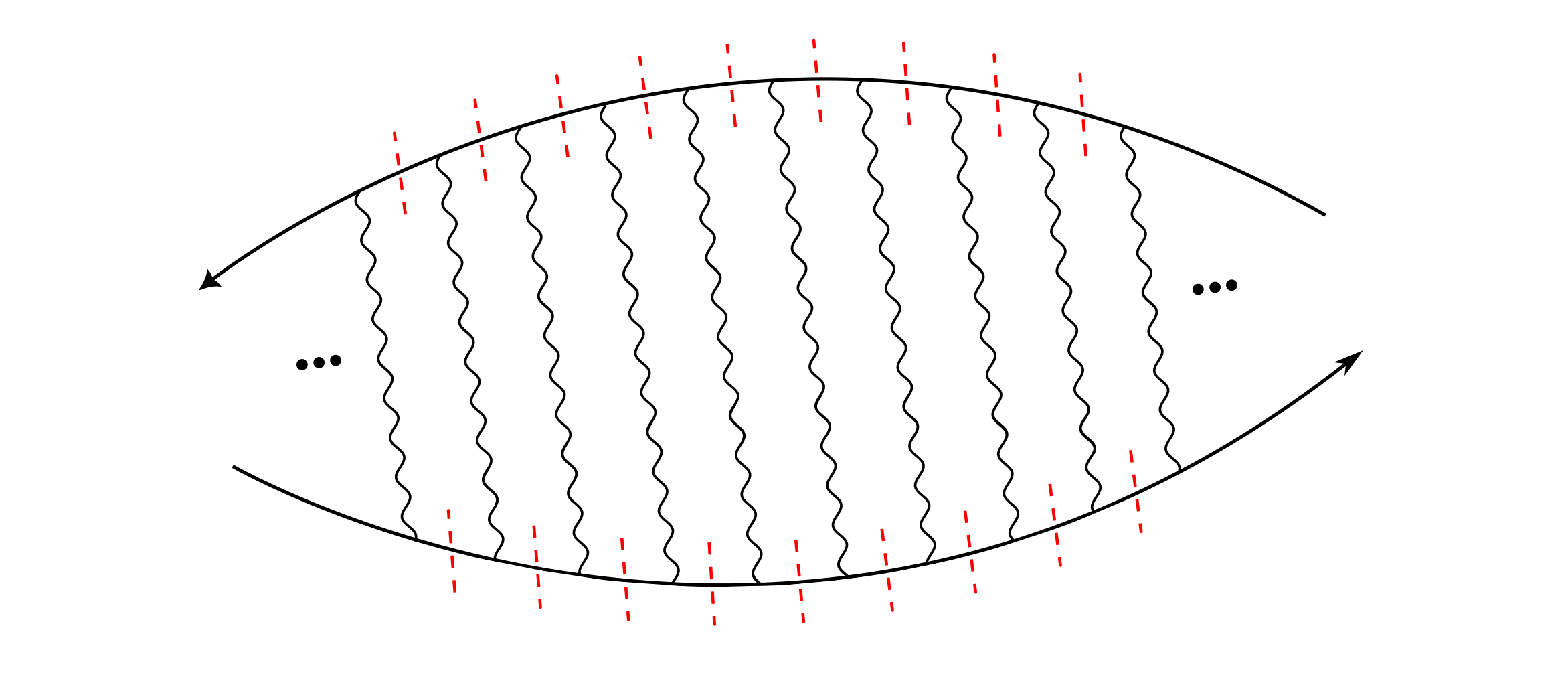}
\caption{A classical gravitational bound state arises from an infinite number of graviton exchanges in QFT, with massive external propagators approximately on-shell.}
\label{fig:boundstate}
\end{figure}

In this paper, we study the classical gravitational bound state two-body problem and its properties (like the binding energy or decay rate) directly from the analytic structure of scattering amplitudes in the classical limit. The Bethe-Salpeter equation~\cite{Salpeter:1951sz,Gell-Mann:1951ooy,Mandelstam:1955sd,Bethe:1957ncq} is a non-perturbative two-body recursion relation for relativistic bound states in QFT which, via the iteration of a suitable kernel, produces a pole in the resummed scattering amplitude corresponding to the energy of the bound state. Our aim is to revisit this equation, for both spinless and spinning gravitationally-coupled massive particles, and derive its classical counterpart using a heavy-particle effective field theory (HEFT) approach~\cite{Georgi:1990um,Manohar:2000dt,Goldberger:2004jt,Damgaard:2019lfh,Aoude:2020onz,Brandhuber:2021eyq}. This is a topic which has been considered many times in the past, often by using a ladder approximation in the Bethe-Salpeter equation~\cite{Wick:1954eu,Cutkosky:1954ru,Schwartz:1966dnp}, but this is not sufficient to establish a direct connection to classical bound states when relativistic effects are relevant\footnote{In particular, cross-ladder diagrams have to be included to develop a consistent relativistic expansion.}. Other approaches (e.g., making use of `generalized ladder' approximations) have gotten closer~\cite{Logunov:1963yc,Blankenbecler:1965gx,Gross:1969rv,Johnson:1971,Wallace:1989nm,Simonov:1993kp,Nieuwenhuis:1993gh,Nieuwenhuis:1996mc}, but all these attempts are at least partially incomplete due to lacking an appropriate classical limit at amplitude level~\cite{Bern:2019crd,Kosower:2018adc} or failing to recover probe limit results~\cite{tHooft:1987vrq,Kabat:1992tb,Cheung:2020gbf,Brandhuber:2021eyq,Kol:2021jjc,Bjerrum-Bohr:2021wwt,Adamo:2021rfq,Adamo:2022rob} at leading order in the resummation.

An important point that we clarify is how a classical gravitational field is represented at the quantum level through an infinite number of graviton exchanges, as shown in Figure~\ref{fig:boundstate}. At leading order, this corresponds to the interaction with a coherent state of virtual gravitons representing the classical field: symmetrization over these gravitons is a direct consequence of this manifestly classical description, unifying the treatment of ladder and crossed-ladder diagrams. Applying this principle to the Bethe-Salpeter equation leads to a new classical recursion relation, which at leading order is equivalent to the eikonal resummation but holds more generally in terms of the iteration of a classical kernel. The solution of this classical Bethe-Salpeter equation can then be written compactly as an exponential in impact parameter space in terms of two-massive-particle-irreducible diagrams, consistent with recently discovered exponential representations of the S-matrix~\cite{Brandhuber:2021eyq,Damgaard:2021ipf}.

The Hamilton-Jacobi action provides a natural description of the two-body dynamics, as emphasized in the effective one-body approach~\cite{Damour:1988mr,Buonanno:1998gg,Damour:1999cr,Damour:2008yg}, and it is natural that it should be connected with the solution of the classical Bethe-Salpeter equation. Indeed, we verify this and provide a direct connection with classical observables for bound orbits, such as the binding energy and the periastron advance, along the lines of~\cite{Kalin:2019rwq,Kalin:2019inp}. This in turn prompts us to understand how bound state properties are encoded in the analytic structure of the resummed classical amplitude in momentum space, such as bound state energy poles on the physical sheet. In particular, we generalize the early work of~\cite{Kabat:1992tb} on the spinless case to higher orders (i.e., to the next-to-leading resummation, or 2PM classical kernel) and extend the analysis of the leading-resummed spinning amplitude of~\cite{Adamo:2021rfq}. Along the way, we show how to recover the leading binding energy and two-body decay rate directly from the poles and residues of these amplitudes. 

Finally, we discuss a new effect of the resummation on the cross section for bound states, which generalizes \emph{Sommerfeld enhancement}~\cite{Sommerfeld:1931qaf} to the relativistic case. Sommerfeld enhancement is a quantum-mechanical effect whereby cross sections in an attractive potential are enhanced relative to the bare cross section, and can be described in terms of the non-relativistic limit of resummations of ladder diagrams~\cite{Iengo:2009ni,Cassel:2009wt}. In the relativistic context, the effect of leading resummation (which is equivalent to scattering one particle in the classical gravitational potential of the other) for scattering orbits is simply to dress the tree-level Born approximation by a phase, so there is no difference between the bare and resummed cross sections. However, for bound orbits we show that the effect of resummation on the potential is non-negligible. In the non-relativistic limit, this effect is an enhancement, in agreement with the Sommerfeld effect, but in the fully relativistic regime the resummation can either enhance or suppress the cross section depending on the kinematics.

The paper is organized as follows: Section~\ref{sec:eikonal} revisits the leading eikonal approximation, showing how the infinite sum of ladder and crossed-ladder diagrams emerges by describing the classical gravitational field generated by one particle as a coherent state of virtual gravitons. Section~\ref{sec:BSE} then begins with a brief review of the Bethe-Salpeter equation before discussing its classical limit for both spinless and spinning external particles. It is shown how to solve the resulting recursive equation by exponentiating a classical kernel in impact parameter space, and this is then connected with the description of bound state observables in Hamilton-Jacobi theory using the boundary-to-bound map. Section~\ref{sec:resummation} explores how bound state data is encoded in the analytic structure of the explicit resummed amplitudes at leading and next-to-leading resummation order in the spinless case and leading resummation in the spinning case. The relativistic analogue of the Sommerfeld effect is also described. Section~\ref{sec:concl} concludes, and Appendix~\ref{appendixA} provides details of the description of coherent states of virtual gravitons in terms of exponentiated three-point, partially off-shell scattering amplitudes.

\paragraph{Conventions}
We work in the mostly minus signature, using relativistic units $c=1$ and the setting $\kappa := \sqrt{32 \pi G_N}$ where $G_N$ is the Newton constant. For convenience, we employ the shorthand notation $\hat{\delta}(\cdot) := (2 \pi) \delta(\cdot)$ and $\hat{\mathrm{d}}^4 q := \mathrm{d}^4 q /(2 \pi)^4$.

%%%%%%%%%%%%%%%%%%%%%%%
%%%%%%%%%%%%%%%%%%%%%%%

\section{Revisiting the leading eikonal resummation}
\label{sec:eikonal}

Before beginning our exploration of classical gravitational bound states, we begin by revisiting a well-studied topic: the leading eikonal approximation in the two-body gravitational scattering problem. The eikonal approximation shares many features with the bound state problem, including an infinite number of graviton exchanges, and the fact that poles of eikonal amplitudes encode information about gravitational bound states~\cite{Abarbanel:1969ek,Levy:1970yn,Brezin:1970zr,Kabat:1992tb}. In the leading eikonal approximation, the infinite sum of ladder and crossed-ladder graviton exchange diagrams can be resummed into an exponential in impact parameter space, whose structure is fully determined by the probe scattering of one particle on a fixed classical background determined by the other~\cite{tHooft:1987vrq,Amati:1987wq,Amati:1990xe,Adamo:2021rfq}. 

While a more general connection of the eikonal S-matrix to classical gravitational observables has been established recently~\cite{DiVecchia:2022piu,DiVecchia:2021bdo,Cristofoli:2021jas}, here we would like to understand the origin of the class of diagrams arising in the eikonal approach. In particular, we provide a new interpretation of the leading eikonal approximation for the classical gravitational two-body problem which clarifies the role of symmetrization over the internal graviton exchanges and its connection to classical physics. To do this, we represent the classical gravitational field generated by one of the bodies as a coherent state of virtual gravitons; the infinite sum of diagrams contributing to the eikonal amplitude are then directly reproduced by using unitarity methods for amplitudes with coherent states. Later, this principle will play an important role in our formulation of the classical Bethe-Salpeter equation.

\medskip

The basic framework is the following: we consider general relativity as an effective field theory below the Planck scale, with two minimally coupled stable massive particles of mass $m_1$ and $m_2$. When these particles are scalars\footnote{The results in this section can be also extended in principle to color-charged particles, following \cite{delaCruz:2020bbn}.} (i.e., non-spinning), we use the action
\begin{align}
S[g,\Phi]= -\frac{2}{\kappa^2} \int \mathrm{~d}^{4} x\, \sqrt{|g|}\, R[g] + \sum_{j=1,2} \frac{1}{2} \int \mathrm{~d}^{4} x\, \sqrt{|g|}\left(g^{\mu \nu}\, \partial_{\mu} \phi_j \,\partial_{\nu} \phi_j-\frac{m_j^{2}}{\hbar^{2}}\phi_j^{2}\right)  \,,
\label{eqn:lagrangian_GRmatter}
\end{align}
where $R[g]$ is the Ricci scalar of the spacetime metric $g_{\mu \nu}$. To perform calculations, we expand in perturbation theory around the trivial QFT vacuum for both the massive fields and the gravitational field, defining $\kappa h_{\mu \nu} := g_{\mu \nu} - \eta_{\mu \nu}$ to be the graviton. For spinning particles, we will use a coupling to gravity which is consistent with a particle-like approximation of the Kerr metric in the classical limit, as in \cite{Bern:2020buy}. To take the classical limit of elastic 4-point amplitudes for $2\to2$ scattering of the massive particles, we use the heavy mass effective field theory (HEFT) of~\cite{Brandhuber:2021eyq,Damgaard:2019lfh,Aoude:2020onz}, as it will be relevant for the explicit resummation of spinless and spinning amplitudes.

We use the following notational conventions: we consider the conservative 4-point elastic amplitude with two massive particles 1 and 2 have incoming (resp. outgoing) momenta $p_1$, $p_2$ (resp. $p_1'$, $p_2'$). A convenient (symmetric) parametrization of the external kinematics for the classical limit is given by~\cite{Parra-Martinez:2020dzs}
\begin{align}
p_1^{\mu} := p_A^{\mu} + \frac{q^{\mu}}{2}\,, \qquad (p_1')^{\mu} := p_A^{\mu} - \frac{q^{\mu}}{2}\,,  \qquad p_1^2 = (p_1')^2 = m_1^2\,, \nonumber  \\
p_2^{\mu} := p_B^{\mu} - \frac{q^{\mu}}{2}\,, \qquad (p_2')^{\mu} := p_B^{\mu} + \frac{q^{\mu}}{2}\,,  \qquad p_2^2 = (p_2')^2 = m_2^2\,,
\label{eq:kinematics2-2}
\end{align}
where we have isolated the contribution of the momentum transfer $q^{\mu}=p_1^{\mu}-(p_1')^{\mu}$. It follows from \eqref{eq:kinematics2-2} that
\begin{align}
p_A \cdot q = p_B \cdot q = 0\,,
\label{eq:trasversality_pbarq}
\end{align}
which implies that $p_A$ and $p_B$ are strictly orthogonal to $q$. The Mandelstam variables are 
\begin{align}
s := (p_1 + p_2)^2 = P^2\,, \qquad t := (p_1 - p_1')^2 = q^2 = -|\vec{q}|^2\,.
\end{align}
With the parametrization in \eqref{eq:kinematics2-2}, we can now restore $\hbar$-dependence, as discussed in~\cite{Kosower:2018adc}, in order to take the classical limit $\hbar \to 0$. In particular, the momentum transfer is expressed in terms of the wavenumber
\begin{align}
\bar{q}^{\mu} := \frac{q^{\mu}}{\hbar}\,,   
\label{eq:wavenumber}
\end{align}
and the relativistic classical 4-velocities of the particles are defined as
\begin{align}
v_A^{\mu} &= \frac{p_A^{\mu}}{m_A}\,, v_B^{\mu} = \frac{p_B^{\mu}}{m_B}\,, \qquad v_A^2 = v_B^2 = 1\,, \qquad m_A^2 = m_1^2 - \frac{q^2}{4}\,,m_B^2 = m_2^2 - \frac{q^2}{4}\,.
\end{align}
A $n$-point, $L$-loop scattering amplitude will be denoted by $\mathcal{M}_n^{(L)}$. A ``classical'' amplitude, $\mathcal{M}_n^{(L),\text{cl}}$, denotes only those contributions from the scattering amplitude which survive in the $\hbar\to0$ limit\footnote{The classical expansion is equivalent to the eikonal expansion where $-t/s \ll1$, given that $-t = |\vec{q}|^2$.}. To denote the scattering amplitude where we keep all the dominant diagrams at leading order in the classical expansion, we use $\mathcal{M}_{n,\text{LR}}^{(L),\text{cl}}$, where ``LR'' stands for \emph{Leading Resummation}. Thus, the fully resummed amplitude in the classical expansion at leading order is
\begin{equation}\label{LRamp}
\mathcal{M}^{\text{cl}}_{n,\text{LR}}=\sum_{L=0}^{\infty}\mathcal{M}_{n,\text{LR}}^{(L),\text{cl}}\,,
\end{equation}
which includes all the superclassical terms at higher loop order. In analogy, next-to-leading, next-to-next-to-leading, etc., contributions are denoted by NLR, N$^2$LR, and so forth.

\subsection{Scalar eikonal amplitudes in the heavy-particle EFT}

\begin{figure}[h!]
\centering
\includegraphics[scale=0.8]{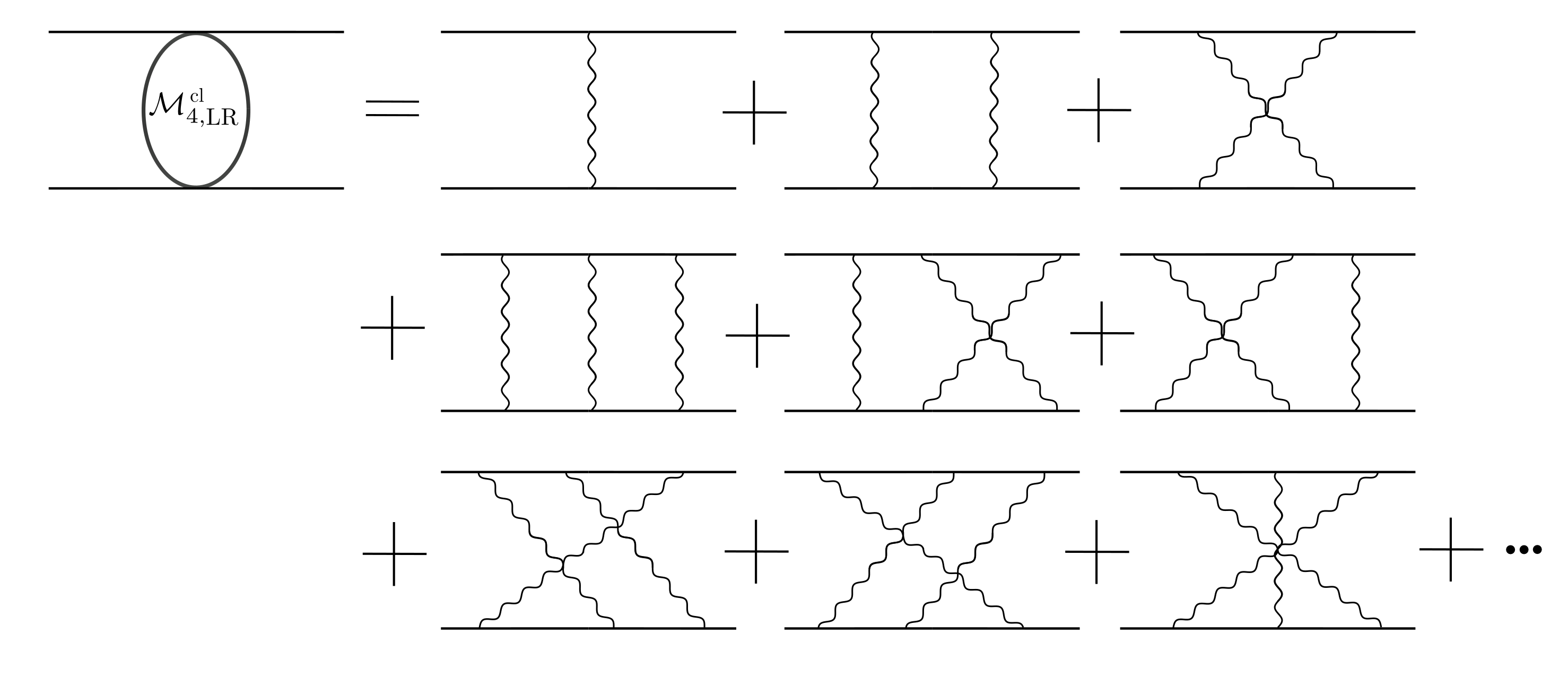}
\caption{The explicit set of diagrams relevant for the leading eikonal approximation.}
\label{fig:Leading_eikonal_diagrams}
\end{figure}

Let us begin with the leading eikonal approximation of the conservative 4-point gravitational scattering amplitude between two massive scalars. This amounts to the well-known resummation of ladder and crossed ladder diagrams, depicted in Fig.~\ref{fig:Leading_eikonal_diagrams} in the classical limit. 

The graviton-scalar vertex defined by \eqref{eqn:lagrangian_GRmatter} is
\begin{align}
V_{\mu \nu}(p,p')= i \frac{\kappa}{2} \left(p_{\mu} p'_{\nu}+p_{\nu} p'_{\mu}-\left(p \cdot p'-m^{2}\right) \eta_{\mu \nu}\right) \,,
\label{eq:gravmatter_3ptvertex}
\end{align}
and the graviton propagator in the de Donder gauge is
\begin{align}
\frac{P^{\mu\nu\alpha\beta}}{q^2+i\epsilon}\,, \qquad P^{\mu \nu \alpha \beta} &:= \frac{1}{2}\left[\eta^{\mu \alpha} \eta^{\nu \beta}+\eta^{\mu \beta} \eta^{\nu \alpha}-\frac{1}{2} \eta^{\mu \nu} \eta^{\alpha \beta}\right] \,.
\end{align}
The tree-level contribution to the leading eikonal amplitude is then given by the $t$-channel single graviton exchange
\begin{align}
i \mathcal{M}_{4,\text{LR}}^{(0)}(p_1,p_2;q)&= \frac{i}{q^2 + i \epsilon}\, V_{\mu \nu}(p_1,p_1')\, P^{\mu \nu \alpha \beta}\, V_{\alpha \beta}(p_2,p_2')\,,
\end{align}
which, in the classical limit, simplifies to 
\begin{align}
i \mathcal{M}_{4, \text{LR}}^{(0),\text{cl}}(p_A,p_B;q)&=  \frac{i}{q^2 + i \epsilon}\, V_{\mu \nu}(p_A)\, P^{\mu \nu \alpha \beta}\, V_{\alpha \beta}(p_B) \,, \qquad V_{\mu \nu}(p) := i\, \kappa\,p_{\mu}\, p_{\nu} \,.
\end{align}
This corresponds to the first term contributing to $\mathcal{M}^{\text{cl}}_{4,\text{LR}}$ in Figure~\ref{fig:Leading_eikonal_diagrams}.

It is convenient to define the impact parameter representation in the classical limit as
\begin{align}
\widetilde{\mathcal{M}}^{\text{cl}}_4(p_A,p_B;x_{\bot}) &:= \int \mathrm{\hat{d}}^4 q \,\hat{\delta}(2 p_A \cdot q) \hat{\delta}(2 p_B \cdot q) e^{ i \frac{q \cdot x_{\bot}}{\hbar}} \mathcal{M}_{4}^{\text{cl}}(p_A,p_B;q) \,, 
\end{align}
which is very natural in the HEFT approach as a consequence of \eqref{eq:trasversality_pbarq}. For the tree-level contribution to the leading eikonal amplitude, this gives
\begin{align}
i \widetilde{\mathcal{M}}_{4, \text{LR}}^{(0),\text{cl}}(p_A,p_B;x_{\bot}) &= \int \mathrm{\hat{d}}^4 q \, e^{i \frac{q \cdot x_{\bot}}{\hbar}}\, \widetilde{R}^{\alpha_1 \beta_1}(p_A,q) \left(\frac{i}{q^2 + i \epsilon}\, V_{\alpha_1 \beta_1}(p_B)\, \hat{\delta}(2 p_B \cdot q)\right)  \,,
\label{eq:eik-clas-final}
\end{align}
where the tensor
\begin{align}
\widetilde{R}^{\alpha \beta}(p_A,q) := P^{\mu \nu \alpha \beta}\, V_{\mu \nu}(p_A)\, \hat{\delta}(2 p_A \cdot q)\,,
\end{align}
contains all of the interactions of the massive particle line with classical momentum $p_A$. At one loop in the leading eikonal approximation, we take the sum over box and crossed-box diagrams in the $t$-channel, averaged over the choices of momenta in the internal graviton exchanges. While this averaging is often dismissed as a ``prescription'' (e.g., \cite{Kabat:1992tb}), we will soon see that it comes directly from requiring the classical gravitational field to be responsible for bound states\footnote{Indeed, in classical physics one cannot physically distinguish labelled diagrams with one internal graviton line configuration from another.}. Therefore, 
\begin{align}
i \mathcal{M}_{4, \text{LR}}^{(1),\text{cl}}(p_A,p_B;q)&= \frac{1}{2!} \int \mathrm{\hat{d}}^4 l_1 \int \mathrm{\hat{d}}^4 l_2 \hat{\delta}^4 (l_1 + l_2 - q) V_{\mu_1 \nu_1}(p_A) V_{\mu_2 \nu_2}(p_A) V_{\alpha_1 \beta_1}(p_B) V_{\alpha_2 \beta_2}(p_B)  \nonumber \\
& \times \left(P^{\mu_1 \nu_1 \alpha_1 \beta_1} P^{\mu_2 \nu_2 \alpha_2 \beta_2}\right) \frac{i}{l_1^2 + i \epsilon} \frac{i}{l_2^2 + i \epsilon} \nonumber \\
& \times \left[ \frac{i}{-2 l_1 \cdot p_A + i \epsilon} + \frac{i}{2 l_1 \cdot p_A + i \epsilon} \right] \left[ \frac{i}{-2 l_1 \cdot p_B + i \epsilon} + \frac{i}{2 l_1 \cdot p_B + i \epsilon} \right] \nonumber \\
&= \frac{1}{2!} \int \mathrm{\hat{d}}^4 l_1 \int \mathrm{\hat{d}}^4 l_2 \hat{\delta}^4 (l_1 + l_2 - q) V_{\mu_1 \nu_1}(p_A) V_{\mu_2 \nu_2}(p_A) V_{\alpha_1 \beta_1}(p_B) V_{\alpha_2 \beta_2}(p_B)  \nonumber \\
& \times \left(P^{\mu_1 \nu_1 \alpha_1 \beta_1} P^{\mu_2 \nu_2 \alpha_2 \beta_2}\right)  \frac{i}{l_1^2 + i \epsilon} \frac{i}{l_2^2 + i \epsilon} \hat{\delta} (2 l_1 \cdot p_A) \hat{\delta} (2 l_1 \cdot p_B)  \,.
\label{eq:1oop-eik}
\end{align}
Here, we pass to the second equality by combining the two diagrams into a single diagram where the massive propagators are cut, using the identity
\begin{equation}
\frac{i}{x+i\epsilon}-\frac{i}{x-i\epsilon}=\hat{\delta}(x)\,,
\end{equation}
which is a consequence of Sokhotsky's formula. This is expected from the fact that classical particles should be nearly on-shell. In \eqref{eq:1oop-eik}, the factor of $\frac{1}{2!}$ arises purely as a consequence of the averaging, but otherwise seems completely \emph{ad hoc}.

In impact parameter space, the one-loop contribution to the eikonal amplitude is
\begin{align}
i \widetilde{\mathcal{M}}_{4, \text{LR}}^{(1),\text{cl}}&(p_A,p_B;x_{\bot})= \int \mathrm{\hat{d}}^4 q \,  e^{ i \frac{q \cdot x_{\bot}}{\hbar}} \int \mathrm{\hat{d}}^4 l_1 \int \mathrm{\hat{d}}^4 l_2 \hat{\delta}^4 (l_1 + l_2 - q)  \widetilde{R}^{\alpha_1 \alpha_2 \beta_1 \beta_2}(p_A,l_1,q)  \nonumber \\
& \times \frac{1}{2!} \left(\frac{i}{l_1^2 + i \epsilon} V_{\alpha_1 \beta_1}(p_B) \hat{\delta} (2 l_1 \cdot p_B)\right) \left(\frac{i}{l_2^2 + i \epsilon} V_{\alpha_2 \beta_2}(p_B) \hat{\delta}(2 l_2 \cdot p_B) \right)  \,,
\label{eq:1oop-eik-clas-final}
\end{align}
where on the support of the delta functions $2 p_B \cdot q = -2 p_B \cdot l_2$ and we have defined
\begin{align}
\hspace{-15pt}\widetilde{R}^{\alpha_1 \alpha_2 \beta_1 \beta_2}(p_A,l_1,q) :=& (P^{\mu_1 \nu_1 \alpha_1 \beta_1} P^{\mu_2 \nu_2 \alpha_2 \beta_2} ) V_{\mu_1 \nu_1}(p_A) \hat{\delta} (2 l_1 \cdot p_A) V_{\mu_2 \nu_2}(p_A) \hat{\delta}(2 p_A \cdot q) \,.
\end{align}
It is now straightforward to generalize the calculation to the $(n-1)^{\text{th}}$ loop order, corresponding to the exchange of $n$ gravitons in the leading eikonal approximation:
\begin{align}
i &\widetilde{\mathcal{M}}_{4, \text{LR}}^{(n-1),\text{cl}}(p_A,p_B;x_{\bot})= \int \mathrm{\hat{d}}^4 q \,  e^{ i \frac{q \cdot x_{\bot}}{\hbar}} \left[ \prod_{i=1}^n \int \mathrm{\hat{d}}^4 l_i \right]  \hat{\delta}^4 \left(\sum_{i=1}^n l_i - q \right)   \nonumber \\
& \qquad \times  \widetilde{R}^{\alpha_1 \alpha_2 ... \alpha_n \beta_1 \beta_2 ... \beta_n}(p_A,l_1,...,l_{n-1},q) \frac{1}{n!} \prod_{i=1}^n\left(\frac{i}{l_i^2 + i \epsilon} V_{\alpha_i \beta_i}(p_B) \hat{\delta} (2 l_i \cdot p_B)\right)  \,,
\label{eq:n1oop-eik-clas-final}
\end{align}
where 
\begin{equation}
\widetilde{R}^{\alpha_1 \alpha_2\cdots\alpha_n \beta_1 \beta_2 \cdots\beta_n}(p_A,l_1,...,l_{n-1},q):=\prod_{i=1}^{n} P^{\mu_i\nu_i\alpha_i\beta_i}\,V_{\mu_i\nu_i}(p_A)\,\hat{\delta}(2l_{i-1}\cdot p_A)\,,
\end{equation}
for $l_{0}\equiv q$. The expression \eqref{eq:n1oop-eik-clas-final} calls for an explanation in terms of a resummed classical structure. As we will see, this is possible by defining a coherent state of virtual gravitons.

%%%%%%%%%%%%%%%%%%%

\subsection{Spinning eikonal amplitudes in the heavy-particle EFT}

This formulation of the eikonal resummation can be extended to describe external spinning particles of mass $m_1$, $m_2$ and spin $s_1$, $s_2$. We define the classical spin vector $a^\mu$ from the expectation value of the Pauli-Lubanski vector $\mathbb{W}^\mu$ on the one-particle massive state $\ket{p_j}$:
\begin{align}
s^\mu_j \equiv \bra{p_j} s^\mu \ket{p_j} =\frac{1}{m_j} \bra{p_j} \mathbb{W}^\mu \ket{p_j} = \frac{1}{2 m_j} \epsilon^{\mu}_{\,\,\,\nu \rho \sigma} p_j^\nu S^{\rho \sigma}(p_j)  \,,
\end{align}
such that finite classical spin vectors $a^\mu_A,a^\mu_B$ are obtained in the classical limit \cite{Maybee:2019jus}:
\begin{align}
\hbar \to 0\,, \qquad s_1,s_2 \to \infty\,, \qquad \hbar s^\mu_1 \to a^\mu_A \,, \qquad \hbar s^\mu_2 \to a^\mu_B \,. 
\end{align}
Going from single particle states to matrix elements between distinct (little group) Hilbert spaces for the external massive particles states requires a technical procedure, called ``Hilbert-space matching'', to extract the spin vector dependence from amplitudes~\cite{Chung:2019duq}. In particular, if we define the little group indices $I,\tilde{I}$ (resp. $J,\tilde{J}$) which run from $1$ to $2 s_1+1$ (resp. $1$ to $2 s_2+1$), then one expects in the classical limit~\cite{Bern:2020buy}
\begin{align}
\frac{\varepsilon_{I}^{*\sigma_A}(p_1) \varepsilon_{J}^{*\sigma_B}(p_2)\,i (\mathcal{M}_{4})^{I \tilde{I} J \tilde{J}}(p_1^{\sigma_A},p_2^{\sigma_B};q) \varepsilon_{\tilde{I}}^{\sigma_A}(p_1) \varepsilon_{\tilde{J}}^{\sigma_B}(p_2)}{\varepsilon^{*\sigma_A}(p_1) \cdot \varepsilon^{\sigma_A}(p_1) \varepsilon^{*\sigma_B}(p_2) \cdot \varepsilon^{\sigma_B}(p_2)} \stackrel{\hbar \to 0}{\to}  i \mathcal{M}_{4}^{\text{cl}}(\{p_A,a_A\},\{p_B,a_B\};q) \,,
\end{align}
where $\varepsilon_I^{\sigma_A}$, $\varepsilon_J^{\sigma_B}$ are the on-shell massive spinning polarization tensors, with $\sigma_A,\sigma_B$ running over the $2s_1+1$, $2s_2+1$ on-shell states.

In general, though, incoming and outgoing massive particle states have different momenta which are connected through a boost: there is a non-trivial Thomas-Wigner rotation which needs to be taken into account~\cite{Chung:2020rrz}. A nice feature of the HEFT framework is that the polarizations can be associated directly to the classical momenta $p_A,p_B$~\cite{Aoude:2020onz}, which, crucially, are the same for each massive particle line. Therefore, the classical spinning amplitude can be defined as
\begin{align}
i &\mathcal{M}_{4}^{\text{cl}}(\{p_A,a_A\},\{p_B,a_B\};q) \nonumber \\
&\qquad\qquad = \Big\langle\hspace{-3pt}\Big\langle \varepsilon_{I}^{*\sigma_A}(p_A) \varepsilon_{J}^{*\sigma_B}(p_B)\,i (\mathcal{M}_{4})^{I\tilde{I}J\tilde{J}}(p_A^{\sigma_A},p_B^{\sigma_B};q) \varepsilon_{\tilde{I}}^{\sigma_A}(p_A) \varepsilon_{\tilde{J}}^{\sigma_B}(p_B) \Big\rangle\hspace{-3pt}\Big\rangle\,,
\end{align}
where the double angle braket notation indicates the classical limit. The diagrams relevant for the leading eikonal approximation of this amplitude have the same topology as in Figure \ref{fig:Leading_eikonal_diagrams}, with the crucial difference that every vertex as well as the massive propagators carries an additional little-group structure associated with the spin of the massive particles. 

The four-point tree amplitude can be written as
\begin{align}
&i \mathcal{M}_{4, \text{LR}}^{(0),\text{cl}}(\{p_A,a_A\},\{p_B,a_B\};q) \nonumber \\
& \,\,\,\,\,= \Big\langle\hspace{-3pt}\Big\langle \varepsilon_{I}^{\sigma_A}(p_A) \varepsilon_{J}^{\sigma_B}(p_B) V^{I\tilde{I}}_{\mu \nu}(p_1,p_1',q) \frac{i P^{\mu \nu \alpha \beta}(q)}{q^2 + i \epsilon} V^{J\tilde{J}}_{\alpha \beta}(p_2,p_2',q)  \varepsilon_{\tilde{I}}^{*\sigma_A}(p_A) \varepsilon_{\tilde{J}}^{*\sigma_B}(p_B) \Big\rangle\hspace{-3pt}\Big\rangle \,,
\end{align}
for $V^{I\tilde{I}}_{\mu \nu}(p_1,p_1',q)$ the 3-point interaction vertex between the incoming and outgoing massive particle of spin $s$ and a single graviton. We are primarily interested in the cubic coupling between the massive particles and a graviton which is relevant to describe Kerr black holes\footnote{In principle we can describe general compact spinning objects by allowing arbitrary Wilson coefficients, but for the purpose of this paper we restrict to the simplest case.}, at least at leading order in the eikonal approximation. For this to be the case, we require that the vertex reproduces the linearized Kerr stress tensor~\cite{Vines:2017hyw}: this implies that in the classical limit
\begin{align}
\hspace{-7pt}\varepsilon_{I}^{\sigma_A}(p_A) & V^{I\tilde{I}}_{\mu \nu}(p_1,p_1',q) \varepsilon_{\tilde{I}}^{*\sigma_A}(p_A) \stackrel{\hbar \to 0}{\longrightarrow} V^{\text{Kerr}}_{\mu \nu}(\{p_A,a_A\};q)\,, \nonumber \\
\hspace{-7pt}V^{\text{Kerr}}_{\mu \nu}(\{p_A,a_A\};q) &= i \kappa \left[\cosh (a_A \cdot q) (p_{A})_{\mu} (p_{A})_{\nu}-\frac{i}{a_A \cdot q} \sinh (a_A \cdot q) q^\rho S_\rho^{(\mu}(p_A) p_A^{\nu)} \right]\,,
\label{Kerrvert}
\end{align}
and similarly for the other particle: $\varepsilon_{J}^{\sigma_B}(p_B)  V^{J\tilde{J}}_{\alpha \beta}(p_2,p_2') \varepsilon_{\tilde{J}}^{*\sigma_B}(p_B) \stackrel{\hbar \to 0}{\to} V^{\text{Kerr}}_{\alpha \beta}(\{p_B,a_B\};q)$. From now on, we assume that we are working with the vertex appropriate for Kerr black holes in the classical limit, and drop the related superscript. In impact parameter space, the associated four-point spinning amplitude for the leading eikonal approximation is
\begin{align}
i \widetilde{\mathcal{M}}_{4, \text{LR}}^{(0),\text{cl}}(\{p_A,a_A\},\{p_B,a_B\};x_{\bot}) &=  \int \mathrm{\hat{d}}^4 q \, e^{ i \frac{q \cdot x_{\bot}}{\hbar}} \widetilde{R}^{\alpha_1 \beta_1}(\{p_A,a_A\},q) \nonumber \\
&\quad \times\left(\frac{i}{q^2 + i \epsilon}\, V_{\alpha_1 \beta_1}(\{p_B,a_B\};q)\, \hat{\delta}(2 p_B \cdot q)\right) \,,
\end{align}
where $\widetilde{R}^{\alpha_1 \beta_1}(\{p_A,a_A\},q)$ is defined as
\begin{align}
\widetilde{R}^{\alpha_1 \beta_1}(\{p_A,a_A\},q) := P^{\mu_1 \nu_1 \alpha_1 \beta_1}\, V_{\mu_1 \nu_1}(\{p_A,a_A\};q)\, \hat{\delta}(2 p_A \cdot q) \,,
\end{align}
with the spinning vertex \eqref{Kerrvert}.

\begin{figure}[h]
\centering
\includegraphics[scale=1.05]{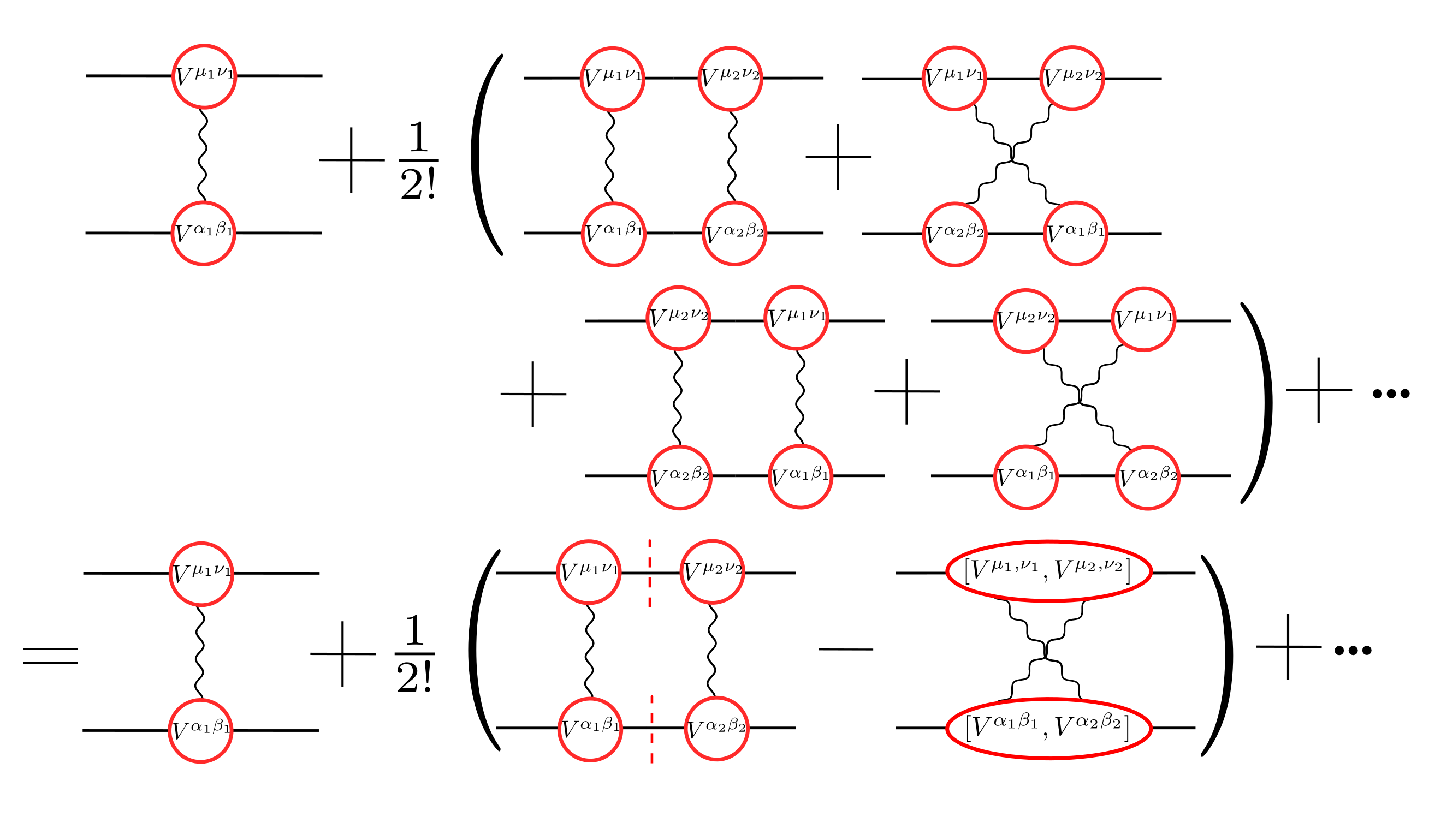}
\caption{The set of diagrams relevant for the leading eikonal approximation with spin shows that, compared to the scalar case, there are extra contributions related to the contraction of double commutators of the vertices. These terms are suppressed in the classical limit for Kerr black holes.}
\label{fig:Spin_exponentiation}
\end{figure}

The one-loop contribution is quite interesting, since the vertices for massive spinning lines contain \emph{non-commuting} spin operators in the classical limit. Classical physics dictates that we should sum over all possible permutations of the vertices, and if we want to rearrange the vertices in order to combine the linearized massive propagators as delta functions\footnote{This is crucial for writing the solution of the Bethe-Salpeter equation as an exponential, which is expected in the geometric optics limit for point particles, as we will see later in section \ref{sec:BSE}.} this implies that in principle there are additional contributions related to the commutator of the vertices. This is illustrated schematically in Figure~\ref{fig:Spin_exponentiation}. The one-loop spinning eikonal amplitude (in the leading resummation) is therefore
\begin{align}
\label{eq:1oop-eik-spin}
i \mathcal{M}_{4, \text{LR}}^{(1),\text{cl}}(\{p_A,a_A\},\{p_B,a_B\};q) &= \frac{1}{2!} \int \mathrm{\hat{d}}^4 l_1 \int \mathrm{\hat{d}}^4 l_2\, \hat{\delta}^4 (l_1 + l_2 - q)\, \frac{i\, P^{\mu_1 \nu_1 \alpha_1 \beta_1}}{l_1^2 + i \epsilon}\, \frac{i\,  P^{\mu_2 \nu_2 \alpha_2 \beta_2}}{l_2^2 + i \epsilon} \nonumber \\
& \times V_{\mu_1 \nu_1}(\{p_A,a_A\},l_1) \, V_{\mu_2 \nu_2}(\{p_A,a_A\},l_2) \, \hat{\delta} (2 l_1 \cdot p_A)  \nonumber \\
& \times V_{\alpha_1 \beta_1}(\{p_B,a_B\},l_1) \, V_{\alpha_2 \beta_2}(\{p_B,a_B\},l_2)  \hat{\delta} (2 l_1 \cdot p_B) \nonumber \\
& +\mathcal{C}^{(1)}_{\text{LR}}(\{p_A,a_A\},\{p_B,a_B\};q) \,.
\end{align}
where we have used the fact that\footnote{Given the completeness relation $N_{\tilde{I}_1 I_2}(p_A) = \sum_{\sigma_A' = 1, \dots, 2 s_1 + 1} \varepsilon_{\tilde{I}_1}^{* \sigma_A'}(p_A) \varepsilon_{I_2}^{\sigma_A'}(p_A)$, the product of matrices in the space of helicities becomes the product of spin-dependent operators in the classical limit.}
\begin{align}
&\Big\langle\hspace{-3pt}\Big\langle \varepsilon_{I_1}^{\sigma_A}(p_A)\, V^{I_1 \tilde{I}_1}_{\mu_1 \nu_1}(p_A,l_1)\, N_{\tilde{I}_1 I_2}(p_A)\, V^{I_2 \tilde{I}_2}_{\mu_2 \nu_2}(p_A,l_2)\, \varepsilon_{\tilde{I}_1}^{*\sigma_A}(p_A) \Big\rangle\hspace{-3pt}\Big\rangle \nonumber \\
&\qquad \qquad \qquad \qquad \qquad  \qquad \qquad = V_{\mu_1 \nu_1}(\{p_A,a_A\},l_1) \, V_{\mu_2 \nu_2}(\{p_A,a_A\},l_2)\,,
\end{align}
and
\begin{align}
\label{eq:1oop-spin-commutator}
\mathcal{C}^{(1)}_{\text{LR}}(\{p_A,& a_A\},\{p_B,a_B\};q) \nonumber \\
&= -\frac{1}{2!} \int \mathrm{\hat{d}}^4 l_1 \int \mathrm{\hat{d}}^4 l_2\, \hat{\delta}^4 (l_1 + l_2 - q)\, \frac{i\, P^{\mu_1 \nu_1 \alpha_1 \beta_1}}{l_1^2 + i \epsilon}\, \frac{i\, P^{\mu_2 \nu_2 \alpha_2 \beta_2}}{l_2^2 + i \epsilon} \nonumber \\
& \times \Big\langle\hspace{-3pt}\Big\langle  \frac{i \, \varepsilon_{I_1}^{\sigma_A}(p_A) \left[V^{I_1 \tilde{I}_1}_{\mu_1 \nu_1}(p_A,l_1)\, N_{\tilde{I}_1 I_2}(p_A),\, V^{I_2 \tilde{I}_2}_{\mu_2 \nu_2}(p_A,l_2)\right] \varepsilon_{\tilde{I}_2}^{*\sigma_A}(p_A)}{2 l_1 \cdot p_A + i \epsilon} \nonumber \\
& \qquad \times\,\frac{i \, \varepsilon_{J_1}^{\sigma_B}(p_B)  \left[V^{J_1 \tilde{J}_1}_{\alpha_1 \beta_1}(p_B,l_1)\, N_{\tilde{J}_1 J_2}(p_B),\, V^{J_2 \tilde{J}_2}_{\alpha_2 \beta_2}(p_B,l_2)\right] \varepsilon_{\tilde{J}_2}^{*\sigma_B}(p_B)}{-2 l_1 \cdot p_B + i \epsilon} \Big\rangle\hspace{-3pt}\Big\rangle \,,
\end{align}
denotes the contribution from commutators.

Note that there are two commutators present in \eqref{eq:1oop-spin-commutator}; the presence of this double commutator changes the classical scaling of this term relative to the first term of \eqref{eq:1oop-eik-spin}. In particular, the two commutators introduce an overall factor of $\hbar^2/(m_A\,m_B)$~\cite{Maybee:2019jus}, which cannot be compensated by the $1/\hbar$ factor coming from the one-loop integral~\cite{Bern:2020buy,Haddad:2021znf,Bellazzini:2022wzv}. Hence, upon taking the classical limit $\hbar\to0$ it follows that $\mathcal{C}^{(1)}_{\text{LR}}\to0$ and only the first contribution in \eqref{eq:1oop-eik-spin} survives. Note that for a generic cubic vertex (i.e., not corresponding to the Kerr metric, including arbitrary Wilson coefficients) this need not be the case: there can be commutator terms which survive the classical limit and in turn spoil eikonal exponentiation in impact parameter space\footnote{As we will discuss, the exponentiation is related to the Hamilton-Jacobi expansion, which is nothing but the geometric optics limit. For general finite size objects, such limit is not consistent~\cite{Chen:2022yxw}.}.

Restricting our attention to the Kerr case (where the commutator term has vanishing classical limit), the resulting one-loop contribution in impact parameter space is
\begin{align}
i &\widetilde{\mathcal{M}}_{4, \text{LR}}^{(1),\text{cl}}(\{p_A,a_A\},\{p_B,a_B\};x_{\bot}) \nonumber \\
&= \int \mathrm{\hat{d}}^4 q \,  e^{ i \frac{q \cdot x_{\bot}}{\hbar}} \int \mathrm{\hat{d}}^4 l_1 \int \mathrm{\hat{d}}^4 l_2\, \hat{\delta}^4 (l_1 + l_2 - q)\, \widetilde{R}^{\alpha_1 \alpha_2 \beta_1 \beta_2}(\{p_A,a_A\},l_1,q)  \nonumber \\
&\quad \times  \frac{1}{2!} \left(\frac{i}{l_1^2 + i \epsilon}\, V_{\alpha_1 \beta_1}(\{p_B,a_B\},l_1)\, \hat{\delta} (2 l_1 \cdot p_B)\right) \left(\frac{i}{l_2^2 + i \epsilon}\, V_{\alpha_2 \beta_2}(\{p_B,a_B\},l_2)\, \hat{\delta}(2 l_2\cdot p_B) \right) \,,
\label{eq:1oop-eik-clas-spin-final}
\end{align}
where we have defined
\begin{align}
&\widetilde{R}^{\alpha_1 \alpha_2 \beta_1 \beta_2}(\{p_A,a_A\},l_1,q) :=  \left(P^{\mu_1 \nu_1 \alpha_1 \beta_1}\, P^{\mu_2 \nu_2 \alpha_2 \beta_2}\right) \nonumber \\
&\qquad\qquad\qquad \times\, V_{\mu_1 \nu_1}(\{p_A,a_A\},l_1) \, \hat{\delta} (2 l_1 \cdot p_A)\, V_{\mu_2 \nu_2}(\{p_A,a_A\},l_2)\, \hat{\delta}(2q\cdot p_A) \,.
\end{align}
Proceeding in this way, the $(n-1)$-loop contribution to the leading eikonal approximation with spin can be written as:
\begin{align}
i \widetilde{\mathcal{M}}_{4, \text{LR}}^{(n-1),\text{cl}}(\{p_A,a_A\},\{p_B,a_B\}&;x_{\bot}) = \int \mathrm{\hat{d}}^4 q \,  e^{ i \frac{q \cdot x_{\bot}}{\hbar}}  \left[ \prod_{i=1}^n \int \mathrm{\hat{d}}^4 l_i \right] \hat{\delta}^4 \left(\sum_{i=1}^n l_i - q \right)\,  \nonumber \\
& \times \, \widetilde{R}^{\alpha_1 \alpha_2 ... \alpha_n \beta_1 \beta_2 ... \beta_n}(\{p_A,a_A\},l_1,...,l_{n-1},q) \nonumber \\
& \times \frac{1}{n!} \prod_{i=1}^n\left(\frac{i}{l_i^2 + i \epsilon}\, V_{\alpha_i \beta_i}(\{p_B,a_B\},l_i)\, \hat{\delta} (2 l_i \cdot p_B)\right) \,,
\label{eq:n1oop-eik-clas-spin-final}
\end{align}
where
\begin{align}\label{spinTensor}
\widetilde{R}^{\alpha_1 \alpha_2\cdots\alpha_n \beta_1 \beta_2\cdots\beta_n}&(\{p_A,a_A\},l_1,\ldots,l_{n-1},q)\nonumber \\
& \qquad = \prod_{i=1}^{n} P^{\mu_i \nu_i \alpha_i \beta_i} \,V_{\mu_i \nu_i}(\{p_A,a_A\},l_i)\, \hat{\delta} (2 l_{i-1} \cdot p_A)\,,
\end{align}
for $l_{0}\equiv q$. This expression gives the spinning generalization of \eqref{eq:n1oop-eik-clas-final}.

%%%%%%%%%%%%%%%%%%%%%%%%%

\subsection{Eikonal amplitudes from unitarity with coherent states}

In the previous sections, we found an interesting representation of the leading eikonal amplitude both for spinless \eqref{eq:n1oop-eik-clas-final} and spinning \eqref{eq:n1oop-eik-clas-spin-final} external massive particles. We now show that these results can be derived from a more physical classical perspective, where one massive particle moves in the classical gravitational field generated by the other, which is represented by a coherent state of gravitons. 

The classical on-shell graviton field in QFT is naturally represented by a coherent state of gravitons\footnote{We always work at zero temperature, like in the standard S-matrix formalism.}, as discussed in~\cite{Cristofoli:2021vyo,Britto:2021pud,Cristofoli:2021jas}. To provide a framework suitable to describing classical gravitational bound states, at leading order the virtual contribution to the graviton field between the massive bodies must be represented by an \emph{off-shell} analogue of the coherent state construction, as shown in~\cite{Cristofoli:2020hnk}. We review this here, referring the reader to appendix \ref{appendixA} for a full derivation from first principles. An on-shell coherent state of gravitons with definite helicity $\sigma$ is defined as 
\begin{align}
\ket{\alpha^{\sigma}} := \exp \left(\int \mathrm{d} \Phi(k) \, (\alpha^{\sigma}(k)\, a_{\sigma}^{\dagger}(k) - \alpha^{*\sigma}(k)\, a_{\sigma}(k)) \right) \ket{0} \,,
\label{eq:on-shell_coherent}
\end{align}
where $\alpha_{\sigma}(k)$ is the wave profile and $a_{\sigma}(k)$ (resp. $a_{\sigma}^{\dagger}(k)$) is the annihilation (resp. creation) operator appearing in the on-shell plane wave expansion of the linearized graviton field
\begin{align}
h_{\mu \nu}(x) = \sum_{\sigma = \pm} \frac{1}{\sqrt{\hbar}} \int \mathrm{d} \Phi(k) \left[\varepsilon^{* \sigma}_{\mu \nu}(k)\, a_{\sigma}(k)\, e^{-\frac{i k \cdot x}{\hbar}} + \varepsilon^{\sigma}_{\mu \nu}(k)\, a^{\dagger}_{\sigma}(k)\, e^{\frac{i k \cdot x}{\hbar}}\right] \,.
\label{eq:gravitonfield_onshell}
\end{align}
Here $\mathrm{d}\Phi(k):=\hat{\mathrm{d}}^4 k\,\hat{\delta}(k^2)\,\Theta(k^0)$ is the Lorentz-invariant on-shell measure and $\varepsilon^{\sigma}_{\mu\nu}(k)$ is the polarization tensor for the graviton field of helicity $\sigma$, on-shell with respect to $k^{\mu}$.

A plane wave expansion for an off-shell graviton field can now be defined as
\begin{align}
H_{\mu \nu}(x) = \sum_{\sigma' = 0,\pm} \frac{1}{\sqrt{\hbar}} \int \frac{\mathrm{\hat{d}}^4 K}{K^2 + i \epsilon} \left[\varepsilon^{* \sigma'}_{\mu \nu}(K)\, A_{\sigma'}(K)\, e^{-\frac{i K \cdot x}{\hbar}} + \varepsilon^{\sigma'}_{\mu \nu}(K)\, A^{\dagger}_{\sigma'}(K)\, e^{\frac{i K \cdot x}{\hbar}}\right] \,.
\label{eq:gravitonfield_offshell}
\end{align}
where $A^{\dagger}_{\sigma}(K)$ (resp. $A_{\sigma}(K)$) creates (resp. annihilates) a virtual graviton state with $K^2 \neq 0$, and $\varepsilon^{\sigma'}_{\mu \nu}(K)$ provides an appropriate polarization basis adapted to the momentum $K^{\mu}$. It is worth stressing that for external physical states, the longitudinal contribution to the basis of polarization vectors in \eqref{eq:virtualH} will eventually decoupled, allowing for a smooth on-shell limit. This formal expansion has a natural place in the Schwinger-Keldysh formulation, as discussed in appendix \ref{appendixA}, and has the advantage that once the graviton is on-shell we recover the standard operator $H_{\mu \nu}(x) \to h_{\mu \nu}(x)$ since
\begin{align}
\hspace{-15pt}\int \frac{\mathrm{\hat{d}}^4 K}{K^2 + i \epsilon} \to \int \mathrm{d} \Phi(k)  \,,\qquad   A_{\sigma}^{\dagger}(K) \to  a_{\sigma}^{\dagger}(k)  \,,\qquad  A_{\sigma}(K) \to  a_{\sigma}(k)\,.
\end{align}
Consequently, we use
\begin{align}
\exp \left(\int \frac{\mathrm{\hat{d}}^4 K}{K^2 + i \epsilon} \, (\alpha^{\sigma}(K) A_{\sigma}^{\dagger}(K) - \alpha^{*\sigma}(K) A_{\sigma}(K)) \right) \ket{0} \,.
\label{eq:off-shell_coherent}
\end{align}
to define a coherent state operator for virtual gravitons.

\begin{figure}[h!]
\centering
\includegraphics[scale=0.8]{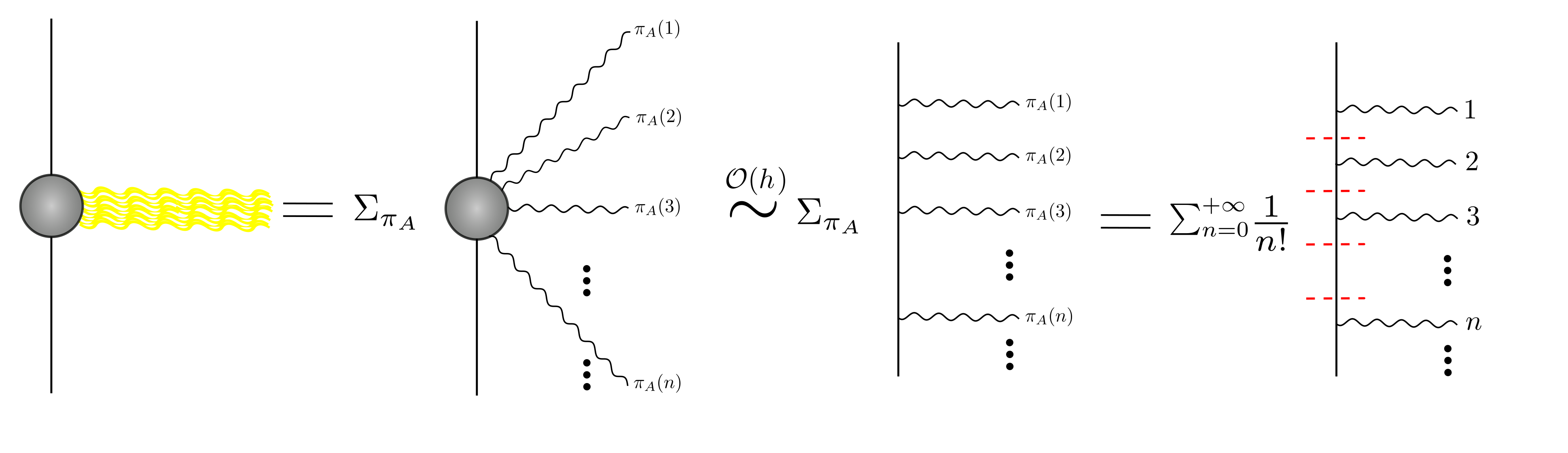}
\caption{The leading classical gravitational field is generated by the 3-point function in Lorentzian signature involving on-shell massive particle lines and off-shell gravitons, where the permutations over all virtual graviton emissions come directly from the virtual coherent state structure.}
\label{fig:Emission_coherent}
\end{figure}

At this point we can consider the quantum state generated by the motion of a massive point particle in general relativity, at leading order in the classical field. In appendix \ref{appendixA}, we review the derivation of such a state by generalizing the calculation in~\cite{Monteiro:2020plf} originally performed in $(2,2)$-signature. The result can be written, in the spinless case, as (see Figure~\ref{fig:Emission_coherent})
\begin{align}
\hspace{-12pt}|\psi^{\sigma}_{\text{LR}}\rangle=\frac{1}{\mathcal{N}} \int \mathrm{d} \Phi(p_1)\, \phi_{v_A}(p_1)\, \exp \left[ \int \frac{\mathrm{\hat{d}}^4 l}{l^2 + i \epsilon} \, \hat{\delta}(2 p_1 \cdot l)\, i \mathcal{M}^{(0),\text{cl}}_{3}(p_1,l^{\sigma}) A_{\sigma}^{\dagger}(l)\right]|p_1\rangle \,,
\label{eq:LR_state_offshell}
\end{align}
where $\mathcal{N}$ is a normalization factor, $\mathrm{d}\Phi(p_1):=\hat{\mathrm{d}}^4p_1\,\hat{\delta}(p_1^2-m_1^2)\,\Theta(p_1^0)$ is the on-shell measure for the massive particle, $\phi_{v_A}(p_1)$ is a wavefunction sharply peaked around the classical momentum $p_A$, and $\mathcal{M}^{(0),\text{cl}}_{3}(p_1,l^{\sigma})$ is the 3-point function of two on-shell massive scalar particles with one off-shell graviton:
\begin{align}
\mathcal{M}^{(0),\text{cl}}_{3}(p_1,l^{\sigma})=-\kappa\, \left(p_1 \cdot \varepsilon_{\sigma}(l)\right)^{2} = i\, V_{\mu \nu}(p_1)\, \varepsilon^{\mu \nu}_{\sigma}(l) \,.
\end{align}
Note that the momentum of the gravitational field is shared democratically between all graviton legs upon expanding the exponential in \eqref{eq:LR_state_offshell}, and symmetrization over the graviton emissions is required by the bosonic symmetry of the quanta of the classical field.

\medskip

Now, let us turn to the four-point leading eikonal amplitude; at $(n-1)$-loops, this can be computed by gluing together, in the spirit of unitarity, the contributions coming from $(n+2)$-point tree-level amplitudes with two massive particles and $n$ gravitons, as illustrated in Figure~\ref{fig:Emission_coherent}. 
\begin{figure}[h!]
\centering
\includegraphics[scale=1.1]{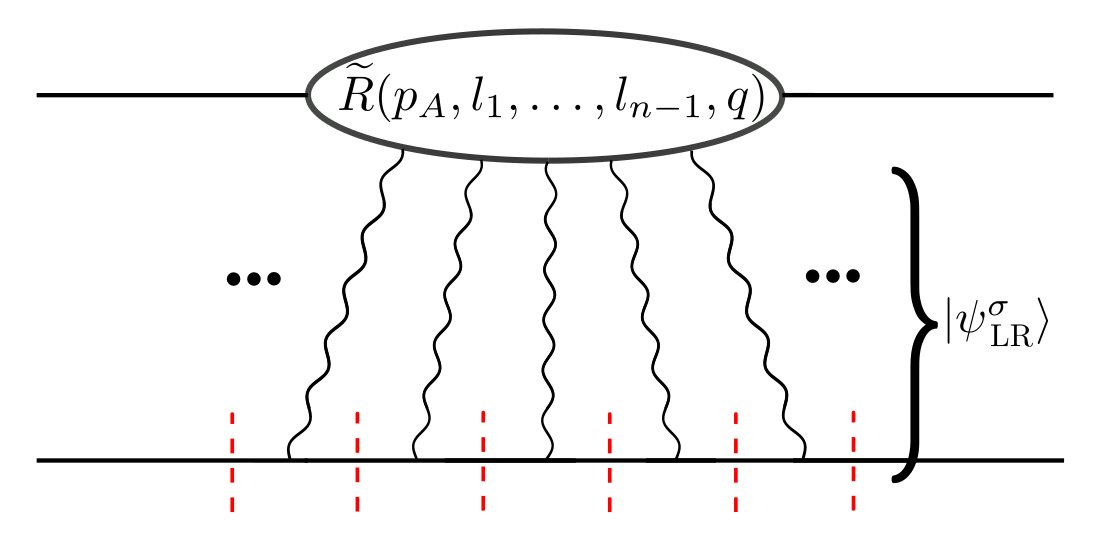}
\caption{The classical 4-point amplitude arises from gluing together off-shell $n$-point graviton emission from the massive particle lines: an infinite number of gravitons and permutation over the internal exchanges is required to have a classical gravitational bound state.}
\label{fig:Eikonal_wave}
\end{figure}
At leading order in the classical expansion, we saw that the eikonal amplitude for massive scalar particles can be written as \eqref{eq:1oop-eik-clas-final}:
\begin{align}
\hspace{-15pt}i \widetilde{\mathcal{M}}_{4, \text{LR}}^{(n-1),\text{cl}}&(p_A,p_B;x_{\bot})= \int \mathrm{\hat{d}}^4 q \,  e^{i \frac{q \cdot x_{\bot}}{\hbar}} \left[ \prod_{i=1}^n \int \mathrm{\hat{d}}^4 l_i \right]  \hat{\delta}^4 \left(\sum_{i=1}^n l_i - q \right)   \nonumber \\
& \times  \widetilde{R}^{\alpha_1 \alpha_2 ... \alpha_n \beta_1 \beta_2 ... \beta_n}(p_A,l_1,...,l_{n-1},q)\, \frac{1}{n!}\, \prod_{j=1}^n\left(\frac{i}{l_j^2 + i \epsilon} V_{\alpha_j \beta_j}(p_B) \hat{\delta} (2 l_j \cdot p_B)\right)  \,,
\end{align}
in impact parameter space. The tensor structure $\widetilde{R}_{\alpha_1 \alpha_2 ... \alpha_n \beta_1 \beta_2 ... \beta_n}(p_A,l_1,...,l_{n-1},q)$ can be equivalently written as
\begin{align}
\widetilde{R}_{\alpha_1 \alpha_2 ... \alpha_n \beta_1 \beta_2 ... \beta_n}(p_A,l_1,...,l_{n-1},q) &=  \left[\prod_{i=1}^{n}  V^{\mu_i \nu_i}(p_A) \hat{\delta} (2 l_i \cdot p_A) \right] \nonumber \\
& \qquad \times \sum_{\sigma_1,\dots,\sigma_n = \pm} \prod_{i=1}^n \left[\varepsilon^{*\sigma_i}_{\mu_i \nu_i}(l_i)\, \varepsilon^{\sigma_i}_{\alpha_i \beta_i}(l_i) \right] \,,
\end{align}
making it possible to rewrite \eqref{eq:1oop-eik-clas-final} as
\begin{align}
i \widetilde{\mathcal{M}}_{4, \text{LR}}^{(n),\text{cl}}(p_A,p_B;x_{\bot})&= \left[ \prod_{i=1}^n \int \mathrm{\hat{d}}^4 l_i \right]  \hat{\delta}^4 \left(\sum_{i=1}^n l_i - q \right)   \nonumber \\
& \times \sum_{\sigma_1,\dots,\sigma_n = \pm}  \left[\prod_{i=1}^{n} V^{\mu_i \nu_i}(p_A)\, \varepsilon^{*\sigma_i}_{\mu_i \nu_i}(l_i)\, \hat{\delta} (2 l_i \cdot p_A) \right] \nonumber \\
& \times\, \frac{1}{n!}\, \prod_{j=1}^n  \left(\frac{i}{l_{j}^2 + i \epsilon}\, V^{\alpha_{j} \beta_{j}}(p_B)\, \varepsilon^{\sigma_{j}}_{\alpha_{j} \beta_{j}}(l_{j})\, \hat{\delta} (2 l_{j} \cdot p_B)\right)  \,.
\end{align}
This representation can manifestly be obtained by gluing the classical tree-level $(n+2)$-point amplitude with massive particle $1$
\begin{align}
&\mathcal{M}^{(0),\text{cl}}_{n+2}(p_A,\{l_i^{\sigma_i}\}) \nonumber \\
&\qquad=\left\{\begin{array}{ll}
  V^{\mu_1 \nu_1}(p_A) \varepsilon^{*\sigma_1}_{\mu_1 \nu_1}(l_1) & \text{if}\quad n = 1\\
     \left[\prod_{i=1}^{n-1} V^{\mu_i \nu_i}(p_A) \varepsilon^{*\sigma_i}_{\mu_i \nu_i}(l_i) \hat{\delta} (2 l_i \cdot p_A) \right] V^{\mu_n \nu_n}(p_A) \varepsilon^{*\sigma_n}_{\mu_n \nu_n}(l_n)  & \text{if} \quad n \geq 1
                   \end{array}\right.\,,
%\begin{cases}
%  V^{\mu_1 \nu_1}(p_A) \varepsilon^{*\sigma_1}_{\mu_1 \nu_1}(l_1) & \text{if}\,\, n = 1\\
%     \left[\prod_{i=1}^{n-1} V^{\mu_i \nu_i}(p_A) \varepsilon^{*\sigma_i}_{\mu_i \nu_i}(l_i) \hat{\delta} (2 l_i \cdot p_A) \right] V^{\mu_n \nu_n}(p_A) \varepsilon^{*\sigma_n}_{\mu_n \nu_n}(l_n)  & \smash{\raisebox{-1.6ex}{if $n \geq 1$}}
%    \end{cases}  \,,
\end{align}
with the $n^{\mathrm{th}}$ term in the expansion of the state \eqref{eq:LR_state_offshell} containing the leading classical gravitational field of definite helicity produced by particle $2$: 
\begin{align}
\exp &\left[\int \frac{\mathrm{\hat{d}}^4 l}{l^2 + i \epsilon} \, \hat{\delta}(2 p_A \cdot l) i \mathcal{M}^{(0),\text{cl}}_{3}(p_A,l^{\sigma}) A_{\sigma}^{\dagger}(l)\right] \Bigg|_{n-th} \nonumber \\
&\qquad = \frac{1}{n!} \prod_{j=1}^n \left[\int \frac{\mathrm{\hat{d}}^4 l_{j}}{l_{j}^2 + i \epsilon} \, \hat{\delta}(2 p_A \cdot l_{j})\, i V^{\mu_{j} \nu_{j}}(p_A)\, \varepsilon_{\mu_{j} \nu_{j}}^{\sigma_{j}}(l_{j})\, A_{\sigma_{j}}^{\dagger}(l_{j})\right] \,.
\end{align}
Indeed, unitarity instructs us to sum over all possible internal states and because of the coherent state we must sum over all permutations of the internal graviton exchanges. 

An analogous calculation can be done for spinning external particles. Indeed, the quantum state generated by a massive spinning particle, provided that the commutator term \eqref{eq:1oop-spin-commutator} is suppressed (as it is for Kerr black holes), can be written as
\begin{align}
\hspace{-10pt}|\psi^{a_A,\sigma}_{\text{LR}}\rangle&=\frac{1}{\mathcal{N}}\, \int \mathrm{d} \Phi(p_1) \,\phi_{v_A}(p_1)\, \nonumber \\
& \times \exp \left[  \int \frac{\mathrm{\hat{d}}^4 l}{l^2 + i \epsilon} \, \hat{\delta}(2 p_1 \cdot l) i \mathcal{M}_{3}^{(0),\text{cl}}(\{p_1,a_A\},l^{\sigma}) A_{\sigma}^{\dagger}(l)  \right]|p_1\rangle \,,
\label{eq:LR_state_offshell-spin}
\end{align}
where the 3-point function with two on-shell massive spinning particles and one off-shell graviton is given in terms of the Kerr vertex \eqref{Kerrvert}
\begin{align}
\hspace{-12pt}\mathcal{M}_{3}^{(0),\text{cl}}(\{p_1,a_A\},l^{\sigma})= i V_{\mu \nu}(\{p_1,a_A\},l)\, \varepsilon^{\mu \nu}_{\sigma}(l) \,.
\label{eq:3pt_spinning}
\end{align}

\medskip

To summarize, at leading order in the classical expansion, the leading eikonal approximation is equivalent to considering the scattering process in terms of the classical motion of one point particle moving in the classical background field generated in the other via a coherent state of virtual gravitons. There is no need to talk about ladder and crossed-ladder diagrams: these are generated dynamically by unitarity from gluing amplitudes with coherent states of virtual gravitons. In particular, one should never be able to distinguish a single quanta which is part of the classical field, and this physical principle dictates the sum over all internal graviton configurations. In Section~\ref{sec:BSE}, we promote this principle to a rule for defining the space of classical amplitudes.

%%%%%%%%%%%%%%%%%%%%%%%%%%
%%%%%%%%%%%%%%%%%%%%%%%%%%

\section{Classical gravitational bound states from the Bethe-Salpeter equation}
\label{sec:BSE}

In this section, we review the Bethe-Salpeter equation in its original formulation~\cite{Salpeter:1951sz} and apply it to classical gravitational bound states. A bound state produces a pole in the S-matrix in the channel where it appears (for us, it will be the $s$-channel). This requires an infinite sum of Feynman diagrams, and is usually studied either numerically or in some approximation. However, in classical physics the dynamics can be solved directly from the equations of motion and in some cases it must be possible to analytically perform this infinite sum. Indeed, we find a new recursion relation for classical amplitudes, reminiscent of the eikonal resummation of Section \ref{sec:eikonal}, and solve it analytically in terms of the classical kernel by working in impact parameter space. As a byproduct, this provides a way to generate a covariant, crossing-symmetric Bethe-Salpeter equation for relativistic bound states.

%%%%%%%%%%%%%%%%%%%%%%%%%%

\subsection{The original Bethe-Salpeter equation}

\begin{figure}[h]
\centering
\includegraphics[scale=0.85]{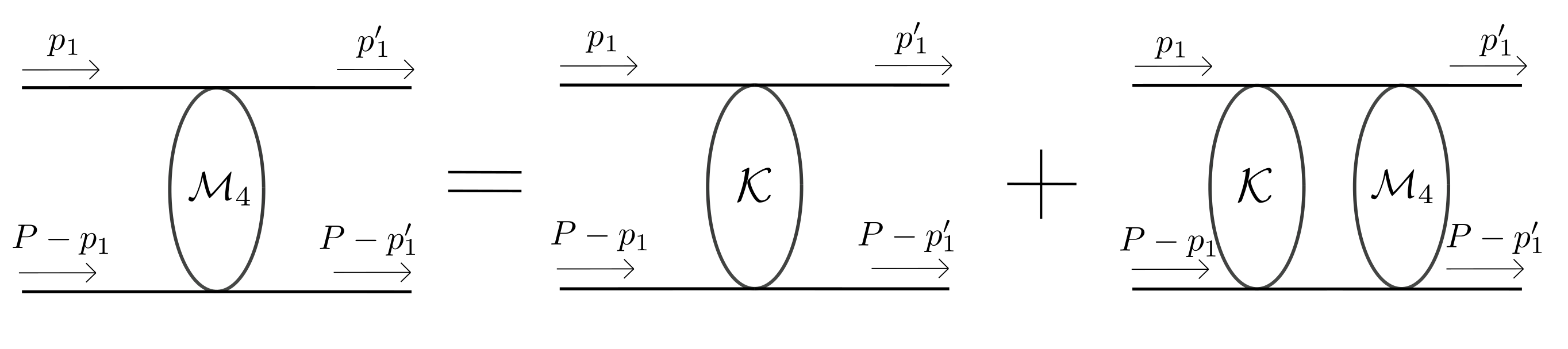}
\caption{The Bethe-Salpeter equation is a recursion relation for amplitudes written in terms of 2MPI diagrams, which are collected in the kernel $\mathcal{K}$.}
\label{fig:Bound_state_recursion}
\end{figure}

In quantum field theory, the Bethe-Salpeter equation is the covariant form of the bound state equation written in terms of amplitudes (see \cite{Itzykson:1980rh,Gross:1993zj,Hoyer:2014gna,Hoyer:2016aew,Hoyer:2021adf} for modern reviews). For external massive scalar fields and focusing on the elastic 4-point amplitude relevant for the two-body problem, the Bethe-Salpeter equation (in the $s$-channel) is a recursion relation\footnote{Similarly to the Lippmann-Schwinger equation ~\cite{Cristofoli:2019neg,Bjerrum-Bohr:2019kec}, this equation is derived from the LSZ reduction of the 4-point Green's function entering the Schwinger-Dyson equations. But in this work we use conventional (rather than ``old-fashioned'') perturbation theory, and note that in principle the Bethe-Salpeter recursion for the Green's function involves both scattering and bound states.}
\begin{align}
\mathcal{M}_4 (p_1,p'_1;P) = \mathcal{K}(p_1,p'_1;P) + \int \mathrm{\hat{d}}^4 l \, \mathcal{K}(p_1,l;P)\, G(l,P)\, \mathcal{M}_4 (l,p'_1;P) \,,
\label{eq:integral_eq}
\end{align}
where $\mathcal{M}_4$ is the 4-point scattering amplitude, $G$ is the two-massive particle propagator 
\begin{align}
G(l,P) = \frac{i}{(P - l)^2 - m_1^2 + i \epsilon}\, \frac{i}{l^2 - m_2^2 + i \epsilon}\,,
\end{align}
and $\mathcal{K}$ is the kernel of the equation which is made of two-massive particle irreducible (2MPI) diagrams. A diagrammatic representation of the equation is given in Figure~\ref{fig:Bound_state_recursion}. Provided the kernel can be treated perturbatively, \eqref{eq:integral_eq} can be solved iteratively 
\begin{align}
\mathcal{M}_4 (p_1,p'_1;P) = \mathcal{K}(p_1,p'_1;P) + \int \mathrm{\hat{d}}^4 l \, \mathcal{K}(p_1,l;P)\, G(l,P)\, \mathcal{K}(l,p'_1;P) + \cdots \,,
\label{eq:integral_eq_exp}
\end{align}
leading to an infinite (integral) series
\begin{align}
\mathcal{M}_4 =\sum_{n=0}^{\infty}(\mathcal{K}\,G)^n\,\mathcal{K}\,, \qquad  (\mathcal{K}\, G)^n\, \mathcal{K}:=\int\left(\prod_{i=1}^{n}\hat{\mathrm{d}}^4 l_{i}\,\mathcal{K}(p_1,l_i;P)\,G(l_i,P)\right)\mathcal{K}(l_n,p'_1;P)\,.
%\mathcal{K} + \mathcal{K} G \mathcal{K}  + \mathcal{K} G \mathcal{K} G \mathcal{K} + \dots +  (\mathcal{K} G)^n \mathcal{K} + \ldots \,,
\label{eq:integral_eq_formal}
\end{align}
At least formally, this can be re-summed to give 
\begin{align}
\mathcal{M}_4 = (1 -  \mathcal{K}\, G)^{-1}\, \mathcal{K} \,,
\label{eq:integral_eq_formal2}
\end{align}
as a geometric series.

\begin{figure}[h]
\centering
\includegraphics[scale=0.85]{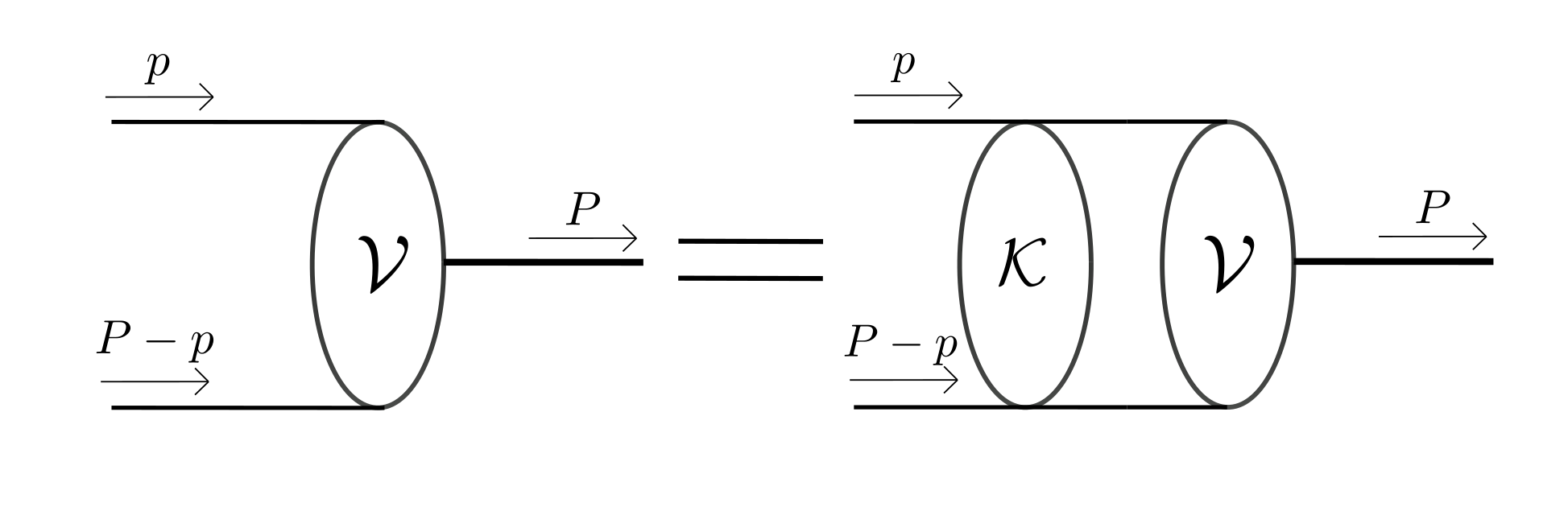}
\caption{Diagrammatic expression of the vertex function, encoding the bound state wavefunction.}
\label{fig:Bethe_salpeter_vertex}
\end{figure}

For there to be a bound state, the solution for $\mathcal{M}_4$ should have a pole in the $s$-channel; how do we phrase this condition in terms of the formal solution \eqref{eq:integral_eq_formal2}? A useful analogy is given by a function of a single complex variable with the geometric series expansion:
\begin{align}
f(z) = 1 + z + z^2 + \dots + z^n + \cdots = \sum_{n=0}^{\infty} z^n \quad \rightarrow \quad f(z) = \frac{1}{1-z} \,,
\end{align}
and the resummation shows that $f(z)$ has a simple pole at $z = 1$. The analogous condition for the formal integral series in \eqref{eq:integral_eq_formal2} is to require the functional $\mathcal{K}\,G$ to have an eigenvector with eigenvalue 1; namely, there should exist a vertex function $\mathcal{V}(p_1,P)$ with (see Figure~\ref{fig:Bethe_salpeter_vertex})
\begin{align}
\int \mathrm{\hat{d}}^4 l \, \mathcal{K}(p_1,l;P)\, G(l,P)\, \mathcal{V}(l,P)= \mathcal{V}(p_1,P) \,.
\label{eq:Bethe-Salpetereq}
\end{align}

\begin{figure}[h]
\centering
\includegraphics[scale=0.85]{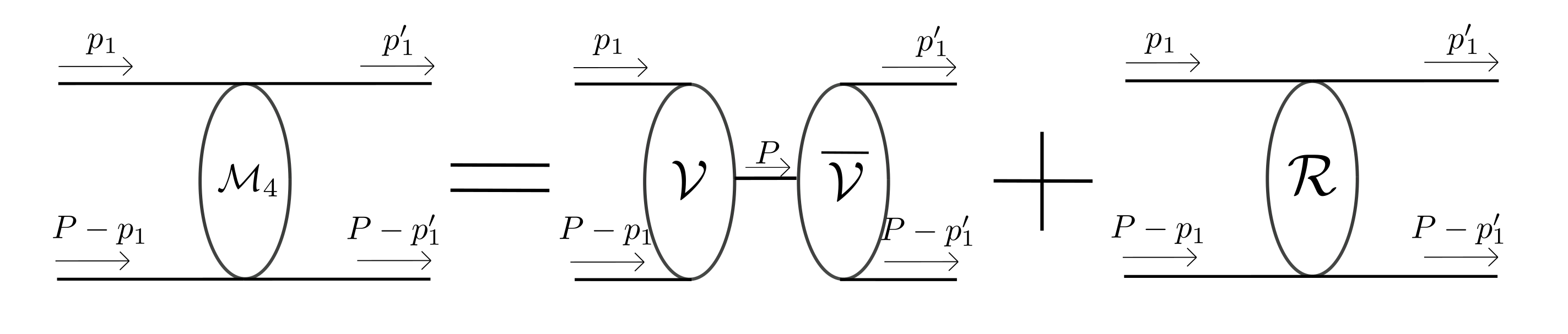}
\caption{The energy of a bound state appear in the S-matrix as a pole in the $s$-channel.}
\label{fig:schannel_factorization}
\end{figure}

This shows that \eqref{eq:Bethe-Salpetereq} is a sufficient condition to have a bound state, and we can also prove that it is necessary. Assume that a bound state exists in the $s$-channel; that is, there exists a factorization of the 4-point amplitude of the form (see Figure~\ref{fig:schannel_factorization})
\begin{align}
\mathcal{M}_4(p_1,p'_1;P) = \mathcal{V}(p_1,P)\, \frac{i}{s-M^2_{\text{bound}} + i \epsilon}\, \bar{\mathcal{V}}(p'_1,P) + R(p_1,p'_1;P) \,,
\label{eq:pole_factorization}
\end{align}
where $R(p_1,p'_1;P)$ is a remainder term which has no pole at $s= M^2_{\text{bound}}$. Clearly, $\mathcal{V}(p_1,P)$ is not defined away from the pole, but all we need is the vertex function evaluated at $P^2 = s = M^2_{\text{bound}}$. 
To prove the condition, one plugs \eqref{eq:pole_factorization} into \eqref{eq:integral_eq}: after extracting the residue at $P^2 = M^2_{\text{bound}}$, one obtains
\begin{align}
\hspace{-10pt}\left[ \mathcal{V}(p_1,P)\, \bar{\mathcal{V}}(p'_1,P)  \right]\Big|_{s=M^2_{\text{bound}}} &= \left[\int \mathrm{\hat{d}}^4 l \, \mathcal{K}(p_1,l;P)\, G(l,P)\, \mathcal{V}(l,P)\, \bar{\mathcal{V}}(p'_1,P) \right]\Big|_{s=M^2_{\text{bound}}} \,,
\label{eq:necessary_cond2}
\end{align}
which reduces to \eqref{eq:Bethe-Salpetereq} after eliminating $\bar{\mathcal{V}}(p'_1,P)$ on both sides of \eqref{eq:necessary_cond2}. 

This shows that the Bethe-Salpeter equation \eqref{eq:Bethe-Salpetereq} is needed to understand bound states in QFT~\cite{Nakanishi:1969ph,Sazdjian:1986aw,Gross:1993zj}. The bound state wavefunction can be defined from the vertex function 
\begin{align}
\Psi_{\text{bound}}(p_1,P) = C\, G(p_1,P)\, \mathcal{V}(p_1,P) \,,
\end{align}
and observables can then be computed from it\footnote{Often, it is important to appropriately fix the normalization $C$, see~\cite{Cutkosky:1964zz,Nakanishi:1969ph,Gross:1993zj} for more details. This is not relevant for our concerns in this paper.}.

%%%%%%%%%%%%%%%%%%%%

\subsection{The classical Bethe-Salpeter equation for scalar particles}

The crucial question is: what is the classical limit of the Bethe-Salpeter equation?

\begin{figure}[h]
\centering
\includegraphics[scale=0.85]{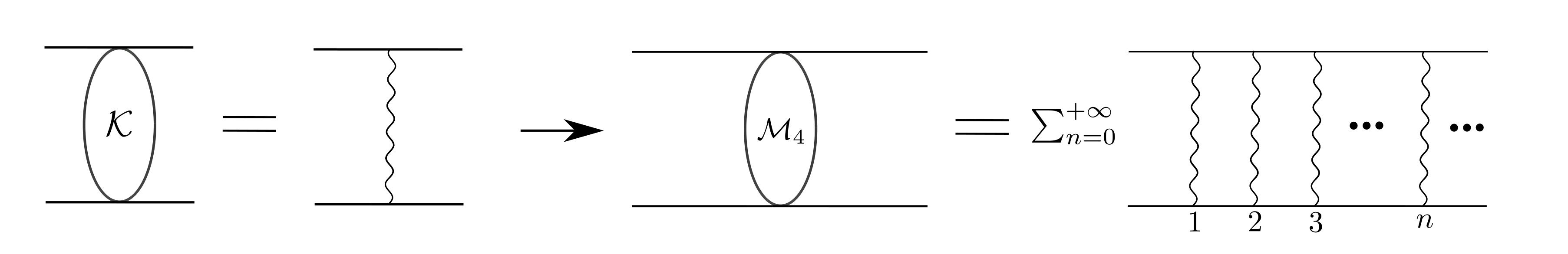}
\caption{Ladder diagrams are generated by iteration of the tree-level exchange.}
\label{fig:Ladder_diagrams}
\end{figure}

\begin{figure}[h]
\centering
\includegraphics[scale=0.85]{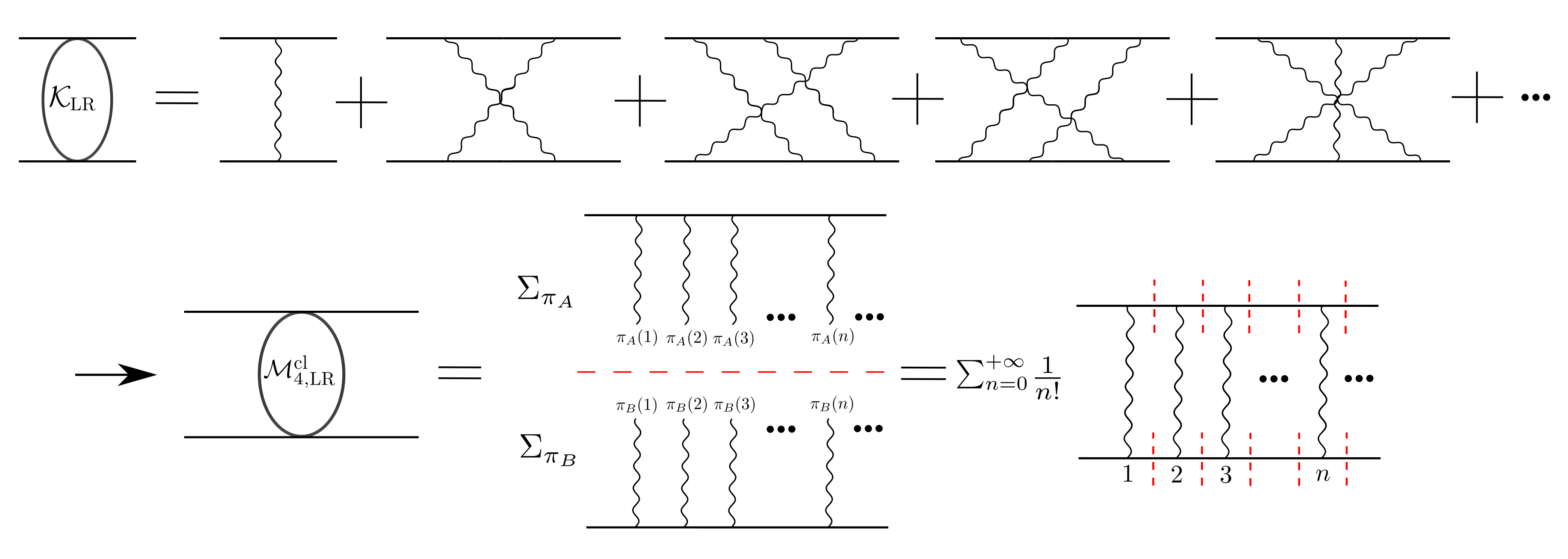}
\caption{The leading eikonal amplitude is generated by a kernel which includes all 2MPI diagrams generated by all possible internal permutations of ladder-type diagrams. This means that at leading order in the classical field, all the crossed-ladder diagrams must be included in the kernel.}
\label{fig:Eikonal_BSE}
\end{figure}

It is easy to see that the leading approximation with a single tree-level graviton exchange for the kernel $\mathcal{K}$ of the Bethe-Salpeter equation generates only the ladder diagrams, as shown in Figure~\ref{fig:Ladder_diagrams}.
However, this misses all the crossed-ladder diagrams which must be included in the classical relativistic approach: the problem is that the kernel has to include all the 2MPI diagrams, and the crossed-ladders are irreducible from this perspectives. Therefore, an infinite sum of crossed-ladder diagrams, generated by permutations over the internal graviton legs of ladder diagrams, must be included -- this is illustrated in Figure~\ref{fig:Eikonal_BSE}. 

The complexity of the kernel for the leading classical expansion is apparent: after summing over all the permutations, we are left with only one non-trivial building block which is again a tree-level amplitude. Indeed, the result \eqref{eq:1oop-eik-clas-final} for the $(n-1)^{\mathrm{th}}$-loop contribution to the classical eikonal amplitude can be rewritten in momentum space as
\begin{align}
i \mathcal{M}_{4, \text{LR}}^{(n-1),\mathrm{cl}}(p_A,p_B;q)&= \frac{1}{n!} \left[ \prod_{i=1}^n \int \mathrm{\hat{d}}^4 l_i \right]  \hat{\delta}^4 \left(\sum_{i=1}^n l_i - q \right) V_{\mu_n \nu_n}(p_A)\,  \frac{i\, P^{\mu_n \nu_n \alpha_n \beta_n}}{l_n^2 + i \epsilon}\, V_{\alpha_n \beta_n}(p_B)  \nonumber \\
& \times  \prod_{i=1}^{n-1} \left[V_{\mu_i \nu_i}(p_A)\,  \frac{i\, P^{\mu_i \nu_i \alpha_i \beta_i}}{l_i^2 + i \epsilon}\, V_{\alpha_i \beta_i}(p_B)\, \hat{\delta} (2 l_i \cdot p_B)\, \hat{\delta} (2 l_i \cdot p_A) \right]   \,,
\label{eq:n1oop-eik-clas-BS}
\end{align}
which is highly reminiscent of a new recursion relation with the new kernel
\begin{align}
\mathcal{K}_{\text{cl}}^{(0)}(p_A,p_B,q) := V_{\mu \nu}(p_A)  \frac{i P^{\mu \nu \alpha \beta}}{l^2 + i \epsilon} V_{\alpha \beta}(p_B)\,, 
\label{eq:kernelcl_eik}
\end{align}
and the two-body propagator replaced by its on-shell version
\begin{align}
G_{\text{cl}}(p_A,p_B,l) := \hat{\delta} (2 l \cdot p_A) \hat{\delta} (2 l \cdot p_B) \,.
\label{eq:twobodycl_eik}
\end{align}
Our immediate goal is to make this precise.

To do this, we define the space of classical, conservative 4-point diagrams
\begin{align}
\mathcal{H}_{4, \text{cl}} := \mathcal{H}_{4} / \Sigma\,, 
\end{align}
as the quotient space of Feynman diagrams contributing to conservative 4-point amplitudes, $\mathcal{H}_{4}$, up to the permutation group $\Sigma$ of all graviton exchanges between a fixed set of interaction vertices along the matter line\footnote{Mathematically, for every set of vertices (interactions) and ``internal'' edges (internal graviton exchanges) we can form a permutation graph. In classical physics we cannot distinguish between such graphs, so we need to average over them. An analogy would be, for unpolarized scattering observables (like the cross-section), the need to average over all helicities of the external states.}
\begin{align}
\varrho(\mathcal{D}_4) = \mathcal{D}_4' \quad &\varrho \in \Sigma\,,
\end{align}
for any two $\mathcal{D}_4,\mathcal{D}_4'\in\mathcal{H}_4$. For instance, at 1-loop the ladder and crossed ladder diagrams are identified by a single graph in the space $\mathcal{H}_{4,\text{cl}}$. More generally, the entire class of ladder and crossed-ladder diagrams (to any loop order) are generated by a single, tree-level building block in the quotient space $\mathcal{H}_{4, \text{cl}}$: namely, the single graviton exchange diagram.

\begin{figure}[h]
\centering
\includegraphics[scale=0.85]{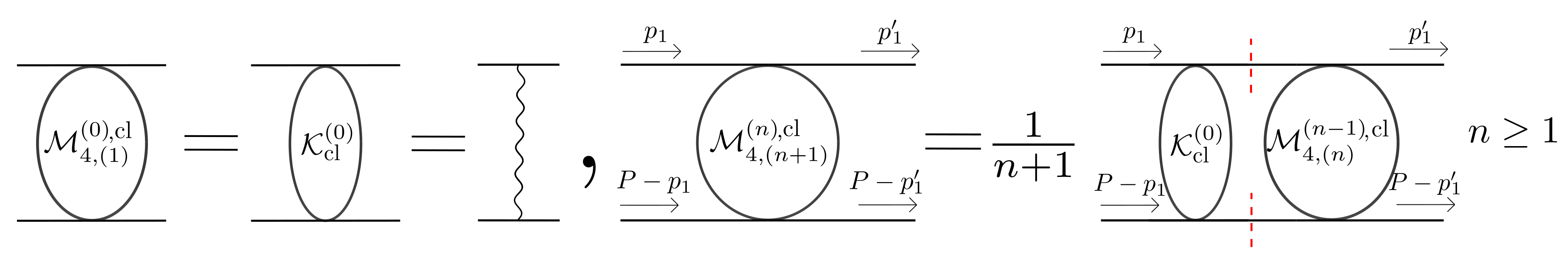}
\caption{The leading eikonal 4-point amplitude is generated by the tree-level recursion relation.}
\label{fig:recursion_tree}
\end{figure}

It is now possible to recast the leading eikonal amplitude as a solution of a recursion relation in the space $\mathcal{H}_{4, \text{cl}}$, in the spirit of the Bethe-Salpeter equation. A subscript $(n)$ on an amplitude will denote the number of classical 2MPI diagrams, so that $\mathcal{M}^{(L),\text{cl}}_{4,(n)}$ is a $L$-loop, 4-point amplitude composed of $n$ classical 2MPI diagrams. We can then rewrite \eqref{eq:n1oop-eik-clas-BS} as a recursion relation\footnote{Only for the leading eikonal amplitude does the number of loops $L$ equal the number of classical 2MPI diagrams (i.e. tree-level exchange diagrams) plus one.}
\begin{align}
\hspace{-10pt}&\mathcal{M}^{(n),\text{cl}}_{4,(n+1)} (p_A,p_B,q) \nonumber \\
\hspace{-10pt}&\quad\quad= \begin{cases}
      \mathcal{K}^{(0)}_{\text{cl}}(p_A,p_B,q) & \text{if}\,\, n = 0\\
      \frac{1}{n+1} \int \mathrm{\hat{d}}^4 l \, \mathcal{K}^{(0)}_{\text{cl}}(p_A,p_B,l) G_{\text{cl}}(p_A,p_B,l) \mathcal{M}^{(n-1),\text{cl}}_{4,(n)} (p_A,p_B,q-l) & \text{if} \,\,n \geq 1\\
    \end{cases}  \,,
\label{eq:integral_eq_cl_tree}
\end{align}
where $\mathcal{K}^{(0)}_{\text{cl}}$ and $G_{\text{cl}}(p_A,p_B,l)$ are defined by \eqref{eq:kernelcl_eik} and \eqref{eq:twobodycl_eik}, respectively. This tree-level recursion relation is represented in Figure~\ref{fig:recursion_tree}, which has the leading classical eikonal 4-point amplitude as a solution. We see that the new classical kernel $\mathcal{K}^{(0)}_{\text{cl}}$ encodes only the tree-level exchange, which is the only non-trivial diagram needed for the leading resummation. Unlike the original Bethe-Salpeter equation, in this case we get a crossing symmetric solution of the recursion relation in \eqref{eq:integral_eq_cl_tree}.

\begin{figure}[h]
\centering
\includegraphics[scale=0.85]{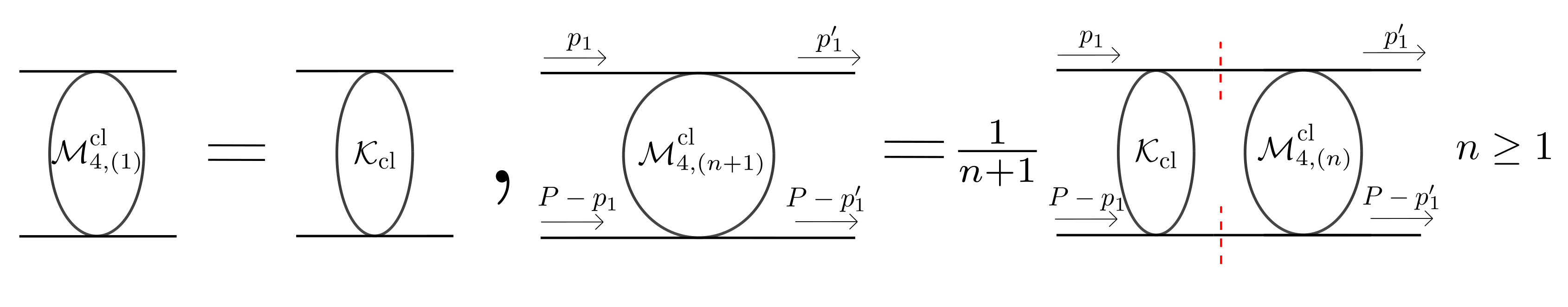}
\caption{The classical Bethe-Salpeter equation, where the kernel $\mathcal{K}_{\text{cl}}$ is defined as the set of all 2MPI diagrams in the space $\mathcal{H}_{4, \text{cl}}$.}
\label{fig:classical_BSE}
\end{figure}

This is teaching us something deeper about classical physics and how one can derive, in full generality, the ``classical Bethe-Salpeter equation.'' First, all iterations must be generated from the standard kernel made of 2MPI diagrams, $\mathcal{K}$ in the space $\mathcal{H}_{4}$ using the original Bethe-Salpeter equation. Then, after quotienting down to the space of classical diagrams $\mathcal{H}_{4, \text{cl}}$, we obtain a new recursion relation in terms of a classical kernel $\mathcal{K}_{\text{cl}}$ defined by the set of 2MPI diagrams in the space $\mathcal{H}_{4, \text{cl}}$. The final result of this procedure is the classical Bethe-Salpeter equation 
\begin{align}
\mathcal{M}_{4,(n+1)}^{\text{cl}} &(p_A,p_B,q) \nonumber \\
&= \begin{cases}
      \mathcal{K}_{\text{cl}}(p_A,p_B,q) & \text{if}\,\, n = 0\\
      \frac{1}{n+1} \int \mathrm{\hat{d}}^4 l \, \mathcal{K}_{\text{cl}}(p_A,p_B,l) G_{\text{cl}}(p_A,p_B,l) \mathcal{M}^{\text{cl}}_{4,(n)} (p_A,p_B,q-l) & \text{if} \,\,n \geq 1\\
    \end{cases}  \,,
\label{eq:integral_eq_cl}
\end{align}
pictorially represented in Figure~\ref{fig:classical_BSE}. 

The advantage of this representation is two-fold. First, we have identified a subset of 2MPI diagrams (the ones which cannot be generated from others by iteration and permutation of the internal graviton legs) which are the only relevant diagrams for the conservative, classical two-body dynamics. Second, the recursion relation in \eqref{eq:integral_eq_cl} has a natural solution in terms of an exponential function~\cite{Brandhuber:2021eyq}. Indeed, if we define the impact parameter representation for a $q$-dependent function $f(q)$ as
\begin{align}
\widetilde{f}(x_{\bot}) := \mathcal{F}_{x_{\bot}}[f(q)] \equiv  \int \mathrm{\hat{d}}^4 q \,\hat{\delta}(2 p_A \cdot q)\, \hat{\delta}(2 p_B \cdot q)\, e^{ i \frac{q \cdot x_{\bot}}{\hbar}}\, f(q) 
\end{align}
then applying this to both sides of the recursion in \eqref{eq:integral_eq_cl} gives
\begin{align}
\mathcal{F}_{x_{\bot}}&[\mathcal{M}^{\text{cl}}_{4,(n+1)} (p_A,p_B,q)] \nonumber \\
& = \begin{cases}
      \mathcal{F}_{x_{\bot}}[\mathcal{K}_{\text{cl}}(p_A,p_B,q)] & \text{if}\,\, n = 0\\
      \frac{1}{n+1}  \mathcal{F}_{x_{\bot}}[\int \mathrm{\hat{d}}^4 l \, \mathcal{K}_{\text{cl}}(p_A,p_B,l) G_{\text{cl}}(p_A,p_B,l) \mathcal{M}^{\text{cl}}_{4,(n)} (p_A,p_B,q-l)] & \text{if} \,\,n \geq 1 \\
    \end{cases}  \,.
\end{align}
Because of the convolution theorem, the Fourier transform of a convolution of two functions becomes the product of the Fourier transforms in the conjugate space. Hence, we find\footnote{$\mathcal{F}_{x_{\bot}}[\int \mathrm{\hat{d}}^2 l_{\perp} \, \mathcal{K}_{\text{cl}}(p_A,p_B,l_{\bot}) \mathcal{M}^{\text{cl}}_{4,(n)} (p_A,p_B,q_{\bot}-l_{\bot})] = \mathcal{\widetilde{K}}_{\text{cl}}(p_A,p_B,x_{\bot}) \mathcal{\widetilde{M}}^{\text{cl}}_{4,(n)} (p_A,p_B,x_{\bot})$.}
\begin{align}
\mathcal{\widetilde{M}}^{\text{cl}}_{4,(n+1)} &(p_A,p_B,x_{\bot}) = \begin{cases}
      \mathcal{\widetilde{K}}_{\text{cl}}(p_A,p_B,x_{\bot}) & \text{if}\,\, n = 0\\
      \frac{1}{n+1} \mathcal{\widetilde{K}}_{\text{cl}}(p_A,p_B,x_{\bot}) \mathcal{\widetilde{M}}^{\text{cl}}_{4,(n)} (p_A,p_B,x_{\bot}) & \text{if} \,\,n \geq 1\\
    \end{cases}  \,.
 \label{eq:BSequation_IPS}
\end{align}
As we will see, this allows us to solve the recursion relation directly in terms of the classical kernel $\mathcal{\widetilde{K}}_{\text{cl}}(p_A,p_B,x_{\bot})$. Before doing this, we first turn to the formulation of the classical Bethe-Salpeter equation with spin.

%%%%%%%%%%%%%%%%%%%%%%%%%%%%%%%%%%%%%%%

\subsection{The classical Bethe-Salpeter equation for spinning particles}

Generalizing to the case of external massive spinning particles of spin $a_A$ and $a_B$, we restrict our attention to the case where spin-commutator terms are suppressed in the classical limit, which (as we argued previously) is relevant for the point-particle approximation of Kerr black holes. In this case, it is conceivable that a classical recursion can be found which is similar to the scalar recursion \eqref{eq:BSequation_IPS}, motivating us to look for a classical, spinning Bethe-Salpeter recursion relation in the space of classical amplitudes $\mathcal{H}_{4,\text{cl}}$. We will discuss further in section \ref{sec:resummation} the possible relevance of the result for bound states of Kerr black holes in general relativity.

We start by considering the original Bethe-Salpeter equation for spinning amplitudes, involving the elastic 4-point amplitude of two external spinning massive fields of spin $s_1$ and $s_2$. This can be written as
\begin{align}
\varepsilon^{*\sigma_A}_{I_1}& \varepsilon^{*\sigma_B}_{I_2}\, (\mathcal{M}_4)^{I_1 J_1 ,I_2 J_2} (p_1,p'_1;P)\, \varepsilon^{\sigma_A}_{J_1} \varepsilon^{\sigma_B}_{J_2} \nonumber \\
&\,\,\,= \varepsilon^{*\sigma_A}_{I_1} \varepsilon^{*\sigma_B}_{I_2}\,  (\mathcal{K})^{I_1 J_1 ,I_2 J_2}(p_1,p'_1;P)\, \varepsilon^{\sigma_A}_{J_1} \varepsilon^{\sigma_B}_{J_2} \nonumber \\
&\,\,\,+ \varepsilon^{*\sigma_A}_{I_1} \varepsilon^{*\sigma_B}_{I_2} \int \mathrm{\hat{d}}^4 l \, (\mathcal{K})^{I_1 K_1, I_2 K_2}(p_1,l;P) \nonumber \\
&\qquad\qquad\qquad\times (G^{(s_1,s_2)})_{K_1 Z_1 ,K_2 Z_2}(l,P)\, (\mathcal{M}_{4})^{Z_1 J_1,Z_2 J_2} (l,p'_1;P)\,  \varepsilon^{\sigma_A}_{J_1} \varepsilon^{\sigma_B}_{J_2} \,,
\label{eq:integral_eq_spinning}
\end{align}
where $I_1,J_1,K_1$ (resp. $I_2,J_2,K_2$) stand for the collection of $s_1$ (resp. $s_2$) little-group indices of particle 1 (resp. 2), the two-body massive spinning propagator is
\begin{align}
G^{(s_1,s_2)}_{I_1 J_1 ,I_2 J_2}(l,P) = \frac{i\, N_{I_1 J_1}(P-l)}{(P - l)^2 - m_1^2 + i \epsilon}\, \frac{i\, N_{I_2 J_2}(l)}{l^2 - m_2^2 + i \epsilon} \,,
\end{align}
and $(\mathcal{K})^{I_1 K_1, I_2 K_2}$ is the spinning kernel.

To take the classical limit, it is instructive (as in the scalar case) to first consider the leading eikonal resummation of ladder and crossed-ladder spinning amplitudes for Kerr, which gives from \eqref{eq:1oop-eik-clas-spin-final}
\begin{align}
\hspace{-5pt}i \mathcal{M}_{4, \text{LR}}^{(n-1),\text{cl}}(\{p_A,a_A\},\{p_B,a_B\};q) & = \frac{1}{n!} \left[ \prod_{i=1}^n \int \mathrm{\hat{d}}^4 l_i \right]  \hat{\delta}^4 \left(\sum_{i=1}^n l_i - q \right) \prod_{i=1}^{n-1} \hat{\delta} (2 l_i \cdot p_B)\, \hat{\delta} (2 l_i \cdot p_A) \nonumber \\
\hspace{-5pt}& \times V_{\mu_n \nu_n}(\{p_A,a_A\},l_n)\, \frac{i\, P^{\mu_n \nu_n \alpha_n \beta_n}}{l_i^2 + i \epsilon}\,  V_{\alpha_n \beta_n}(\{p_B,a_B\},l_n) \nonumber \\
\hspace{-5pt}& \times  \prod_{i=1}^{n-1} \left[V_{\mu_i \nu_i}(\{p_A,a_A\},l_i)\, \frac{i\, P^{\mu_i \nu_i \alpha_i \beta_i}}{l_i^2 + i \epsilon}\, V_{\alpha_i \beta_i}(\{p_B,a_B\},l_i) \right] \,.
\label{eq:n1oop-eik-clas-spin-BS}
\end{align}
Written in this form, one observes that this calls for a new recursion relation with the leading-order kernel
\begin{align}
\mathcal{K}^{(0)}_{\text{cl}}(\{p_A,a_A\},\{p_B,a_B\},q) := V_{\mu_i \nu_i}(\{p_A,a_A\},q)\, \frac{i P^{\mu_i \nu_i \alpha_i \beta_i}}{q^2 + i \epsilon}\, V_{\alpha_i \beta_i}(\{p_B,a_B\},q) \,,
\label{eq:kernelcl_eik-spin}
\end{align}
and the on-shell two-body propagator 
\begin{align}
G^{(s_1,s_2)}_{\text{cl}}(p_A,p_B,l) &:= \hat{\delta}\left(2 p_A \cdot l\right) \,\hat{\delta}\left(2 p_B \cdot l\right) \,.
\label{eq:twobodycl_eik-spin}
\end{align}
We now recast the leading spinning eikonal amplitude as a solution of a recursion relation. The classical Bethe-Salpeter equation for bound states of massive spinning particles is 
\begin{align}
\label{eq:integral_eq_cl_spinning}
\mathcal{M}^{\text{cl}}_{4,(n+1)}(\{p_A,a_A\},\{p_B,a_B\},q) \hspace{440pt} & \nonumber \\
= \begin{cases}
  \mathcal{K}_{\text{cl}}(\{p_A,a_A\},\{p_B,a_B\},q) & \text{if}\,\, n = 0\\
      \frac{1}{n+1} \int \mathrm{\hat{d}}^4 l \, \mathcal{K}_{\text{cl}}(\{p_A,a_A\},\{p_B,a_B\},l) & \smash{\raisebox{-1.6ex}{if $n \geq 1$}}\\
       \qquad \qquad\times G^{(s_1,s_2)}_{\text{cl}}(p_A,p_B,l) \mathcal{M}^{\text{cl}}_{4,(n)} (\{p_A,a_A\},\{p_B,a_B\},q-l) \\
    \end{cases} \hspace{200pt}& \,,
    \raisetag{36pt}
\end{align}
defined in the space of classical amplitudes $\mathcal{H}_{4,\text{cl}}$. As in the scalar case, the subscript $(n)$ on the amplitude denotes the number of classical 2MPI diagrams contributing. Since the recursion relation \eqref{eq:integral_eq_cl_spinning} involves the convolution of products of matrices in momentum space, it is convenient to recast the recursion in impact parameter space 
\begin{align}
\label{eq:BSequation_IPS-spin}
\mathcal{\widetilde{M}}^{\text{cl}}_{4,(n+1)} (\{p_A,a_A\},\{p_B,a_B\},x_{\bot})\hspace{430pt} & \nonumber \\
= \begin{cases}
\mathcal{\widetilde{K}}_{\text{cl}}(\{p_A,a_A\},\{p_B,a_B\},x_{\bot})  & \text{if}\,\, n = 0\\
      \frac{1}{n+1}  \int \mathrm{\hat{d}}^2 q_{\bot} e^{i \bar{q}_{\bot} \cdot x_{\bot}} \int \mathrm{\hat{d}}^2 l_{\bot} \, \mathcal{K}_{\text{cl}}(\{p_A,a_A\},\{p_B,a_B\},l_{\bot}) & \smash{\raisebox{-1.6ex}{if $n \geq 1$}}\\
       \qquad \qquad \qquad \qquad \qquad \times \mathcal{M}^{\text{cl}}_{4,(n)} (\{p_A,a_A\},\{p_B,a_B\},q_{\bot} -l_{\bot}) \\
    \end{cases} \hspace{200pt}& \,. 
    \raisetag{36pt}
\end{align}
Note that the objects appearing in this spinning recursion relation are not \emph{a priori} commutative due to the presence of spin-dependent operators.

%%%%%%%%%%%%%%%%%%%%%%%%%

\subsection{Exponentiation of the two-body classical kernel: an exact solution}

We are now ready to solve the classical Bethe-Salpeter equation for both scalar \eqref{eq:integral_eq_cl} and spinning \eqref{eq:integral_eq_cl_spinning} point particles. By working in impact parameter space, we found that the 4-point classical scalar amplitude in the space of classical diagrams $\mathcal{H}_{4,\text{cl}}$ obeys \eqref{eq:BSequation_IPS}
\begin{align}
\mathcal{\widetilde{M}}_{4,(n+1)}^{\text{cl}} &(p_A,p_B,x_{\bot}) = \begin{cases}
      \mathcal{\widetilde{K}}_{\text{cl}}(p_A,p_B,x_{\bot}) & \text{if}\,\, n = 0\\
      \frac{1}{n+1} \mathcal{\widetilde{K}}_{\text{cl}}(p_A,p_B,x_{\bot}) \mathcal{\widetilde{M}}_{4,(n)}^{\text{cl}} (p_A,p_B,x_{\bot}) & \text{if} \,\,n \geq 1\\
    \end{cases}  \,,
\end{align}
and therefore by iteration one obtains, schematically,
\begin{align}
\mathcal{\widetilde{M}}_{4,(n+1)}^{\text{cl}} & = \mathcal{\widetilde{K}}_{\text{cl}} + \frac{1}{2!} \mathcal{\widetilde{K}}_{\text{cl}}^2 + \dots + \frac{1}{(n+1)!} \mathcal{\widetilde{K}}_{\text{cl}}^{n+1}\,.
\end{align}
This is easily recognized as an exponential series, giving a solution to the classical recursion
\begin{align}
\boxed{\mathcal{\widetilde{M}}_{4}^{\text{cl}} (p_A,p_B,x_{\bot}) = \sum_{n=0}^{\infty} \mathcal{\widetilde{M}}^{\text{cl}}_{4,(n+1)} (p_A,p_B,x_{\bot}) = e^{\mathcal{\widetilde{K}}_{\text{cl}}(p_A,p_B,x_{\bot})} -1\,.}
\label{eq:BS-solution-IPS}
\end{align}
Remarkably, the amplitude exponentiates exactly in impact parameter space! Going back to the momentum space amplitude, one finds 
\begin{align}
\mathrm{i} \mathcal{M}_4^{\text{cl}}(p_A,p_B;q_{\bot})=& 
\frac{4 \sqrt{(p_A \cdot p_B)^{2}-m_A^2 m_B^2}}{\hbar^{2}} \int \mathrm{d}^{2} x^{\bot}\, e^{-i \bar{q}_{\bot} \cdot x_{\bot} }\left(e^{\mathcal{\widetilde{K}}_{\text{cl}}(p_A,p_B,x_{\bot})}-1\right) \,,
\label{eq:BS-solution-momentum}
\end{align}
which means that the analytic structure of classical gravitational bound states is entirely encoded in the 2MPI classical kernel. From a purely bound state perspective, it is also interesting to notice that the Fourier transform in \eqref{eq:BS-solution-momentum} encodes the bound state poles in a way that is more compact and completely crossing symmetric compared to the original Bethe-Salpeter equation.

A similar analysis can be done for classical spinning amplitudes for Kerr black holes, assuming that extra commutators are suppressed in the classical limit as they are in the leading-order contribution. Using the recursion relation \eqref{eq:BSequation_IPS-spin}
\begin{align} \label{eq:iteration-sol}
\mathcal{\widetilde{M}}_{4,(n+1)}^{\text{cl}} (\{p_A,a_A\},\{p_B,a_B\},x_{\bot}) \hspace{440pt}  &  \nonumber  \\
= \begin{cases}
   \mathcal{\widetilde{K}}_{\text{cl}}(\{p_A,a_A\},\{p_B,a_B\},x_{\bot})  & \text{if}\,\, n = 0\\
      \frac{1}{n+1} \mathcal{\widetilde{K}}_{\text{cl}}(\{p_A,a_A\},\{p_B,a_B\},x_{\bot}) \mathcal{\widetilde{M}}_{4,(n)}^{\text{cl}} (\{p_A,a_A\},\{p_B,a_B\},x_{\bot}) & \text{if} \,\,n \geq 1 \\
    \end{cases} \hspace{200pt} \,, 
    \raisetag{26pt}
\end{align}
the iteration can be written as a multiplication of spin-dependent operators \cite{Bern:2020buy},
\begin{align}
\mathcal{\widetilde{M}}^{\text{cl}}_{4,(n+1)}&  = \mathcal{\widetilde{K}}_{\text{cl}} + \frac{1}{2!} \mathcal{\widetilde{K}}_{\text{cl}} \cdot \mathcal{\widetilde{K}}_{\text{cl}}  + \dots + \frac{1}{(n+1)!} {\underbrace{\left(\mathcal{\widetilde{K}}_{\text{cl}}\cdot \mathcal{\widetilde{K}}_{\text{cl}}\cdot \dots \mathcal{\widetilde{K}}_{\text{cl}}\right)}_{n+1}} \,.
\end{align}
A solution of the recursion relation can then be written as
\begin{align}
\boxed{ \mathcal{\widetilde{M}}^{\text{cl}}_{4}(\{p_A,a_A\},\{p_B,a_B\},x_{\bot}) = e^{\mathcal{\widetilde{K}}_{\text{cl}}(\{p_A,a_A\},\{p_B,a_B\},x_{\bot})} - 1 \,,}
\label{eq:BS-solution-IPS-spin}
\end{align}
where it is worth stressing that $[\mathcal{\widetilde{K}}_{\text{cl}}, \mathcal{\widetilde{K}}_{\text{cl}}] \neq 0$\footnote{This is the reason why it was crucial to study the double (and higher) commutator terms in section \ref{sec:eikonal}.}.
Finally, the massive spinning classical amplitude in momentum space is given at all orders in terms of the classical kernel by
\begin{align}
&\mathcal{M}^{\text{cl}}_{4}(\{p_A,a_A\},\{p_B,a_B\},q_{\bot}) \nonumber \\
&\qquad =\frac{4 \sqrt{(p_A \cdot p_B)^{2}-m_A^2 m_B^2}}{\hbar^{2}} \int \mathrm{d}^{2} x^{\bot}\, e^{-i \bar{q}_{\bot} \cdot x_{\bot}}\left(e^{\mathcal{\widetilde{K}}_{\text{cl}}(\{p_A,a_A\},\{p_B,a_B\},x_{\bot})}-1\right) \,,
\label{eq:BS-solution-momentum-spin}
\end{align}
so once again the classical kernel encodes all of the analytic structure of the full amplitude.

%%%%%%%%%%%%%%%%%%%%%%%%%%%%%%%%%%%%%

\subsection{Hamilton-Jacobi action and bound state observables}
\label{sec:HJ}

We now discuss explicit results for the classical kernel and their relation to the classical Hamilton-Jacobi action. This discussion will be particularly useful in preparation for Section~\ref{sec:resummation}, where we recover some classical observables directly from the analytic structure of the resummed amplitudes in momentum space. In order to be consistent with the explicit results presented later, we limit our attention to the 2PM calculation in the spinless case and 1PM in the spinning case, both of which have been well-studied in the literature.

We start with the classical dynamics of spinless particles. The motion is restricted to a plane and one defines the conserved quantities
\begin{align}
E=\frac{m_A m_B \sqrt{y^2-1}}{p_{\infty}} \,, \qquad J=p_{\infty} |x_{\bot}| \,.
\label{eq:EJ_standard}
\end{align}
where $p_{\infty}$ is the center-of-mass energy at infinity and $y = v_A \cdot v_B$ is the rapidity. A key variable to consider is the energy above threshold
\begin{align}
\mathcal{E} := \frac{E - m_A - m_B}{\mu} \,, \qquad \mu = \nu (m_A + m_B) = \frac{m_A m_B}{(m_A + m_B)}\,,
\end{align}
where $\mu$ is the reduced mass and $\nu$ is the symmetric mass ratio. We notice that $\mathcal{E}$ is positive ($\mathcal{E} > 0$) for scattering orbits and negative for bound orbits ($\mathcal{E} < 0$). In terms of the rapidity, this implies
\begin{align}\label{eq:yanalyticcont}
\text{scattering:} \,\, y > 1 \,, \qquad \text{bound:} \,\, -\frac{m_A^2 + m_B^2}{2 m_A m_B} < y < 1 \,. 
\end{align}
We introduce the superscript $>$ to denote an expression valid for scattering orbits and $<$ to denote an expression valid for a bound orbit.

The classical two-body kernel for scattering orbits up to 2PM is~\cite{Akhoury:2013yua,Bjerrum-Bohr:2016hpa,Luna:2016idw,KoemansCollado:2019ggb}  
\begin{align}\label{eq:K_2PM}
\mathcal{\widetilde{K}}_{\text{cl}}^{>}(p_A,p_B,x_{\bot}) =& \frac{i}{\hbar} \Bigg[-2 G_N\, \log(\mu_{\text{IR}} |x_{\bot}|)\, m_A m_B\, \frac{2 y^2 - 1}{\sqrt{y^2 - 1}} \nonumber \\
&+ \frac{3 \pi}{4}\, G_N^2\, m_A m_B\, (m_A + m_B)\, \frac{5 y^2 - 1}{\sqrt{y^2 - 1}}\, \frac{1}{|x_{\bot}|} \Bigg] \,,
\end{align}
where $\mu_{\text{IR}}$ is an infrared regulator for the divergent Coulomb phase in 4-dimensions. This can be written in terms of the (conservative) Hamilton-Jacobi action of the system
\begin{align}
 \mathcal{\widetilde{K}}^{>}_{\text{cl}}(p_A,p_B;x_{\bot})= \frac{i}{\hbar} I^{>}\left(\mathcal{E},J\right) \,,
 \label{eq:HJ-action}
\end{align}
which is equivalent to the ``amplitude-action'' relation originally found in \cite{Bern:2021dqo} and further developed in \cite{Kol:2021jjc,Damgaard:2021ipf}. This is particularly convenient, since the analogous result for bound orbits requires an analytic continuation in the conserved charges $(\mathcal{E},J)$ of the Hamiltonian dynamics~\cite{Kalin:2019rwq,Kalin:2019inp}. 

For this system the regularized Hamilton-Jacobi action\footnote{Technically, since the integral is infrared divergent as $r \to +\infty$, we should employ a regulator $\sim 1/\mu_{\text{IR}}$ for the final expression, consistent with the classical kernel obtained with amplitude techniques \eqref{eq:K_2PM}.} admits an expansion in terms of the radial action
\begin{align}
I^{>}\left(\mathcal{E},J\right) = \oint_{\mathcal{C}_{>}} \mathrm{d} r \, p_{r}(r,\mathcal{E},J) + J\, \pi \,,
\label{eq:radialaction}
\end{align}
where $p_{r}$ is the radial momentum in the center-of-mass frame and (for the case of interest) the contour $\mathcal{C}_{>}$ is defined in terms of the motion from infinity ($r=\infty$) to the classical turning point $r_{m}(\mathcal{E},J)$, defined by $p_r(r_m,\mathcal{E},J)=0$, of the equations of motion and back:
\begin{align}\label{eq:contour-scatt}
\oint_{\mathcal{C}_{>}} \mathrm{d} r = 2 \int_{r_{m}(\mathcal{E},J)}^{\infty} \mathrm{d} r \,.
\end{align}
The natural scattering observable is the deflection angle $\chi$, which can then be computed in a straightforward way from \eqref{eq:radialaction}
\begin{align}\label{eq:scattering-angle}
\chi(\mathcal{E},J) &= - \frac{\partial I^{>}\left(\mathcal{E},J\right)}{\partial J}\,, \qquad \mathcal{E} > 0 \,,
\end{align}
which gives
\begin{align}
\chi(\mathcal{E},J)= \frac{2\,G_N}{J}\, m_A m_B\,\frac{2 y^2-1}{\sqrt{y^2-1}}+\frac{G_N^2}{J^2}\, \frac{3 \pi}{4 E}\, m_A^2 m_B^2\left(m_A+m_B\right)\left(5 y^2-1\right) \,.
\end{align}

For bound states, we must analytically continue the Hamilton-Jacobi action \eqref{eq:HJ-action} from $\mathcal{E} > 0$ to $\mathcal{E} < 0$. The contour of the integration in the radial action representation \eqref{eq:radialaction} now depends on the two radial turning points $r_{\pm}(\mathcal{E},J)$ of the equation of motion
\begin{align}\label{eq:contour-bound}
\oint_{\mathcal{C}_{<}} \mathrm{d} r = 2 \int_{r_{-}(\mathcal{E},J)}^{r_{+}(\mathcal{E},J)} \mathrm{d} r \,,
\end{align}
and a careful analysis of their relation with $(\mathcal{E},J)$ shows that, remarkably, they are both determined from $r_{m}(\mathcal{E},J)$ via the analytic continuation $r_{\pm}(\mathcal{E},J) = r_{m}(\mathcal{E}, \mp J)$ for $\mathcal{E} < 0$.\qquad As a consequence, for the spinless conservative system there is a simple map between boundary and bound dynamics~\cite{Kalin:2019rwq}:
\begin{align}
I^{<}\left(\mathcal{E},J\right) = I^{>}\left(\mathcal{E}<0,J\right) -  I^{>}\left(\mathcal{E}<0,-J\right) \,.
\end{align}
Using \eqref{eq:K_2PM} and \eqref{eq:HJ-action}, this gives
\begin{align}\label{eq:I2PM}
I^{<}\left(\mathcal{E},J\right) = 2 \pi\, G_N\, m_A m_B\, \frac{2 y^2 - 1}{\sqrt{1 - y^2}} + \frac{3 \pi}{2 }\, G_N^2\, \frac{m_A^2 m_B^2 (m_A + m_B) (5 y^2 - 1)}{\sqrt{m_A^2+2 m_A m_B y+m_B^2}}\,  \frac{1}{J}\,, 
\end{align}
from which all natural bound state observables evaluated along a period of the orbit can be obtained, up to 2PM accuracy. 

One of the simplest such observables is the periastron advance of the binary system, which is essentially related to the (ratio of) the frequencies of the motion in the azimuthal and radial direction. This is given by~\cite{Damour:1988mr}
\begin{align}
\Delta \Phi(\mathcal{E},J) &= - \frac{\partial I^{<}\left(\mathcal{E},J\right)}{\partial J} = \chi(\mathcal{E},J) + \chi(\mathcal{E},-J)\,, \qquad \mathcal{E} < 0 \,,
\end{align}
and explicitly 
\begin{align}
\Delta \Phi(\mathcal{E},J)  &= \frac{3 \pi\, G_N^2\, m_A^2 m_B^2 (m_A+m_B) \,\left(5 y^2-1\right)}{2 J^2 \sqrt{m_A^2+2 m_A m_B\, y+m_B^2}}  \,.
\end{align}
For us, one of the most interesting bound state observables is the binding energy for circular orbits, which is determined implicitly by the vanishing of the Hamilton-Jacobi action. From \eqref{eq:HJ-action} we obtain the angular momentum $J$ as a function of $\mathcal{E}$
\begin{align}
\oint_{\mathcal{C}_{<}} \mathrm{d} r \, p_{r}(r,\mathcal{E},J) = 0 \rightarrow J_{\text{bind}} = J_{\text{bind}}(\mathcal{E}) \,,
\label{eq:binding_HJ}
\end{align}
which can be made explicit using \eqref{eq:I2PM}. Because of the virial theorem, it is natural to evaluate the action in the non-relativistic expansion at small velocities: this allows to make contact with the Post-Newtonian (PN) expansion. We introduce the variable $x$, related to the orbital frequency of circular orbits $\Omega$, via
\begin{align}
x := [G_N (m_A + m_B) \Omega]^{\frac{2}{3}} = \left(\frac{1}{G_N m_A m_B}\frac{\mathrm{d} J_{\text{bind}}(\mathcal{E})}{\mathrm{d} \mathcal{E}}\right)^{-\frac{2}{3}} \,,
\end{align}
and expand for $\mathcal{E} \ll m_A , m_B$ consistently with the non-relativistic expansion. Finally, we can express $x$ in terms of the reduced binding energy $\epsilon = - 2 \mathcal{E}$, 
\begin{align}
\epsilon(x) = x - \frac{1}{12} x^2 (9 + \nu) + \mathcal{O}(x^3)\,,
\label{eq:binding_energy}
\end{align}
as expected (cf., \cite{Blanchet:2013haa,Kalin:2019rwq}).

\medskip

For the spinning case, we restrict our analysis for simplicity to the case of aligned spin scattering\footnote{A generalization to misaligned spin configurations has been recently discussed in~\cite{Jakobsen:2022zsx}.}, where we can again identify a plane for the orbital motion. The classical spinning two-body kernel at 1PM reads
\begin{align}
\mathcal{\widetilde{K}}_{\text{cl}}&(\{p_A,a_A\},\{p_B,a_B\},x_{\bot})  \label{eq:Kspin_1PM}\\
&=\frac{i}{\hbar} \Bigg[-2 G_N \sum_{\eta=\pm} m_A m_B\, \frac{(y + \eta \sqrt{y^2 - 1})^2}{2 \sqrt{y^2 - 1}}\, \log\left[\mu_{\text{IR}} (|x_{\bot}| - \eta (a_A + a_B))\right] \Bigg] \,,\nonumber
\end{align}
where $a_A$ and $a_B$ are the components of the spin vectors projected along the spin axis, which is the same as the orbital angular momentum in the aligned case. As in the spinless case, we can still identify the kernel with the regularized Hamilton-Jacobi action
\begin{align}
 \mathcal{\widetilde{K}}^{>}_{\text{cl}}(\{p_A,a_A\},\{p_B,a_B\};x_{\bot})= \frac{i}{\hbar}\, I^{>}\left(\mathcal{E},L,a_A,a_B\right) \,,
 \label{eq:HJ-action-spin}
\end{align}
but this requires the introduction of the canonical orbital angular momentum, $L$ \cite{Vines:2017hyw}
\begin{align}
\label{eq:canonicalangular}
L := p_{\infty} |x_{\bot}| + \frac{\nu\, \mathcal{E}}{2}& (m_A + m_B) \left(a_+ - \frac{\sqrt{1-4 \nu}}{1+ \nu \mathcal{E}}\, \frac{m_A - m_B}{|m_A-m_B|}\, a_-\right)\,, \\
 & \quad a_{\pm} := a_A \pm a_B\,, \nonumber  
\end{align}
which differs from the covariant version in \eqref{eq:EJ_standard}. In the scattering region, the scattering angle can be computed at leading order in $G_N$ but all orders in spin: 
\begin{align}
\chi(\mathcal{E},L,a_{\pm}) &= - \frac{\partial I^{>}\left(\mathcal{E},L,a_{\pm}\right)}{\partial L}\,, \nonumber \\
&= G_N m_A m_B \left[\frac{(y + \sqrt{y^2 - 1})^2}{\sqrt{y^2 - 1}} \frac{1}{|x_{\bot}| + a_+} + \frac{(y - \sqrt{y^2 - 1})^2}{\sqrt{y^2 - 1}} \frac{1}{|x_{\bot}| - a_+}\right] \,.
\label{spinscatta}
\end{align}
Taking advantage of \eqref{eq:Kspin_1PM} and \eqref{eq:canonicalangular}, we can then write the action in terms of the canonical radial momentum $P_{r}$ ~\cite{Kalin:2019inp,Liu:2021zxr}
\begin{align}\label{eq:Kspin}
I^{>}\left(\mathcal{E},L,a_{\pm}\right) = \left(\oint_{\mathcal{C}_{>}} \mathrm{d} r \, P_{r}(r,\mathcal{E},a_{\pm},L) + L\, \pi\right) \,,
\end{align}
where compared to \eqref{eq:contour-scatt} the turning point $r_m(\mathcal{E},J)$ depends on the canonical total angular momentum $J = L + m_A a_A + m_B a_B$.

The existence of the representation \eqref{eq:Kspin} allows the analytic continuation from scattering to bound orbits to be performed directly as in the spinless case. In particular, by imposing $\mathcal{E} < 0$ we find $\mathcal{C} = \mathcal{C}_{<}$ where the endpoints of the elliptic motion $r_{\pm}(\mathcal{E},J)$ still depend only on the conserved charges $\mathcal{E},J$ and obey $r_{\pm}(\mathcal{E},J) = r_{m}(\mathcal{E},\mp J)$. Therefore, the same boundary-to-bound map applies in the conservative aligned spin case~\cite{Kalin:2019inp}:
\begin{align}
\Delta \Phi(\mathcal{E},L,a_{\pm}) &= \chi(\mathcal{E},L,a_{\pm}) + \chi(\mathcal{E},-L,-a_{\pm}) \,, \qquad \mathcal{E} < 0 \,,
\end{align}
implying that $\Delta \Phi=0$ to 1PM accuracy using \eqref{spinscatta}. Indeed, the Hamilton-Jacobi action is the same as for the spinless case: this, as we will see, has consequences for the analytic structure of the amplitude in momentum space for $\mathcal{E}<0$. We can also compute the binding energy of circular orbits, which (for the same reason) is not affected by spin effects at this order: using the two-body classical kernel, we obtain the leading result 
\begin{align}
\epsilon(x,a_{\pm}) = x + \mathcal{O}(x^2)\,,
\label{eq:binding_energy_spin}
\end{align}
matching the leading term in the spinless expression \eqref{eq:binding_energy} for the binding energy.

%%%%%%%%%%%%%%%%%%%%%%%%%%%%%%%%%%%%%%
%%%%%%%%%%%%%%%%%%%%%%%%%%%%%%%%%%%%%%

\section{Analytic all-order resummation of classical amplitudes}
\label{sec:resummation}

In this section, we discuss the resummation of the conservative 4-point classical amplitude for the gravitational scattering of two point particles in general relativity. So far, most of the focus in extracting classical physics from amplitudes has been on isolating the Hamilton-Jacobi action, which contains all the information about the classical trajectory. Nevertheless, there is some physics encoded in the explicit resummation of the amplitudes themselves in momentum space. For instance, there is a new way to look at bound state properties -- like the binding energy -- from a pure amplitude perspective. Moreover, it is possible to extract an infrared-finite wavefunction which provides a solution of the classical (conservative) Hamiltonian below threshold and which contains all the data about classical bound states. In particular, its residue at the binding energy poles is directly connected to the bound state wavefunction. Finally, we will discuss also another physical effect from the resummation, which is the relativistic analogue of the Sommerfeld enhancement.

As shown in Section~\ref{sec:BSE}, the resummed classical amplitude can be always related to the two-body classical kernel using \eqref{eq:BS-solution-momentum} which we are going to compute at leading and next-to-leading order. Even though \eqref{eq:BS-solution-momentum} captures only classical conservative effects, dynamical\footnote{Static contributions can affect observables like the angular momentum, but this is not relevant here.} radiative effects appear only at $G_N^3$, so the cases we consider actually account for the full dynamics up to those orders.

%%%%%%%%%%%%%%%%%%%%%%%%%%%%%%%%%%%%%%%

\subsection{Gravitational bound states of scalar particles}

We begin with the leading resummation (LR) contribution to the classical scattering of scalar massive particles. This is defined by the tree-level, 1PM contribution to the classical kernel, which is the first line of \eqref{eq:K_2PM}; feeding this into the expression for the momentum space amplitude \eqref{eq:BS-solution-momentum}, the transverse space integrals can be performed to obtain~\cite{Levy:1970yn,Brezin:1970zr,tHooft:1987vrq,Kabat:1992tb,Irizarry-Gelpi:2013psg}:
\begin{align}
i\mathcal{M}^{\text{cl},>}_{4,\text{LR}}(p_A,p_B;q_{\bot})&= \frac{4 \pi\, m_A m_B\, \sqrt{y^2-1}}{\hbar^2\, \mu_{\text{IR}}^2}\, \frac{\Gamma (1-A_0^{>})}{ \Gamma (A_0^{>})}  \left(\frac{4 \mu_{\text{IR}} ^2}{\bar{q}_{\bot}^2}\right)^{1-A_0^{>}}\,,
\label{eq:leading_gamma}
\end{align}
for scattering orbits, where
\begin{align}\label{eq:A0plusdef}
A_0^{>} &:= i\, \frac{G_N}{\hbar}\, m_A m_B\, \frac{2 y^2-1}{\sqrt{y^2-1}}\,.
\end{align}
In order to discuss the analytic continuation of $A_0^{>}$ from scattering orbits $y > 1$ to bound orbits $y < 1$ (recall \eqref{eq:yanalyticcont}), we must define the branch cut for the square root $\sqrt{y^2 - 1}$. We take it just below the real axis, so that the following boundary-to-bound map holds:
\begin{align}\label{eq:B2By}
\text{scattering}: \sqrt{y^2-1} \quad \longrightarrow \quad \text{bound}: i\, \sqrt{1-y^2} \,.
\end{align}
In turn, this corresponds to a choice of the branch cut for the center of mass energy $s$, 
\begin{align}\label{eq:A0plusschannel}
A_0^{>} &= i\,\frac{G_N}{\hbar}\, \frac{\left(s-m_A^2-m_B^2\right)^2-2 m_A^2 m_B^2}{\sqrt{\left(s-(m_A+m_B)^2\right) \left(s-(m_A-m_B)^2\right)}}\,,
\end{align}
which defines the meaning of physical sheet for the amplitude. We then find
\begin{align}
i\mathcal{M}^{\text{cl},<}_{4,\text{LR}}(p_A,p_B;q_{\bot})&= i\, \frac{4 \pi\, m_A m_B\, \sqrt{1-y^2}}{\hbar^2\, \mu_{\text{IR}}^2}\, \frac{\Gamma (1-A_0^{<})}{ \Gamma (A_0^{<})}  \left(\frac{4 \mu_{\text{IR}}^2}{\bar{q}_{\bot}^2}\right)^{1-A_0^{<}}\,,
\end{align}
for bound orbits, where we now have the real function
\begin{align}\label{eq:A0minusdef}
A_0^{<} := \frac{G_N}{\hbar} m_A m_B \frac{2 y^2-1}{\sqrt{1-y^2}}\,.
\end{align}
Note that for bound states, the kinematics is no longer Lorentzian-real, as expected~\cite{Kabat:1992tb}.

We now turn to establish a connection with the classical scattering wavefunction, closely following some ideas from~\cite{Kabat:1992tb}. As proved earlier in Section~\ref{sec:eikonal} (see also Appendix~\ref{appendixA}), the leading-resummation problem is equivalent to the scattering of a scalar probe particle of mass $m_A$ in the linearized Schwarzschild potential of mass $m_B >\hspace{-5pt}> m_A$:
\begin{align}
\mathrm{d} s^2 = \left(1 - \frac{2 G_N\, m_B}{r}\right) \mathrm{d} t^2 - \left(1 + \frac{2 G_N\, m_B}{r}\right) (\mathrm{d} r^2 + r^2\, \mathrm{d} \theta^2 + r^2\, \sin^2(\theta)\, \mathrm{d} \phi^2) \,.
\end{align}
Thus, the scalar probe motion is described by the Klein-Gordon equation 
\begin{align}
\left[ \left(1 + \frac{2 G_N\, m_B}{r} \right) \frac{\partial^2}{\partial t^2} - \left(1 - \frac{2 G_N\, m_B}{r} \right) \nabla^2 + \frac{m_A^2}{\hbar^2} \right] \phi(x) = 0 \,.
\label{eq:KG-Schw}
\end{align}
Using the separation of variables $\phi(t,\vec{x}) = e^{-i E t / \hbar}\, \phi(\vec{x})$, we obtain\footnote{We need to further expand $\phi(\vec{x})$ to linear order in $G_N$ to get the consistent linearized solution.} from \eqref{eq:KG-Schw}
\begin{align}
\left[\hbar^2 \nabla^2 + (E^2 - m_A^2)  + \frac{2 G_N\, m_B}{r} (2 E^2 - m_A^2) \right] \phi(\vec{x}) = 0 \,.
\end{align}
The solution of such wave equation has the following asymptotic structure \cite{Sakurai:2011zz}
\begin{align}
\phi(\vec{x}) = \left(\frac{k_A \, (r - z)}{\hbar}\right)^{-A_0^{>}}\, e^{\frac{i k_A z}{\hbar}} + \left(\frac{2 k_A\, r}{\hbar}\right)^{A_0^{>}}\, \frac{f^{>}(y,q_{\bot})}{r}\, e^{\frac{i k_A r}{\hbar}} \,,
\end{align}
where $k_A := m_A \sqrt{y^2 - 1}$ and the (infrared-finite) scattering wavefunction is
\begin{align}
i\, f^{>}_{\text{LR}}(y,q_{\bot}) &= \frac{\hbar}{2 k_A}\, \frac{\Gamma (1-A_0^{>})}{ \Gamma (A_0^{>})}  \left(\frac{4 k_A^2}{\hbar^2 \bar{q}_{\bot}^2}\right)^{1-A_0^{>}} \nonumber \\
&= \frac{\hbar}{8 \pi\, m_B} \left(\frac{\hbar^2 \mu^2_{\text{IR}}}{k_A^2}\right)^{A_0^{>}}\, i\, \mathcal{M}^{\text{cl},>}_{4,\text{LR}}(p_A,p_B;q_{\bot}) \,.
\end{align}
Interestingly, the final result is exactly proportional to the scattering amplitude from \eqref{eq:leading_gamma}! The matching coefficient has two separate contributions: a kinematic matching coefficient $\hbar/(8 \pi m_B) \sim \hbar/(8 \pi E)$ and an infrared-divergent piece $\left(\hbar^2 \mu^2_{\text{IR}}/ k_A^2\right)^{A_0^{>}} \sim \left(\hbar^2 \mu^2_{\text{IR}}/ p_{\infty}^2\right)^{A_0^{>}}$, which is due to the long-range contribution to the asymptotic solution. For general kinematics, we can compare a gauge invariant observable like the cross-section to find the general matching between the conservative 4-point amplitude and the wavefunction solution\footnote{This general map holds both for the real and imaginary part of the wavefunction, as shown earlier.}~\cite{Fried1983}:
\begin{align}
f^{>}(y,q_{\bot}) &= \frac{\hbar}{8 \pi E} \left(\frac{\hbar^2 \mu_{\text{IR}}^2}{p_{\infty}^2}\right)^{A_0^{>}} \mathcal{M}^{\text{cl},>}_{4}(p_A,p_B;q_{\bot}) \,.
\end{align}
Therefore, if we perform an analytic continuation below threshold we obtain
\begin{align}
f^{<}_{\text{LR}}(y,q_{\bot}) &= \frac{\hbar}{8 \pi E} \left(\frac{\hbar^2 \mu_{\text{IR}}^2}{p_{\infty}^2}\right)^{A_0^{<}} \mathcal{M}^{\text{cl},<}_{4,\text{LR}}(p_A,p_B;q_{\bot}) \nonumber \\
&= \frac{2 m_A m_B \sqrt{1- y^2}}{E \,\hbar \,\bar{q}_{\bot}^2} \frac{\Gamma (1-A_0^{<})}{ \Gamma (A_0^{<})}  \left(\frac{4 p_{\infty}^2}{\hbar^2 \bar{q}_{\bot}^2}\right)^{-A_0^{<}} \,,
\label{eq:bound_matching}
\end{align}
which will act as the classical wavefunction of the system below threshold.

\begin{figure}[h]
 \centering
 \includegraphics[width=0.78\textwidth]{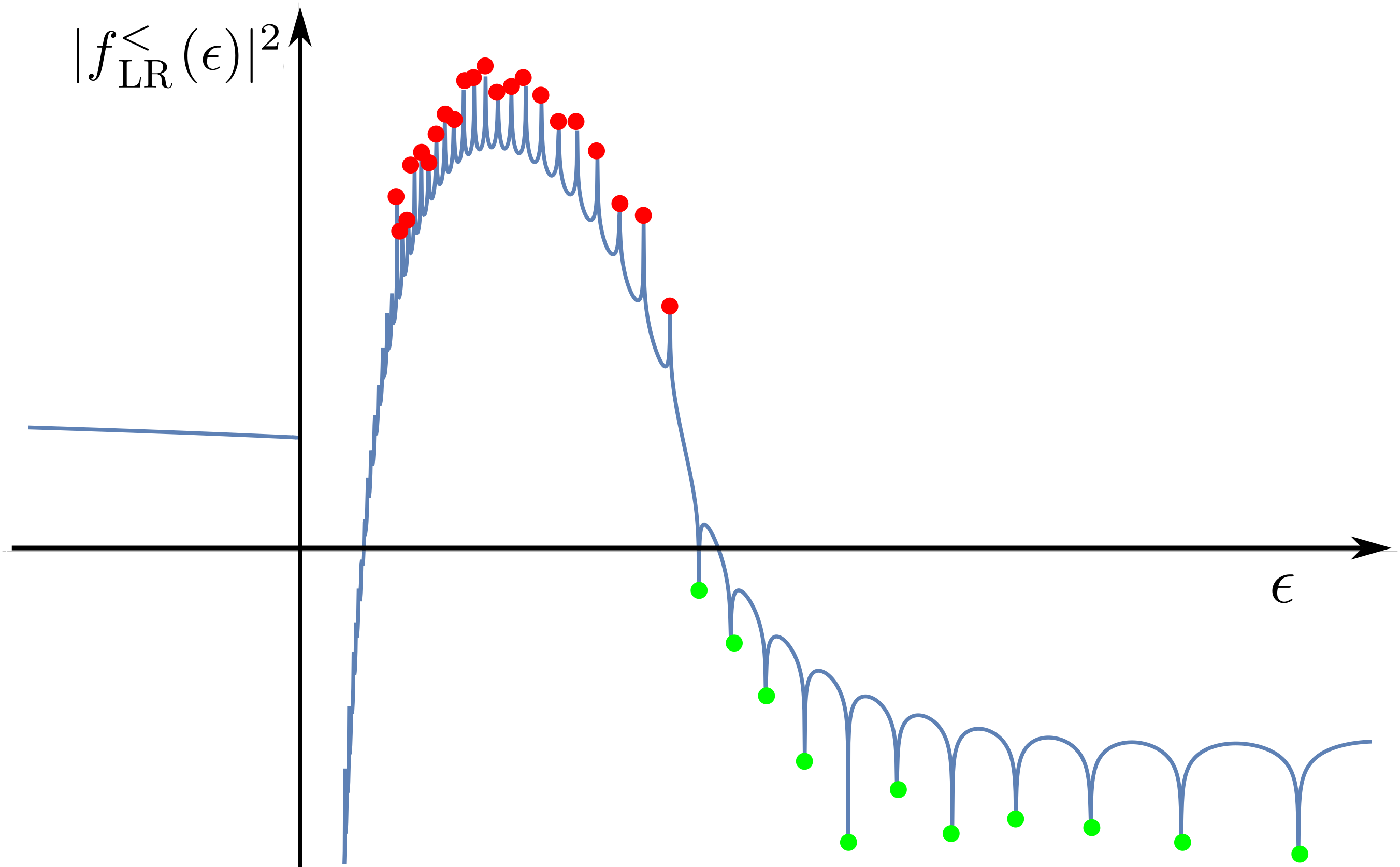} % first figure itself
 \label{fig:analytic-spinless}
  \caption{The cross-section integrand $|f^{<}_{\text{LR}}(\epsilon)|^2$ as a function of the binding energy $\epsilon$ in a logarithmic scale. The LR contribution has both poles (in red) and zeros (in green) for $\epsilon>0$, which are determined analytically.}
\end{figure}

Before delving deeper into the analytic structure of this wavefunction, it is worth studying the behaviour of the cross section integrand $\propto |f^{<}_{\text{LR}}(\epsilon)|^2$ as function of the binding energy $\epsilon$ when all other parameters are kept fixed in an appropriate kinematic regime. To do so, we have represented $|f^{<}_{\text{LR}}(\epsilon)|^2$ in a logarithmic scale in Figure~\ref{fig:analytic-spinless}, where some of the features that we will discuss are already completely manifest (like poles and zeros).

Now, the analytic structure of the leading resummed classical amplitude \eqref{eq:leading_gamma} has been studied before in the literature (cf., \cite{tHooft:1987vrq,tHooft:1988oyr,Verlinde:1991iu,Kabat:1992tb,Adamo:2021rfq}). The amplitude has simple poles on the physical sheet whenever
\begin{align}
1- A_0^{<} = 1-n \,, \qquad n \in \mathbb{Z}_{> 0}\,,
\end{align}
which implies
\begin{align}
s^{(1),\pm}_{n,\text{poles}} = m_A^{2}+m_B^{2} \pm \sqrt{2 m_A^{2} m_B^{2}-\frac{\hbar^{2} n^{2}}{2 G_N^{2}}+\sqrt{\frac{\hbar^{4} n^{4}}{4 G_N^{4}}+\frac{2 m_A^{2} m_B^{2} \hbar^{2} n^{2}}{G_N^{2}}}} \,.
\end{align}
The $+$ sign corresponds to the two-particle bound state, while the $-$ sign is the particle-antiparticle bound state solution; this can be seen by mapping $s^{(1),-}_{n,\text{poles}} \mapsto 2 (m_A^2 + m_B^2) - s^{(1),-}_{n,\text{poles}} =s^{(1),+}_{n,\text{poles}}$.\footnote{In a relativistic QFT, the Klein-Gordon equation for scalar fields allows both particle and antiparticle to propagate. A crossing symmetric amplitude implies the existence of particle-antiparticle bound states together with two-particle bound states.} These solutions are in direct correspondence with the energies of the linearized $1/r$ potential~\cite{Kabat:1992tb}, as expected from the fact that for the leading resummation the situation is equivalent to one particle moving in the gravitational background of the other (see Section~\ref{sec:eikonal}). It is convenient to write the two-particle bound state $s^{(1),+}_{n,\text{poles}}$ in a more compact way as:
\begin{align}
s^{(1),+}_{n,\text{poles}} = m_A^{2}+m_B^{2} + \frac{m_A\, m_B}{\sqrt{2}}\, \sqrt{4 -\xi_n^{2} +\xi_n\, \sqrt{8+\xi_n^2}} \,,
\end{align}
in terms of the dimensionless variable~\cite{Damour:2008yg}
\begin{align}
\xi_n := \frac{\hbar\, n}{G_N\, m_A m_B} \,.
\label{eq:xi_def}
\end{align}
The binding energy for the two-body bound state, corresponding to the peaks in Figure~\ref{fig:analytic-spinless}, can be expressed as\footnote{It is worth noticing that  $\mathcal{\epsilon}^{(1)}_{n} \to 0$ as $m_A \to 0$ or $m_B \to 0$; this is consistent with the fact that we do not expect bound states for massless particles in a $1/r$ potential.}
\begin{align}
\mathcal{\epsilon}^{(1)}_{n, \text{poles}} = \frac{2}{\mu} \left[m_A + m_B - \sqrt{m_A^{2}+m_B^{2} + \frac{m_A m_B}{\sqrt{2}}\,\sqrt{4 -\xi_n^{2} +\xi_n \sqrt{8+\xi_n^2}}} \right]\,,
\label{eq:binding-energy-pole}
\end{align}
which should be compared with the first term in \eqref{eq:binding_energy}. From the equivalence principle, the classical limit should correspond to $\hbar \to 0$ and $n \to +\infty$ in such a way that their product is finite (cf., \cite{Damour:2008yg}). We can actually use the relation found in \eqref{eq:binding_HJ} between $J$ and $\mathcal{E}$ to link $n$ with $J$: interestingly, this gives
\begin{align}
\hbar\, n \stackrel{\hbar \to 0}{\longrightarrow} J \,,
\end{align}
which means that we can recast the pole structure in terms of $(\mathcal{E},J)$ in the classical limit:
\begin{align}\label{eq:xitol}
 \xi_n \to  \xi(J) = \frac{J}{G_N m_A m_B}  \,, \qquad s^{(1),\pm}_{n,\text{poles}} \to  s^{(1),\pm}_{\text{poles}}(J)\,.
\end{align}
We now compute the classical binding energy at leading order in the PN expansion,
\begin{align}
\epsilon^{(1)}(J) = \frac{G_N^2 m_A^2 m_B^2}{\hbar^2 n^2} = \frac{G_N^2 m_A^2 m_B^2}{J^2} \,,
\end{align}
which matches the leading term in \eqref{eq:binding_energy}\footnote{Note that in the non-relativistic limit, from the Hamilton-Jacobi action one gets $J_{\text{bind}}(\epsilon) = \frac{G_N m_A m_B}{\sqrt{\epsilon}}$.}.

There is also a set of zeros in the physical sheet which appear for $A_0 = - n \,, n \in \mathbb{Z}_{> 0}\,$: 
\begin{align}
s^{(1),\pm}_{n,\text{zeros}} = m_A^{2}+m_B^{2} \pm \frac{m_A m_B}{\sqrt{2}}\,  \sqrt{4 -\xi_n^{2} -\xi_n \sqrt{8+\xi_n^2}} \,,
\end{align}
which are the mirror images of the poles, as dictated by S-matrix unitarity in the case of a spherically symmetric potential (see also ~\cite{Frautschi:1963}). We can compute the binding energy for these zeros, corresponding to the valleys in Figure~\ref{fig:analytic-spinless},
\begin{align}
\mathcal{\epsilon}^{(1),\pm}_{n, \text{zeros}} = \frac{2}{\mu} \left[m_A + m_B - \sqrt{m_A^{2}+m_B^{2} + \frac{m_A m_B}{\sqrt{2}}\,  \sqrt{4 -\xi_n^{2} -\xi_n \sqrt{8+\xi_n^2}}} \right]\,.
\end{align}
As expected by analyticity arguments, the zeros in the physical sheet become the poles in the second Riemann sheet (and viceversa). 

\medskip

As in the original Bethe-Salpeter equation, the residue of the amplitude in the $s-$channel bound state pole contains some information about the bound state properties, and in particular its decay rate. Indeed, near the bound state pole, the amplitude is expected to behave like
\begin{align}
f^{<}_{\text{LR}} \stackrel{s \sim s^{(1),+}_{n,\text{poles}}(J)}{\sim} \frac{1}{s-s^{(1),+}_{\text{poles}}(J)}\, \Gamma^{(1)}(E,J)\,,
\end{align}
where $\Gamma^{(1)}(E,J)$ is defined as 
\begin{align}
\Gamma^{(1)}(E,J) :=\text{Res}_{s = s^{(1),+}_{\text{poles}}(J)}\left[ \frac{2 m_A m_B \sqrt{1- y^2}}{E \, \hbar \, \bar{q}_{\bot}^2} \frac{\Gamma (1-A_0^{<})}{ \Gamma (A_0^{<})}  \left(\frac{4 p_{\infty}^2}{\hbar^2 \bar{q}_{\bot}^2}\right)^{-A_0^{<}} \right] \,.
\end{align}
This residue is expected to encode the angular distribution of the two-body decay of the classical gravitational bound state; see~\cite{Anchordoqui:2008hi,Feng:2011qc} for a discussion in similar contexts. Using the well-known fact
\begin{align}
\text{Res}_{x = n} \frac{\Gamma(1-x)}{\Gamma(x)} = \frac{(-1)^{n}}{\left[(n-1)!\right]^2} \,,
\label{eq:poleGamma}
\end{align}
one obtains from \eqref{eq:leading_gamma}
\begin{align}
\label{eq:residuespinless}
\Gamma^{(1)}(E,J)=& \frac{(-1)^{n(J)}}{\left[(n(J)-1)!\right]^2}\, \left(\frac{\hbar^2 \bar{q}_{\bot}^2}{4 p_{\infty}^2}\right)^{n(J)} \nonumber \\
& \times\,\frac{m_A m_B \left(\xi(J)^2- \xi(J)\,\sqrt{\xi(J)^2+8} +4\right)^{2} (\xi(J) + \sqrt{\xi(J)^2+8})}{8 \sqrt{2} G_N E \bar{q}^2 \,\left(\xi(J)^2+ \xi(J)\,\sqrt{\xi(J)^2+8} +8\right) \, \sqrt{4-\xi(J)^2+\xi(J) \sqrt{\xi(J)^2+8}} }  \,, 
\end{align}
where we have taken advantage of the change of variables to parametrize $n$ and $\xi_n$ in terms of $J$. The meaning of $\Gamma^{(1)}(E,J)$ is clear: it is the decay rate for the classical gravitational bound state to break into two free particle states of momenta $p_A$ and $p_B$\footnote{If we associate states to operators, this can be thought as the (absolute square of) the OPE coefficient for the bound state operator (labelled by the energy $E$ and by the quantum number $n$) in the OPE of the external scalar operators. For black-holes in AdS, this can be made completely rigorous by studying the contribution of double-twist operators (associated to the bound state with an anomalous dimension $\Delta$ and quantum numbers $n,l$) to the OPE of the external scalars on the CFT side \cite{Dodelson:2022eiz}.}
\begin{align}
\Gamma^{(1)}(E,J) \propto \left|\Psi^{(1),<}_{\text{bound}}(E,J)\right|^2 = \left|\braket{p_A p_B}{\Psi^{(1),<}_{\text{bound},n(J)}}\right|^2 \,,
\end{align}
where $\Psi^{(1),<}_{\text{bound}}(E,J)$ is the bound state wavefunction, which is the classical counterpart of the solution of the vertex function equation in \eqref{eq:Bethe-Salpetereq}. It is an open problem to find the classical vertex function equation, which would allow us to bypass the extraction of the residue in order to find the bound state wavefunction directly from the classical kernel.

\medskip

At the next-to-leading resummation (NLR) order, the analytic structure of the amplitude becomes more complicated, but the integral in \eqref{eq:BS-solution-momentum} can still be performed analytically with the 2PM classical kernel in \eqref{eq:K_2PM}. In the scattering region we obtain:
\begin{align}
\mathrm{i} \mathcal{M}^{\text{cl},>}_{4,\text{NLR}}&(p_A,p_B;q_{\bot}) = \frac{8 \pi\, m_A m_B\, \sqrt{y^2-1}}{\hbar^2}  \\
& \times \Bigg[2\, \frac{\Gamma \left(1 - A_0^{>}\right)}{\Gamma \left(A_0^{>}\right)}\, \frac{1}{\bar{q}_{\bot}^2} \left(\frac{4 \mu_{\text{IR}}^2 }{\bar{q}_{\bot}^2}\right)^{-A_0^{>}} \, _0F_3\left(;\frac{1}{2},A_0^{>}, A_0^{>};-\frac{(A_1^{>})^2 \bar{q}_{\bot}^2}{16}\right) \nonumber \\
&+\frac{A_1^{>}\, \Gamma \left(\frac{1}{2}- A_0^{>}\right)}{\Gamma \left(\frac{1}{2} + A_0^{>}\right)}\, \frac{1}{\bar{q}_{\bot}} \left(\frac{4 \mu_{\text{IR}}^2}{\bar{q}_{\bot}^2}\right)^{-A_0^{>}}  \, _0F_3\left(;\frac{3}{2},\frac{1}{2}+A_0^{>},\frac{1}{2}+A_0^{>};-\frac{(A_1^{>})^2 \bar{q}_{\bot}^2}{16 }\right)\nonumber \\
&+\left(-A_1^{>}\right)^{2-2 A_0^{>}}\, \mu_{\text{IR}}^{-2 A_0^{>}}\,  \Gamma \left(-2+2 A_0^{>}\right) \, _0F_3\left(;1,\frac{3}{2}-A_0^{>},2- A_0^{>};-\frac{(A_1^{>})^2 \bar{q}_{\bot}^2}{16}\right)\Bigg]\,,\nonumber
\label{eq:subleading_gamma}
\end{align}
with $A_0^{>}$ defined as in \eqref{eq:A0plusdef} and
\begin{align}
A_1^{>}:= i\,\frac{3 \pi}{4 \hbar}\, G_N^2\, m_A m_B\, (m_A + m_B)\, \frac{5 y^2 - 1}{\sqrt{y^2 - 1}} \,.
\end{align}
To our knowledge, this is the first time that this closed-form expression for $\mathcal{M}^{\text{cl},>}_{4,\text{NLR}}$ has appeared in the literature.

\begin{figure}[h!]
 \centering
 \includegraphics[width=0.7\textwidth]{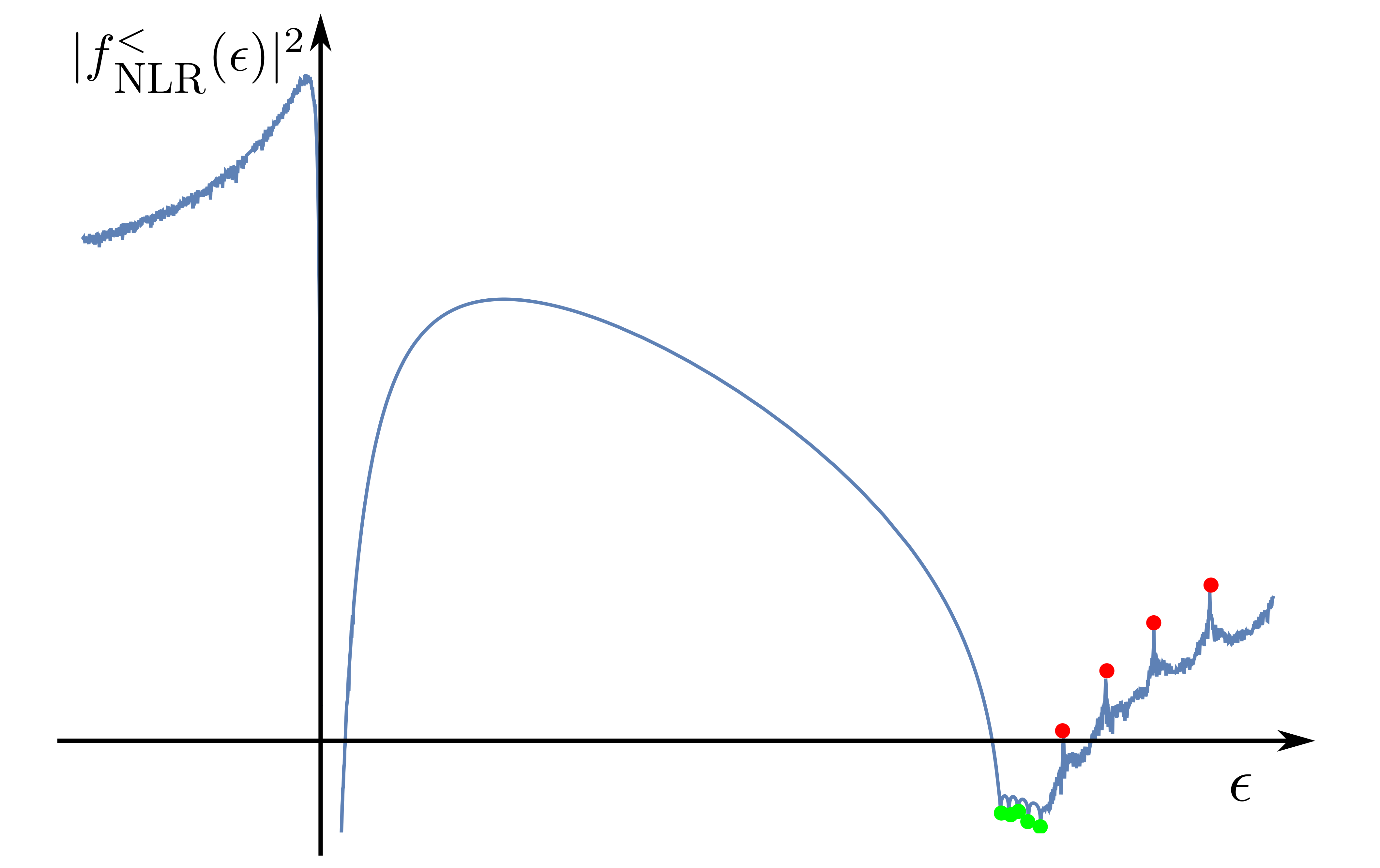} % second 
 \label{fig:analytic-spinless-loop}
  \caption{The cross-section integrand $|f^{<}_{\text{NLR}}(\epsilon)|^2$ as a function of the binding energy $\epsilon$. While the analytic structure is complicated, the presence of a set of poles (in red) and zeros (in green) for $\epsilon>0$ has been identified numerically.}
\end{figure}

The classical wavefunction can then be computed from the matching in \eqref{eq:bound_matching} after the analytic continuation \eqref{eq:B2By}, i.e.
\begin{align}
f^{<}_{\text{NLR}}(y,q_{\bot}) &= \frac{m_A m_B \sqrt{1- y^2}}{\hbar E} \\
& \times \Bigg[2 \frac{\Gamma \left(1 - A_0^{<}\right)}{\Gamma \left(A_0^{<}\right)} \frac{1}{\bar{q}_{\bot}^2} \left(\frac{4 p_{\infty}^2}{\hbar^2 \bar{q}_{\bot}^2}\right)^{-A_0^{<}} \, _0F_3\left(;\frac{1}{2},A_0^{<}, A_0^{<};-\frac{(A_1^{<})^2 \bar{q}_{\bot}^2}{16}\right) \nonumber \\
&+\frac{A_1^{<} \Gamma \left(\frac{1}{2}- A_0^{<}\right)}{\Gamma \left(\frac{1}{2} + A_0^{<}\right)} \frac{1}{\bar{q}_{\bot}} \left(\frac{4 p_{\infty}^2}{\hbar^2 \bar{q}_{\bot}^2}\right)^{-A_0^{<}}  \, _0F_3\left(;\frac{3}{2},\frac{1}{2}+A_0^{<},\frac{1}{2}+A_0^{<};-\frac{(A_1^{<})^2 \bar{q}_{\bot}^2}{16}\right)\nonumber \\
&+\left(-A_1^{<}\right)^{2-2 A_0^{<}} \left(\frac{p_{\infty}^2}{\hbar^2}\right)^{- A_0^{<}}  \Gamma \left(-2+2 A_0^{<}\right) \, _0F_3\left(;1,\frac{3}{2}-A_0^{<},2- A_0^{<};-\frac{(A_1^{<})^2 \bar{q}_{\bot}^2}{16}\right)\Bigg]\,,\nonumber 
\end{align}
with 
\begin{align}\label{eq:A1minusdef}
A_1^{<}:= \frac{3 \pi}{4 \hbar}\, G_N^2\, m_A m_B\, (m_A + m_B)\, \frac{5 y^2 - 1}{\sqrt{1 - y^2}} \,.
\end{align}

The structure is significantly more complicated than the LR amplitude, as it involves (products of) non-trivial hypergeometric functions, and we have not been able to obtain analytic expressions for poles or zeroes of the NLR amplitude. Nevertheless, we have been able to show numerically -- as illustrated in Figure~\ref{fig:analytic-spinless-loop} -- that there are both poles and zeros on the physical sheet. Moreover, the branch cut structure is exactly the same as for the LR amplitude, i.e. it is only related to the physical threshold at this order. We hope to return to a full analysis of this interesting amplitude in the near future.

%%%%%%%%%%%%%%%%%%%%%%%%%%%%%%%

\subsection{Gravitational bound state of spinning particles}

At least to leading order in the PM expansion, Kerr black holes can be treated as classically spinning point particles~\cite{Vines:2017hyw}. As both spinning particles and black holes are uniquely identified by their set of charges (i.e., mass and spin), at least some aspects of this correspondence are expected to extend beyond leading order after imposing a suitable localization; there has been progress on this idea from various directions in recent years~\cite{Guevara:2018wpp,Guevara:2019fsj,Siemonsen:2019dsu,Bern:2020buy,Bautista:2021wfy,Chiodaroli:2021eug,Aoude:2022trd,Aoude:2022thd,Saketh:2022wap,Bern:2022kto,Cangemi:2022bew,Bautista:2022wjf}. Studying classical gravitational bound states of spinning particles is therefore relevant to the long range bound state dynamics of Kerr black holes.

Define the complex quantities
\begin{align}
a := (a_+)_1 + i (a_+)_2 \,, \qquad q:= q_1 + i q_2\,,
\end{align}
in terms of which we write the leading resummation of the classical spinning amplitude~\cite{Adamo:2021rfq}:
\begin{align}
\mathrm{i} \mathcal{M}^{\text{cl},>}_{4,\text{LR}}&(\{p_A,a_A\},\{p_B,a_B\};q)\nonumber \\
&= -\frac{4 \pi\, m_A m_B\, \sqrt{y^2-1}}{\hbar^2} \,(\mu_{\text{IR}}^2)^{-A_0^{<}} \, |2 a|^{2 + 2(\alpha^{>} + \beta^{>})}\, e^{-\frac{i}{2} (q a + q^* a^*)} \nonumber \\
& \quad \times \Big[\sin (\pi (\alpha^{>} + \beta^{>}))\, I_2 (2 q^* a^*)\, I_3(2 q a) \nonumber \\
& \quad \qquad \qquad  + \sin (\pi \alpha^{>})\, I_1 (2 q^* a^*)\, I_3(2 q a) + \sin (\pi \beta^{>})\, I_1 (2 q a)\, I_2 (2 q^* a^*) \Big]\,,
\label{eq:leading_gamma-spin}
\end{align} 
with the hypergeometric integrals
\begin{align}
I_{1}(q) &= \int_{0}^{1} \mathrm{~d} w\, e^{-i\, q w}\, w^{\alpha}\,(1-w)^{\beta}\nonumber \\
&=\frac{\Gamma(1+\alpha) \Gamma(1+\beta)}{\Gamma(\beta+\alpha+2)} \,M(1+\alpha, 2+\alpha+\beta,-i q) \,,\nonumber \\
I_{2}(q) &= \int_{1}^{\infty} \mathrm{d} w\, e^{-i\, q w}\, w^{\alpha}\,(w-1)^{\beta} \nonumber \\
&=\frac{\pi}{\sin (\pi(\alpha+\beta))} \Bigg(\frac{\Gamma(1+\beta)}{\Gamma(\alpha+\beta+2) \Gamma(-\alpha)}\, M(1+\alpha, 2+\alpha+\beta,-i q) \nonumber \\
& \qquad \qquad \qquad \qquad \qquad \qquad \qquad -\frac{(i q)^{-\alpha-\beta-1}}{\Gamma(-\alpha-\beta)}\, M(-\beta,-\alpha-\beta,-i q) \Bigg)\,, \nonumber \\
I_{3}(q) &=\int_{-\infty}^{0} \mathrm{~d} w\, e^{-i\, q w}\,(-w)^{\alpha}\,(1-w)^{\beta} \nonumber \\
&=\Gamma(\alpha+1)\, U(\alpha+1, \alpha+\beta+2,-i q)\,,
\label{eq:hyper_integrals-spin}
\end{align}
defined in terms of the confluent hypergeometric functions of Kummer and Tricomi ($M$ and $U$, respectively) and the coefficients
\begin{align}
\alpha^{>} = -i\,\frac{G_N}{\hbar}\, m_A m_B\, \frac{(y + \sqrt{y^2 - 1})^2}{2\, \sqrt{y^2 - 1}} \,, \qquad \beta^{>} = -i\,\frac{G_N}{\hbar}\, m_A m_B\, \frac{(y - \sqrt{y^2 - 1})^2}{2\, \sqrt{y^2 - 1}}\,.
\end{align}

\begin{figure}
    \centering
        \includegraphics[width=0.55\textwidth]{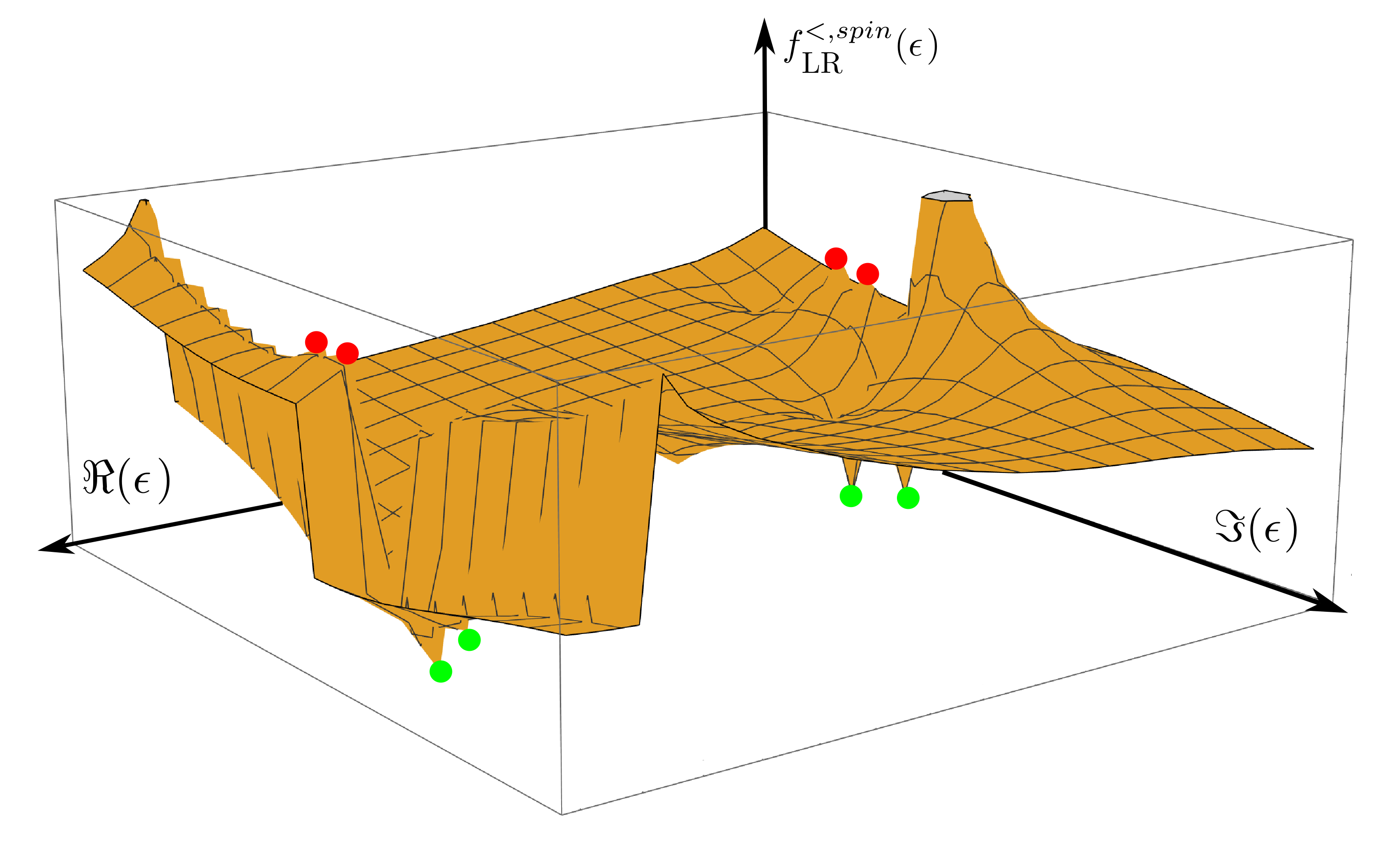} 
    \caption{The analytic structure of the leading resummed spinning amplitude as a function of the complex binding energy $\epsilon$; poles are highlighted in red and zeros in green.}
    \label{fig:analytic-spin}
\end{figure}

Analytic continuation of the amplitude below the threshold, using \eqref{eq:B2By}, and using the matching condition in \eqref{eq:bound_matching}, gives
\begin{align}\label{eq:leading_gamma-spinB}
&f^{<,\text{spin}}_{\text{LR}}(y,q,a) = - \frac{\, m_A m_B\, \sqrt{1 - y^2}}{2 \hbar E}\, \left(p_{\infty}^2/\hbar^2\right)^{-A_0^{<}} |2 a|^{2 + 2(\alpha^{<} + \beta^{<})}\, e^{-\frac{i}{2} (q a + q^* a^*)} \nonumber \\
& \qquad\qquad \qquad \quad \times \Big[\sin (\pi (\alpha^{<} + \beta^{<}))\, I_2 (2 q^* a^*)\, I_3(2 q a) \nonumber \\
& \qquad \qquad \qquad  \qquad \quad + \sin (\pi \alpha^{<})\, I_1 (2 q^* a^*)\, I_3(2 q a) + \sin (\pi \beta^{<})\, I_1 (2 q a)\, I_2 (2 q^* a^*) \Big]\,,
\end{align}
with 
\begin{align}
\alpha^{<} = -\frac{G_N}{\hbar}\, m_A m_B\, \frac{(y + i \sqrt{1 - y^2})^2}{2\, \sqrt{1- y^2}} \,, \qquad \beta^{<} = -\frac{G_N}{\hbar}\, m_A m_B\, \frac{(y - i \sqrt{1 - y^2})^2}{2\, \sqrt{1 - y^2}}\,,
\end{align}
and the hypergeometric integrals now understood to be defined using $\alpha^<$, $\beta^<$.  We have represented the cross section integrand $|f^{<,\text{spin}}_{\text{LR}}(\epsilon)|^2$ associated to the spinning amplitude as a function of the complex binding energy $(\Re (\epsilon),\Im (\epsilon))$ in Figure~\ref{fig:analytic-spin}, where the location of the of poles and zeros have been highlighted in red and green, respectively. 

The wavefunction \eqref{eq:leading_gamma-spinB} has simple poles whenever
\begin{align}
\alpha^{<} = -n_{\alpha}\,,\,\,\,n_{\alpha}\in \mathbb{Z}_{> 0}  \qquad \text{or} \qquad \beta^{<} = -n_{\beta}\,,\,\,\,n_{\beta} \in \mathbb{Z}_{> 0}\,,
\end{align}
which means that there are two sequences of s-channel poles: 
\begin{align}
s^{(1)}_{n_{\alpha},\text{poles}} &= m_A^2+m_B^2-\frac{m_A m_B}{3}  \left[i \xi_{n_{\alpha}} -\frac{i \sqrt[3]{2 \xi_{n_{\alpha}} } \left(\xi_{n_{\alpha}} ^2-12\right)}{\sqrt[3]{P(\xi_{n_{\alpha}})}}-\frac{i \sqrt[3]{P(\xi_{n_{\alpha}})}}{ \sqrt[3]{2 \xi_{n_{\alpha}} }}\right]\,, \nonumber \\
s^{(1)}_{n_{\beta},\text{poles}} &= m_A^2+m_B^2+\frac{m_A m_B}{3} \left[i \xi_{n_{\beta}} + \frac{(-1)^{5/6} \sqrt[3]{2 \xi_{n_{\beta}} } \left(\xi_{n_{\beta}} ^2-12\right)}{ \sqrt[3]{P(\xi_{n_{\beta}})}}+\frac{\sqrt[6]{-1} \sqrt[3]{P(\xi_{n_{\beta}})}}{\sqrt[3]{2 \xi_{n_{\beta}} }} \right]\,,  \nonumber \\
P(\xi_{n}) &= -2 \xi_{n}^4-6 \left(\sqrt{12 \xi_{n}^2+81}+12\right) \xi_{n}^2-3 \left(\sqrt{12 \xi_{n}^2+81}-9\right)\,,
\label{eq:poles_spinning}
\end{align}
written in terms of the dimensionless variable $\xi_n$ introduced in \eqref{eq:xi_def}. One may wonder why there are two sequences of poles; these arise from each $\eta=\pm$ contribution to \eqref{eq:Kspin_1PM}, and are a consequence of the peculiar structure of spinning amplitudes for Kerr black holes, which (at least at this order) can be obtained using a worldsheet-type chiral description~\cite{Guevara:2020xjx}. Thus, the classification of the poles is naturally done within these ``chiral'' sectors.

In addition to the poles, there are also four sequences of zeros:
\begin{align}
\alpha^{<} &= n_{\alpha}\,,\,\,\,n_{\alpha}\in \mathbb{Z}_{> 0}  \to s^{(1)}_{n_{\alpha},\text{zeros}}\,, s^{(2)}_{n_{\alpha},\text{zeros}} \nonumber \\ 
\beta^{<} &= n_{\beta}\,,\,\,\,n_{\beta}\in \mathbb{Z}_{> 0}  \to  s^{(1)}_{n_{\beta},\text{zeros}}\,, s^{(2)}_{n_{\beta},\text{zeros}} 
\end{align}
two of which are analytic continuations of the poles on the physical sheet
\begin{align}
s^{(1)}_{n_{\alpha},\text{zeros}} = s^{(1)*}_{n_{\beta},\text{poles}}\,, \qquad s^{(1)}_{n_{\beta},\text{zeros}} = s^{(1)*}_{n_{\alpha},\text{poles}} \,.
\end{align}
while $s^{(2)*}_{n_{\alpha},\text{zeros}} = s^{(2)}_{n_{\beta},\text{zeros}}$. While there are exact formulae for these zeros, they are not very illuminating so we omit them here.

The physical meaning of the poles for the spinning amplitude is slightly subtle. As noted in~\cite{Adamo:2021rfq}, these poles are \emph{complex}, which clashes with the expectation that the two-body system is purely conservative at this order (dissipative effects, like superradiance, show up only at higher orders in perturbation theory \cite{Detweiler:1980uk}). Furthermore, the poles in \eqref{eq:poles_spinning} do not scale with the spins $a_A$, $a_B$: this is because at 1PM order, the observables for bound state dynamics do not depend on spin and, accordingly, the location of bound state poles on the physical sheet should be spin-independent. Only the first sequence of poles in \eqref{eq:poles_spinning} correspond to the two-particle bound state, with binding energy
\begin{align}
\mathcal{\epsilon}^{(1)}_{n_{\alpha}} = \frac{2}{\mu} \left(m_A + m_B - \sqrt{s^{(1)}_{n_{\alpha},\text{poles}}} \right) \,,
\label{eq:boundstatespin}
\end{align}
while the other set of poles corresponds to the particle-antiparticle bound state energy\footnote{Indeed,$s^{(1)}_{n_{\beta},\text{poles}}$ is mapped to $s^{(1)}_{n_{\alpha},\text{poles}}$ via $s^{(1)}_{n_{\beta},\text{poles}} \to 2 (m_A^2 + m_B^2) - s^{(1)}_{n_{\alpha},\text{poles}}$ for $n_{\alpha} = n_{\beta}$.} $\mathcal{\epsilon}^{(1)}_{n_{\beta}} = 2 \left(m_A + m_B - \sqrt{s^{(1)}_{n_{\beta},\text{poles}}}\right)/\mu$. 

We believe that the complex-valued nature of the poles is an accident arising from using a relativist, amplitude-based description for finite-size objects: the leading PN contribution to the binding energy is real-valued and can be recovered directly from the poles. Taking the classical limit amounts to requiring $n_{\alpha}, n_{\beta} \to \infty$ while $\hbar \to 0$ in such a way that\footnote{This map can be obtained, as in the spinless case, by using the relation for the binding energy $\epsilon(J)$ coming from the bound Hamilton-Jacobi action. This is where the additional factor $1/2$ comes from.}
\begin{align}
\hbar\, n_{\alpha}\,, \hbar\, n_{\beta} \to \frac{J}{2} + \mathcal{O}\left(\frac{G_N}{J}\right)\,, \qquad s^{(1)}_{n_{\alpha},\text{poles}} \to s_{\text{poles}}^{(1)}(J)\,,\quad s^{(1)}_{n_{\beta},\text{poles}} \to s_{\text{poles}}^{(1)}(J) \,,
\end{align}
where, compared to the spinless case, there are corrections to the map from the quantum numbers $n_{\alpha}, n_{\beta}$ to $J$ and there is a factor of $1/2$ in the leading term. Taking the leading PN contribution of the two-particle bound state in \eqref{eq:boundstatespin}, we obtain 
\begin{align}
\epsilon^{(1)}(J) = \frac{G_N^2\, m_A^2 m_B^2}{4 \hbar^2\, n_{\alpha}^2}  = \frac{G_N^2\, m_A^2 m_B^2}{J^2} \,,
\end{align}
which is consistent with the earlier result \eqref{eq:binding_energy_spin} for the bound state binding energy. 

%%%%%%%%%%%%%%%%%%%%%%%%%%%%%

\subsection{A relativistic analogue of Sommerfeld enhancement}

\begin{figure}[h]
\centering
\includegraphics[scale=0.82]{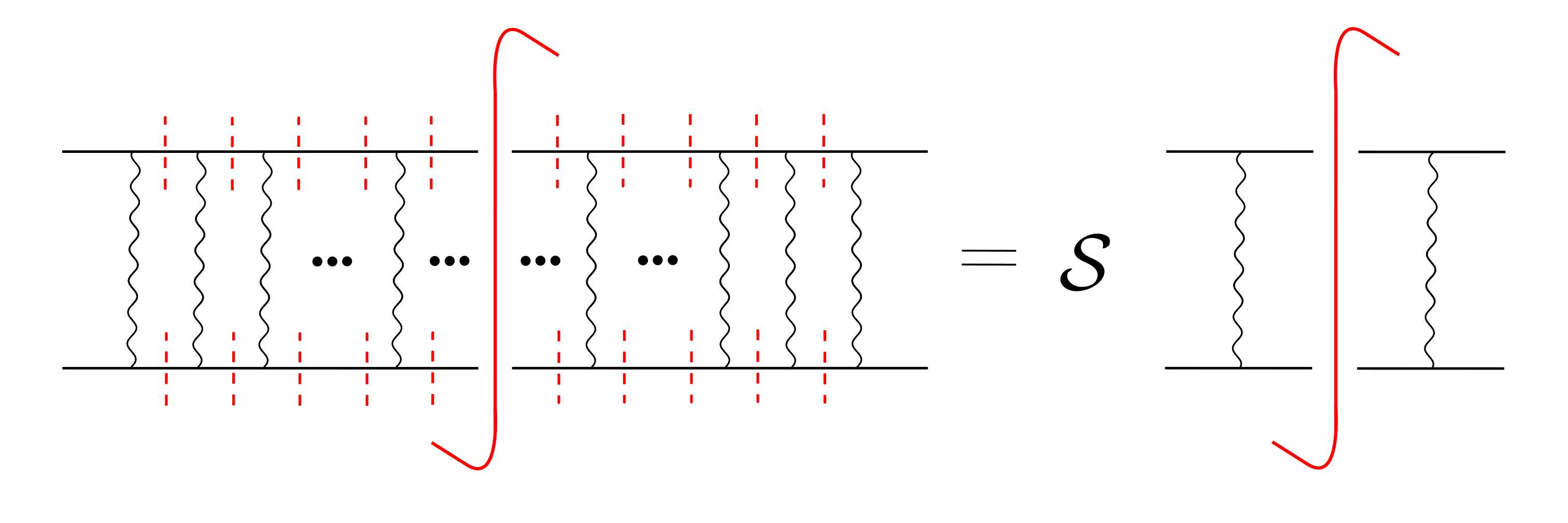}
\caption{The Sommerfeld factor $\mathcal{S}$ can be defined as the ratio of the cross section for resummed classical amplitudes (defined via iteration of a kernel) to the cross section for the single kernel contribution. The figure illustrates this for the leading resummation (i.e., tree-level kernel).}
\label{fig:sommerfeld}
\end{figure}

It is natural to ask what the effect of resummation is on the elastic cross-section below threshold; at leading order, this amounts to asking whether the sum of ladder and crossed-ladder diagrams gives some additional (observable) effects compared to the tree-level Born approximation (see Figure~\ref{fig:sommerfeld}). In non-relativistic quantum mechanics, it is well-known that scattering cross sections in an attractive potential are enhanced relative to the ``bare'' cross section, a phenomenon known as \emph{Sommerfeld enhancement}~\cite{Sommerfeld:1931qaf}. This has a wide array of applications in modern physics, including the phenomenology of non-relativistic dark matter bound states~\cite{Arkani-Hamed:2008hhe,Slatyer:2009vg}. In a nutshell, Sommerfeld enhancement in non-relativistic physics is the enhancement of the cross-section due the resummation of ladder-type diagrams from the Bethe-Salpeter equation; this is a simple consequence of solving the Schr\"odinger equation in the case of an attractive potential~\cite{Cassel:2009wt,Iengo:2009ni}. In relativistic physics, there is still an effect on cross-section-like observables, but (as we will see) this is not always an enhancement.

Consider the elastic cross section for the leading resummed amplitude, and define a Sommerfeld factor as
\begin{align}
\mathcal{S}_{\text{LR}}(y) : = \frac{\sigma^{\text{cl}}_{\text{LR}}}{\sigma^{(0),\text{cl}}_{\text{LR}}} = \frac{\int \mathrm{d} \Omega \, | f_{\text{LR}}(y,q_{\bot}(\theta))|^2}{\int \mathrm{d} \Omega \, |f^{(0)}_{\text{LR}}(y,q_{\bot}(\theta))|^2} \,,
\label{eq:Sommerfeld_factor}
\end{align}
where $f^{(0)}_{\text{LR}}$ represents the classical wavefunction arising from only the tree-level, single graviton exchange (i.e., the Born approximation). Above threshold, for scattering orbits, the resummed amplitude \eqref{eq:leading_gamma} is simply the tree-level term times a phase, so $\mathcal{S}^{>}_{\text{LR}} = 1$ and there is no difference in between the cross sections. However, below threshold (for bound orbits), we find
\begin{align}
\mathcal{S}^{<}_{\text{LR}}(y) = \left[\frac{\Gamma(1-A_0^{<}(y))}{\Gamma(1+A_0^{<}(y))}\right]^2 \,,
\end{align}
where $A_0^{<}$ is defined in \eqref{eq:A0minusdef}. 

In the non-relativistic limit we obtain the Newtonian result
\begin{align}
A_0^{<} |_{\text{Newton}} = \frac{G_N}{\hbar}\, \frac{m_A m_B}{\sqrt{\epsilon}} \,,
\label{eq:PN_bound_alpha}
\end{align}
which, due to its (positive) monotonic behaviour in $\epsilon$, implies the remarkable inequality 
\begin{align}
\mathcal{S}^{<}_{\text{LR}}(y) |_{\text{Newton}} \geq 1\,.
\end{align}
This is precisely the Sommerfeld enhancement
\begin{align}
\sigma^{\text{cl}}_{\text{LR}}|_{\text{Newton}} > \sigma^{(0),\text{cl}}_{\text{LR}}|_{\text{Newton}} \,,
\end{align}
for the resummed cross section, which is expected on general grounds for the Newtonian $1/r$ attractive potential.

\begin{figure}[h]
\centering
\includegraphics[scale=0.55]{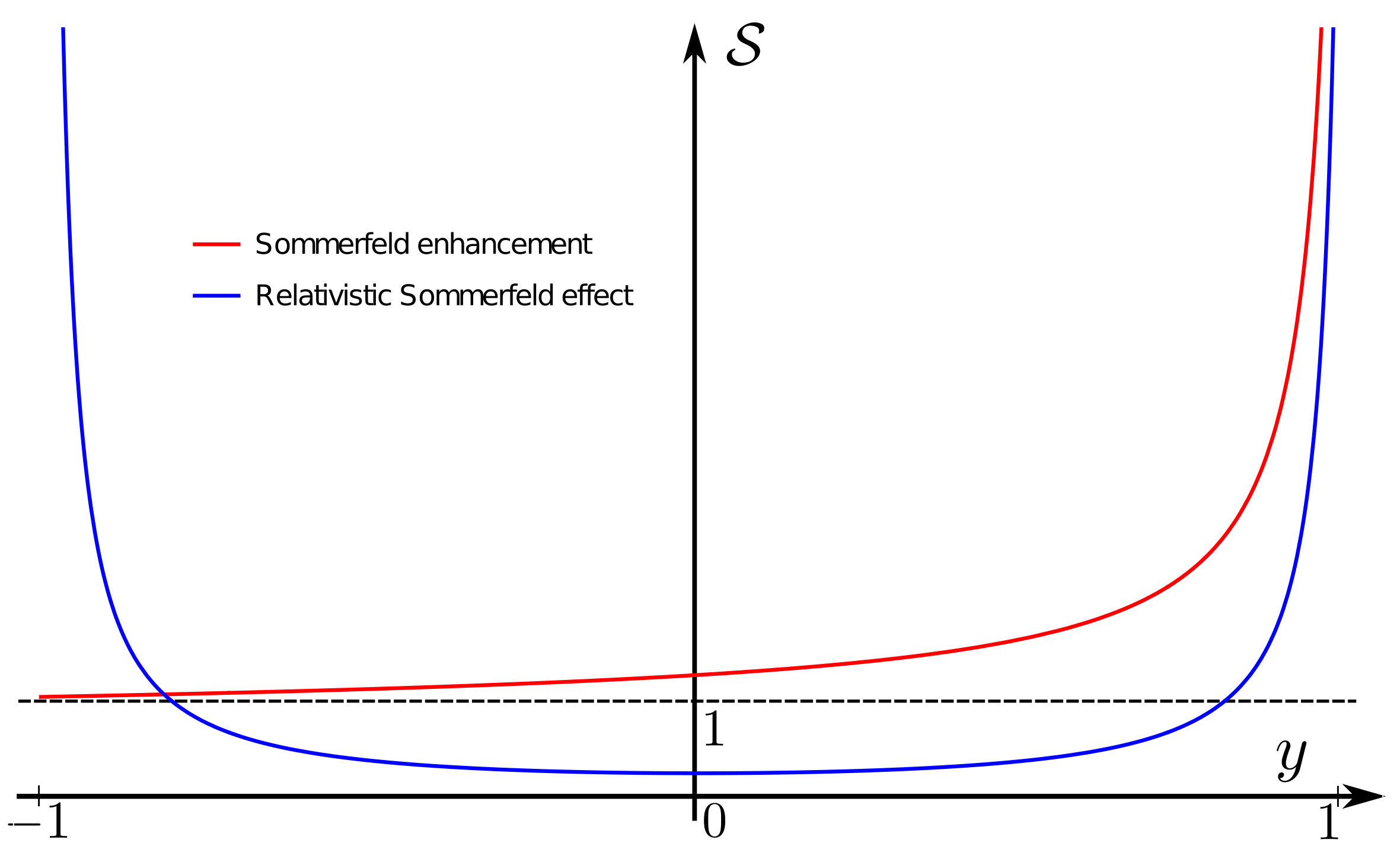}
\caption{The Sommerfeld factor $\mathcal{S}_{\text{LR}}(y)$ for the leading resummed classical amplitude is shown as a function of the rapidity, both for the 1PM and the Newtonian potential.}
\label{fig:sommerfeld_enhancement}
\end{figure}

In the relativistic case, there is a dependence on the rapidity $y$ in the coefficient of the 1PM potential which makes it repulsive at large $\epsilon > 0$ (i.e. for small $y \ll 1$), and therefore the Sommerfeld factor will not be an enhancement anymore in this kinematic region\footnote{R.G. would like to thank T. Slatyer for comments about the possible phenomenological relevance of this effect for classical relativistic bound states.}. This is illustrated in Figure~\ref{fig:sommerfeld_enhancement}, where both the relativistic and non-relativistic Sommerfeld factors $\mathcal{S}_{\text{LR}}$ have been represented as a function of the rapidity in the bound state region: while $\mathcal{S}^{<}_{\text{LR}}$ goes below 1 for $y \ll 1$, we notice that $\mathcal{S}^{<}_{\text{LR}} |_{\text{Newton}} \geq 1$ for all $y$. A similar effect can be found numerically for the NLR resummed amplitude \eqref{eq:leading_gamma}. We leave a detailed analysis of this effect and its implications for future work.

%%%%%%%%%%%%%%%%%%%%%%%%%%
%%%%%%%%%%%%%%%%%%%%%%%%%%

\section{Conclusion}
\label{sec:concl}

The imperatives for studying bound states in quantum field theory are stronger today than ever, due to their relevance in the effective field theory description of the gravitational binary dynamics of compact objects. In this paper, we have taken some initial steps in improving the theory of relativistic effects in bound states by deriving a new classical version of the Bethe-Salpeter equation to describe two-body bound states of point particles in general relativity.

We began by revisiting the well-studied eikonal resummation at leading order for two minimally coupled spinless or spinning massive particles. We showed that a fully classical interpretation can be given to such an amplitude whereby the ladder and crossed-ladder diagrams are all generated by unitarity from a single coherent state of virtual gravitons, which represents the gravitational field at the quantum level. This supports a new physical principle: for classical physics to emerge, one must perform a symmetrization over (real and virtual) graviton exchanges. Indeed, classically one should never be able to distinguish any single graviton for a classical gravitational field.

We then applied this principle to the original (relativistic) Bethe-Salpeter equation, obtaining a new crossing-symmetric recursion relation for classical amplitudes where the two-body massive propagator is replaced by its on-shell version and the kernel consists of 2MPI diagrams in the space of classical diagrams. While this is straightforward for spinless particles, it requires the suppression of double (and higher) commutator terms for spinning particles in the classical limit, which is expected to happen for a set of minimal couplings like those for Kerr black holes\footnote{Finite size effects could still be studied in this way, but might require additional contributions to the classical recursion relation.}. As a direct consequence of the new recursion relation, we showed that a single tree-level classical kernel generates, via iteration, both ladder and crossed-ladder diagrams, consistent with expectations. Then, using the fact that convolutions in momentum space become products in impact parameter space, we diagonalized the recursion relation and solved it exactly by exponentiating the classical kernel at all orders in perturbation theory.

Having obtained a compact solution of the classical Bethe-Salpeter equation in impact parameter space, we then connected it to physical bound state observables in the conservative regime. We showed that the classical kernel is essentially the Hamilton-Jacobi action, which encodes both scattering and bound observables in a natural way, recovering many of the results in the literature. Building on this, we explored the analytic structure of the resummed classical amplitudes in momentum space, where bound states appear as energy poles below threshold in the physical sheet. We then provided a direct connection between such amplitudes and the classical wavefunction of the system, which directly encodes the information about the bound states. At leading order, poles (and zeros) can be computed analytically from the amplitude, while at the next-to-leading order we found new and more intricate hypergeometric structure where the location of the poles (and zeros) could only be determined numerically. We were able to show explicitly how, at leading order, the standard expression of the binding energy in terms of the angular momentum is recovered from the classical limit of the bound state poles: in the correspondence limit, a combination of the Planck constant $\hbar$ and of the quantum number $n$ labelling the states becomes the classical angular momentum $J$. Moreover, we showed that the residue at the poles is related to the two-body bound state wavefunction, as expected from the Bethe-Salpeter equation. Finally, we commented on an effect which arises purely from resummation for any cross section in the gravitational two-body problem, which generalizes the well-known Sommerfeld enhancement to the relativistic setting. 

We are just scratching the surface of a fascinating topic in quantum field theory, which we believe deserves more attention. There are a number of open questions on which we hope to return in the near future. First of all, it would be interesting to explore a more direct connection of the classical wavefunction with bound state observables beyond the Hamilton-Jacobi formulation. This would require understanding how periodic functions arise from the pole structure after analytically continuing both in the energy and the angular momentum of the system. Furthermore, an immediate generalization of our approach would be to include radiation to account for the full bound state dynamics: the 5-point amplitude (including the emission of a graviton) is expected to enter in the recursion, giving some finite width to the poles due to radiative effects. Since the Bethe-Salpeter equation comes from the LSZ reduction of the Schwinger-Dyson equation for the 4-point Green's function, a natural approach would be to consider the recursion relation for the relevant 5-point Green's function. We hope that such extensions might suggest a new, general perspective on how to connect bound state observables to scattering ones, alleviating some of the problems in the analytic continuation for the radiative sector. Alternatively, a better connection with perturbation theory for bound states must be established. 

Another important open problem is to push the study of the analytic structure of resummed amplitudes in momentum space, as they provide important (analytic) examples of how bound states are encoded in perturbation theory. For example, we still do not fully understand the analytic structure of classical spinning amplitudes, which are particularly relevant for Kerr black holes. Finally, this work might have other phenomenological applications, like for bound states of dark matter, in settings where relativistic effects must be accounted for. A deeper understanding of all these aspects might provide a new way to attack the problem of bound states in quantum field theory, which is crucial for phenomenological and theoretical applications.

\acknowledgments

We thank Uri Kol for initial collaboration on this project. We would like to thank A. Cristofoli, L. de la Cruz, A. Ilderton, Z. Liu, D. O'Connell, T. Slatyer, P. Tourkine, C. White and M. Zeng for interesting discussions. TA is supported by a Royal Society University Research Fellowship and by the Leverhulme Trust (RPG-2020-386). RG would like to thank the organizers of the ``High-Precision Gravitational Waves'' program at the Kavli Institute for Theoretical Physics (KITP) for providing a stimulating environment where part of these ideas have been developed. This research was supported in part by the National Science Foundation under Grant No. NSF PHY-1748958.

\appendix

%%%%%%%%%%%%%%%%%%%
%%%%%%%%%%%%%%%%%%%

\section{Exponentiation of the 3-pt function for the classical Lorentzian metric}
\label{appendixA}

The goal of this appendix is to show how the linearized Lorentzian metric of a point-like background, such as Schwarzschild or Kerr, can be recovered from a 3-point function with two massive on-shell scalars and one off-shell graviton in the classical limit. This was first done in~\cite{Cristofoli:2020hnk}, and is also closely related to the analogous calculation in $(2,2)$-signature~\cite{Monteiro:2020plf}. Essentially, we compute the classical expectation value of the linearized graviton operator $H_{\mu \nu}$ in the state generated by the time-evolution of a massive particle (with or without spin). The classical limit of the in-in formalism has been related to the KMOC formalism~\cite{Kosower:2018adc} in~\cite{Britto:2021pud}, and we refer the reader to those papers for further details and references. 

We start with the scalar case, which is simpler. Our initial state is a massive on-shell particle state $\ket{p}$ in general relativity,
\begin{align}
\ket{\phi} =  \int \mathrm{d} \Phi(p)\, \phi_{v}(p)\, \ket{p} \,,
\end{align}
where $\phi_{v}(p)$ is a wavepacket localized around the classical momentum $p \sim M v^{\mu}$ as $\hbar \to 0$. Given this state, we would like to compute the expectation value of the graviton field operator $H_{\mu \nu}(x)$ in the time-evolved state
\begin{align}
\langle H_{\mu \nu}(t,\vec{x}) \rangle=\left\langle \phi \left|\bar{T} e^{-i \int_{-\infty}^t \mathrm{~d} t^{\prime} \mathcal{L}_{\text{int}}(t^{\prime})} H_{\mu \nu}(t,\vec{x}) T e^{i \int_{-\infty}^t \mathrm{~d} t^{\prime} \mathcal{L}_{\text{int}}(t^{\prime})}\right| \phi \right\rangle \,,
\end{align}
where $T$ (resp. $\bar{T}$) denotes the time-ordering (resp. anti-time ordering) and the interaction Lagrangian is the one for minimally coupled massive particles in general relativity (see Section~\ref{sec:eikonal}). At leading order in perturbation theory, one obtains
\begin{align}\label{eq:graviton_LO}
\langle H_{\mu \nu}(t,\vec{x}) \rangle= 2 \operatorname{Re} i \left(\left\langle \phi \left|H_{\mu \nu}(t,\vec{x}) \int_{-\infty}^t \mathrm{~d} t^{\prime} \, \mathcal{L}_{\text{int}}(t^{\prime})\right| \phi \right\rangle\right) \,.
\end{align}
We wish to evaluate this expectation value at large times $t \to +\infty$, which allows to make contact with the standard flat space Feynman rules with a definite $i \epsilon$ prescription. The metric generated by a single point-particle is naturally a (gauge-dependent) off-shell quantity, since the gravitons emitted from point-like sources are virtual gravitons with $|K_0| > |\vec{K}|$. 

\begin{figure}[h]
\centering
\includegraphics[scale=0.77]{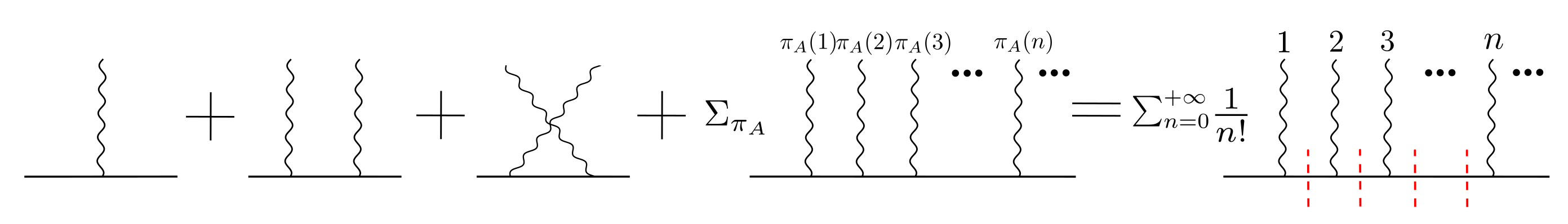}
\caption{The linearized classical metric for point-like backgrounds arises naturally from the tree-level emission of virtual gravitons from an on-shell massive particle line.}
\label{fig:coherent}
\end{figure}

In a plane wave basis, the graviton field can be expressed as 
 \begin{align}
H_{\mu \nu}(x) = \sum_{\sigma' = 0,\pm} \frac{1}{\sqrt{\hbar}} \int \frac{\mathrm{\hat{d}}^4 K}{K^2 + i \epsilon} \left[\varepsilon^{* \sigma'}_{\mu \nu}(K)\, A_{\sigma'}(K)\, e^{-\frac{i K \cdot x}{\hbar}} + \varepsilon^{\sigma'}_{\mu \nu}(K)\, A^{\dagger}_{\sigma'}(K)\, e^{i \frac{K \cdot x}{\hbar}}\right] \,,
\label{eq:virtualH}
\end{align}
where $1/(K^2 + i \epsilon)$ denotes the external graviton propagator and the operators $A_{\sigma'}(K),  A^{\dagger}_{\sigma'}(K)$ provide a convenient way of reproducing the Schwinger-Keldysh diagrammatics where a virtual graviton is annihilated or created in the diagram\footnote{A more careful treatment of the $i \epsilon$ prescription is needed at higher orders.}. While in \eqref{eq:virtualH} we need to consider the sum over both the transverse and the longitudinal polarization vectors in the basis, the latter will eventually decouple in the final expectation value. From \eqref{eq:graviton_LO}, one then finds
\begin{align}
\langle H_{\mu \nu}(t,\vec{x}) \rangle&= \frac{1}{\sqrt{\hbar}} 2 \operatorname{Re} \sum_{\sigma' = 0,\pm} \int \frac{\mathrm{\hat{d}}^4 K}{K^2 + i \epsilon}\, \varepsilon^{* \sigma'}_{\mu \nu}(K)\, e^{-\frac{i K \cdot x}{\hbar}}\, \nonumber \\
&\qquad \qquad \qquad \qquad \qquad \qquad \times i \left(\left\langle \phi \left|A_{\sigma'}(K) \, \int_{-\infty}^{+\infty} \mathrm{~d} t^{\prime} \mathcal{L}_{\text{int}}(t^{\prime})\right| \phi \right\rangle\right) \,,
\end{align}
which amounts to computing the diagrams in Figure~\ref{fig:coherent}, where both massive particle lines are on-shell while the graviton is off-shell. 

The next steps follow easily by generalizing a recent work~\cite{Monteiro:2020plf}, where the authors did a fully on-shell calculation but in (2,2)-signature. By evolving the initial state using the Schwinger-Keldysh formalism, one obtains at leading order 
\begin{align}
(i \hat{M}_3) \ket{\phi} &:= i \left(\int_{-\infty}^{+\infty} \mathrm{~d} t^{\prime} \mathcal{L}_{\text{int}}(t^{\prime}) \right) \ket{\phi} \Bigg|_{\mathcal{O}(\kappa)}  \,, \nonumber \\
i \hat{M}_3 \ket{\phi} \Big|_{\text{cl}} &= \sum_{\sigma} \int \mathrm{d} \Phi(p)\, \phi_v(p) \int \frac{\mathrm{\hat{d}}^4 l}{l^2 + i \epsilon}\,  \hat{\delta}(2 p \cdot l)\, i \mathcal{M}_3^{(0),\text{cl}}(p,l^{\sigma})\, A^{\dagger}_{\sigma}(l) \ket{\phi} \,,
\end{align}
which means the classical linearized metric is given by 
\begin{align}
H_{\mu \nu}^{\text{cl}}(x) &= 2 \operatorname{Re} \sum_{\sigma = \pm} \kappa \int \mathrm{d} \Phi(p)\, \phi_v(p) \nonumber \\
&\qquad\qquad\qquad\times \int \frac{\mathrm{\hat{d}}^4 \bar{K}}{\bar{K}^2 + i \epsilon}\, \varepsilon^{* \sigma}_{\mu \nu}(\bar{K})\, e^{-i \bar{K} \cdot x}\,  \hat{\delta}(2 p \cdot \bar{K})\, i \mathcal{M}_3^{(0),\text{cl}}(p,\bar{K}^{\sigma})  \,.
\end{align}
We can then consider the leading classical resummation of tree-level diagrams, in the same way as we did for the eikonal resummation in Section~\ref{sec:eikonal}. This gives
\begin{align}
i \left(\int_{-\infty}^{+\infty} \mathrm{~d} t^{\prime}\, \mathcal{L}_{\text{int}}(t^{\prime}) \right) \ket{\phi} = (i \hat{M}_3 + i \hat{M}_4 + \dots + i \hat{M}_{n+2}) \ket{\phi} \,,
\end{align}
where the $n$-th term is given by
\begin{align}
\hspace{-10pt} i \hat{M}_{n+2} \ket{\phi} \Big|_{\text{cl}} = \frac{1}{n!} \sum_{\sigma} \int \mathrm{d} \Phi(p)\, \phi_v(p) \left(\int \frac{\mathrm{\hat{d}}^4 l}{l^2 + i \epsilon}\,  \hat{\delta}(2 p \cdot l)\, i \mathcal{M}_3^{(0),\text{cl}}(p,l^{\sigma})\, A^{\dagger}_{\sigma}(l) \right)^n \ket{\phi} \,.
\end{align}
This implies that we can write the final state as a coherent state of virtual gravitons
\begin{align}
\hspace{-12pt}|\psi^{\sigma}_{\text{LR}}\rangle=\frac{1}{\mathcal{N}} \int \mathrm{d} \Phi(p)\, \phi_v(p)\, \exp \left[ \int \frac{\mathrm{\hat{d}}^4 l}{l^2 + i \epsilon} \, \hat{\delta}(2 p \cdot l)\, i \mathcal{M}^{(0),\text{cl}}_{3}(p,l^{\sigma}) A_{\sigma}^{\dagger}(l)\right]|p\rangle \,,
\end{align}
and that the final classical (linearized) metric will be, in de Donder gauge,
\begin{align}
H_{\mu \nu}^{\text{cl}}(x) &= \bra{\psi^{\sigma}_{\text{LR}}}\,\kappa H_{\mu \nu}(x)\, \ket{\psi^{\sigma}_{\text{LR}}} = -4\, G_N M\, P_{\mu \nu \alpha \beta} \,\left(\frac{v^{\alpha} v^{\beta}}{r}\right)\,. 
\end{align}
Clearly, this reproduces the standard form of the linearized Schwarzschild metric in the rest frame where $v^{\mu} = (1,0,0,0)$.

\begin{figure}[h]
\centering
\includegraphics[scale=0.9]{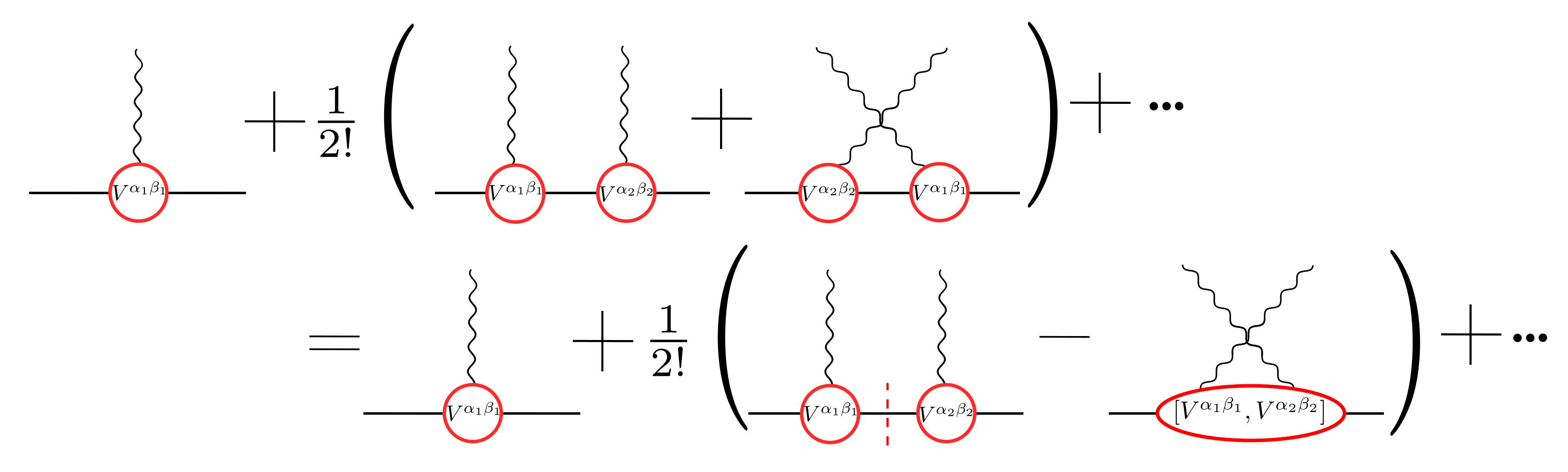}
\caption{The exponentiation of the Kerr metric requires that commutator terms should be suppressed in the classical limit.}
\label{fig:Kerr_exp}
\end{figure}

For the spinning case, we consider as an initial state a massive spinning particle
\begin{align}
\ket{\phi^a} =  \int \mathrm{d} \Phi(p)\, \phi_{v}(p)\, \ket{\{p,a\}} \,,
\end{align}
and follow the same steps as before. While the diagrammatic expansion is the same as in Figure~\ref{fig:coherent}, one again encounters spin commutator terms which arise from the reorganization of the spin-dependent vertex operators, as shown in Figure~\ref{fig:Kerr_exp}.
For the case of Kerr black holes, those commutator terms are suppressed by one power of $\hbar$ in the classical resummation, and therefore we obtain 
\begin{align}
\hspace{-12pt}|\psi^{a,\sigma}_{\text{LR}}\rangle=\frac{1}{\mathcal{N}} \int \mathrm{d} \Phi(p)\, \phi_v(p)\, \exp \left[ \int \frac{\mathrm{\hat{d}}^4 l}{l^2 + i \epsilon} \, \hat{\delta}(2 p \cdot l)\, i \mathcal{M}^{(0),\text{cl}}_{3}(\{p,a\},l^{\sigma})\, A_{\sigma}^{\dagger}(l)\right]|\{p,a\}\rangle \,,
\end{align}
which generalizes the previous discussion to the spinning case. The expectation value of the metric in de Donder gauge is, accordingly, 
\begin{align}
H_{\mu \nu}^{\text{cl}}(x) &= \bra{\psi^{a,\sigma}_{\text{LR}}} \kappa H_{\mu \nu}(x) \ket{\psi^{a,\sigma}_{\text{LR}}} = -4\, G_N M\, \,P_{\mu \nu \alpha \beta} \, \exp\left(a * \partial \right)^{\alpha}{}_{\gamma} \left(\frac{v^{\gamma} v^{\beta}}{r}\right)\,, 
\end{align}
where $(a * \partial)^{\mu}{}_{\nu} \equiv(*(a \wedge \partial))^{\mu}{}_{\nu} =\epsilon^{\mu}{}_{\nu \alpha \beta} a^\alpha \partial^\beta$, consistent with~\cite{Vines:2017hyw}.

% BIBLIOGRAPHY
% use BIBTEX if you want
\bibliographystyle{JHEP-2}
\bibliography{references}

\end{document}